\documentclass[10pt,aps,showpacs,nofootinbib,prd,aps,epsf,floats,
               amsmath,amssymb,amsfonts,axodraw,twocolumn,floatfix]{revtex4}
\usepackage{amsmath, amssymb}
\newcommand{\mathsym}[1]{{}}

\usepackage{graphicx}
\usepackage{amsmath}
\usepackage{amssymb}
\usepackage{bm}
\newcommand{\be}{\begin{equation}}
\newcommand{\ee}{\end{equation}}
\newcommand{\bea}{\begin{eqnarray}}
\newcommand{\eea}{\end{eqnarray}}

\newcommand{\rem}[1]{}
\newsavebox{\PSLASH}
 \sbox{\PSLASH}{$p$\hspace{-1.8mm}/}
 
\renewcommand{\theequation}{\thesection.\arabic{equation}}
\newcounter{saveeqn}
\newcommand{\add}{\addtocounter{equation}{1}}
\newcommand{\alpheqn}{\setcounter{saveeqn}{\value{equation}}%
\setcounter{equation}{0}%
\renewcommand{\theequation}{\mbox{\thesection.\arabic{saveeqn}{\alph{equation}}}}}
\newcommand{\reseteqn}{\setcounter{equation}{\value{saveeqn}}%
\renewcommand{\theequation}{\thesection.\arabic{equation}}}

 \newsavebox{\notrightarrow}
 \sbox{\notrightarrow}{$\to$\hspace{-4mm}/}
 
 \newsavebox{\PARTIALSLASH}
 \sbox{\PARTIALSLASH}{$\partial$\hspace{-1.6mm}/}
 
 \newsavebox{\ASLASH}
 \sbox{\ASLASH}{$A$\hspace{-2.1mm}/}
 
 \newsavebox{\KSLASH}
 \sbox{\KSLASH}{$k$\hspace{-1.8mm}/}
 
 \newsavebox{\LSLASH}
 \sbox{\LSLASH}{$\ell$\hspace{-1.8mm}/}
 
 \newsavebox{\QSLASH}
 \sbox{\QSLASH}{$q$\hspace{-1.8mm}/}
 
 \newsavebox{\DSLASH}
 \sbox{\DSLASH}{$D$\hspace{-2.2mm}/}
 
 \newsavebox{\DbfSLASH}
 \sbox{\DbfSLASH}{${\mathbf D}$\hspace{-2.8mm}/}
 
 \newsavebox{\DELVECRIGHT}
 \sbox{\DELVECRIGHT}{$\stackrel{\rightarrow}{\partial}$}
 
 \newcommand{\blue}{\IfColor{\textCadetBlue}{}}
\newcommand{\black}{\IfColor{\textBlack}{}}
\newcommand{\red}{\IfColor{\textRed}{}}
\newcommand{\green}{\IfColor{\textOliveGreen}{}}
\newcommand{\lila}{\IfColor{\textRedViolet}{}}

\begin{document}
\title{
Contribution of plasminos to the shear viscosity of a \\ hot and
dense Yukawa-Fermi gas}
\author{N. Sadooghi}\email{sadooghi@physics.sharif.ir}
\author{F. Taghinavaz}\email{taghinavaz@physics.sharif.ir}
\affiliation{Department of Physics, Sharif University of Technology,
P.O. Box 11155-9161, Tehran-Iran}
\begin{abstract}
We determine the shear viscosity of a hot and dense Yukawa-Fermi
gas, using the standard Green-Kubo relation, according to which the
shear viscosity is given by the retarded correlator of the traceless
part of viscous energy-momentum tensor. We approximate this retarded
correlator using a one-loop skeleton expansion, and express the
bosonic and fermionic shear viscosities, $\eta_{b}$ and $\eta_{f}$,
in terms of bosonic and fermionic spectral widths, $\Gamma_{b}$ and
$\Gamma_{\pm}$. Here, the subscripts $\pm$ correspond to normal and
collective (plasmino) excitations of fermions. We study, in
particular, the effect of these excitations on thermal properties of
$\eta_{f}[\Gamma_{\pm}]$. To do this, we determine first the
dependence of $\Gamma_{b}$ and $\Gamma_{\pm}$ on momentum $p$,
temperature $T$, chemical potential $\mu$ and $\xi_{0}\equiv
m_{b}^{0}/m_{f}^{0}$, in a one-loop perturbative expansion in the
orders of the Yukawa coupling. Here, $m_{b}^{0}$ and $m_{f}^{0}$ are
$T$ and $\mu$ independent bosonic and fermionic masses,
respectively. We then numerically determine $\eta_{b}[\Gamma_{b}]$
and $\eta_{f}[\Gamma_{\pm}]$, and study their thermal properties. It
turns out that whereas $\Gamma_{b}$ and $\Gamma_{+}$ decrease with
increasing $T$ or $\mu$, $\Gamma_{-}$ increases with increasing $T$
or $\mu$. This behavior qualitatively changes by adding thermal
corrections to $m_{b}^{0}$ and $m_{f}^{0}$, while the difference
between $\Gamma_{+}$ and $\Gamma_{-}$ keeps increasing with
increasing $T$ or $\mu$. Moreover, $\eta_{b}$ ($\eta_{f}$) increases
(decreases) with increasing $T$ or $\mu$. We show that the effect of
plasminos on $\eta_{f}$ becomes negligible with increasing
(decreasing) $T$ ($\mu$).
\end{abstract}
  \maketitle
\section{Introduction}\label{introduction}\label{sec1}
\par\noindent
One of the main goals of the modern experiments of
ultra-relativistic heavy ion collisions is to clarify the nature of
the phase transition of quantum chromodynamics (QCD). As predicted
from numerical computations on the lattice, at a temperature of
about $150$ MeV, quark matter undergoes a phase transition, during
which hadrons melt and a new state of matter, a plasma of quarks and
gluons is built. There are strong evidences for the creation of the
Quark-Gluon Plasma (QGP) in heavy ion experiments at the
Relativistic Heavy-Ion Collider (RHIC) and the Large Hadron Collider
(LHC) \cite{STARcollab}. The experimental results show that the
elliptic flow, $v_{2}$, describing the azimuthal asymmetry in
momentum space, is the largest ever seen in heavy ion collisions
\cite{kapusta2008}. The elliptic flow $v_{2}$ is proportional to the
initial eccentricity $\epsilon_{2}\equiv |\langle
r^{2}e^{2i\phi}\rangle|/\langle r^{2}\rangle$ of a given collision,
that describes the asymmetric region of overlap in a collision
between two nuclei and results in an anisotropy in the transverse
density of the system at the early stages of the collision
\cite{ollitrault2012}. The collective response of the system, well
described by viscous hydrodynamics, transforms this spatial
anisotropy into a momentum anisotropy. Thus, $v_{2}$ is proportional
to  $\epsilon_{2}$, with the proportionality factor depending on the
shear viscosity $\eta$ of the medium \cite{ollitrault2012}. The
latter characterizes the diffusion of momentum transverse to the
direction of propagation. The comparison between the experimentally
measured $v_{2}$ and the results arising from second order viscous
hydrodynamics has suggested that the new state of matter created at
RHIC and LHC is an almost perfect fluid, having a very small shear
viscosity to entropy density ratio $\eta/s$
\cite{heinz2001,romatschke2007} (see also \cite{song2012} for a
recent review on the status of $\eta/s$). However, as is reported in
\cite{song2012}, in all hydrodynamic simulations performed so far,
the shear viscosity is assumed to be temperature independent.
\par
The shear viscosity is one of the transport coefficients, which
describe the properties of a system out of equilibrium, and can
theoretically be determined using two different approaches: The
kinetic theory approach, based on the Boltzmann equation for the
corresponding momentum distribution function \cite{kineticapproach,
redlich2008,ghosh2013}, and the Green-Kubo approach in the framework
of linear response theory \cite{green-kubo}, in which all transport
coefficients are formulated in terms of retarded correlators of the
energy-momentum tensor \cite{jeon1994,kuboapproach}. The advantage of the
second method is, that it provides a framework, where the transport
coefficients can be computed using equilibrium thermal field theory.
Other alternative methods to compute transport coefficients are
direct numerical simulations on a space-time lattice
\cite{aartslattice}, using two-Particle Irreducible (2PI) effective
action \cite{aarts2PI}, and holographic models \cite{kovtun2004}. A
novel diagrammatic method is also presented in \cite{hidaka2010}.
The aim of the most of these computations is to determine the
dependence of $\eta$ on temperature and chemical potential
\cite{nardi2009, lang2012, lang2013} or on external electromagnetic
fields \cite{nam2013}.
\par
In this paper, we use the Green-Kubo formalism to determine the
dependence of the shear viscosity of a Yukawa-Fermi gas on
temperature, chemical potential, and bosonic and fermionic masses.
Thermal corrections to the masses of bosons and fermions will be
considered too and their effect on the shear viscosity will be
scrutinized. Our approach is similar to what is recently presented
by Lang et al. in \cite{lang2012,lang2013}. In \cite{lang2012}, an
appropriate skeleton expansion is used to approximate the retarded
correlators appearing in the Kubo relation for the shear viscosities
of a real $\lambda\varphi^{4}$ theory and an interacting pion gas.
Using the standard K\"allen-Lehmann representation of retarded
two-point Green's function in term of interacting bosonic spectral
function, $\rho_{b}$, the shear viscosity of the scalar and
pseudo-scalar bosons, $\eta_{b}$, is then expressed in terms of the
real and imaginary part of the retarded two-point Green's function.
The latter, denoted by $\Gamma_{b}$, defines, in particular, the
spectral width of the bosons and is inversely proportional to their
mean free-path. To approximate the bosonic correlators, a systematic
Laurent expansion of $\eta_{b}$ in the orders of $\Gamma_{b}$ is
performed. The series is then truncated in the leading
$\Gamma_{b}^{-1}$ order. Computing then $\Gamma_{b}$ perturbatively
in the orders of the small coupling constant of the theory, up to
the first non-vanishing contribution, the $T$ dependence of bosonic
shear viscosity is numerically determined. In \cite{lang2013},
almost the same method is used to determine the fermionic shear
viscosity, $\eta_{f}$, of a strongly interacting quark matter,
described by a two-flavor Nambu--Jona-Lasinio (NJL) model
\cite{NJL}, that consists of a four-fermion interaction with no
gluons involved. To do this, $\eta_{f}$ is first expressed in terms
of fermionic spectral function, $\rho_{f}$, and then working, as in
\cite{iwasaki2006}, in a quasiparticle approximation, a generalized
Breit-Wigner shape for the fermionic spectral function is used to
formulate $\eta_{f}$ in terms of quasiparticle mass $M$ and width
$\Gamma_{f}$. Using then four different parameterizations for
$\Gamma_{f}$, the thermal properties of $\eta_{f}$ is explored.
Eventually, the constant quasiparticle mass $M$ is replaced with $T$
and $\mu$ dependent, dynamically generated constituent quark mass of
the NJL model, and the thermal properties of $\eta_{f}$ are
qualitatively studied in the vicinity of the chiral transition
point.
\par
In the present paper, we will compute the shear viscosity of an
interacting boson-fermion system with Yukawa coupling. In this
theory, the shear viscosity consists of a bosonic and a fermionic
part. Following the method presented in \cite{lang2012}, we will
first derive $\eta_{b}$ in term of $\Gamma_{b}$ in a systematic
Laurent expansion up to ${\cal{O}}(\Gamma_{b}^{0})$. Performing then
a one-loop perturbative expansion in the orders of the Yuwaka
coupling, we will determine $\Gamma_{b}$ as a function of momentum
$p$, temperature $T$, chemical potential $\mu$ and $\xi_{0}\equiv
m_{b}^{0}/m_{f}^{0}$, where $m_{b}^{0}$ and $m_{f}^{0}$ are constant
bosonic and fermionic masses. Using $\eta_{b}[\Gamma_{b}]$, we will
study the $T$ and $\mu$ dependence of the bosonic shear viscosity
for various $\xi_{0}$. We will then add the thermal masses of bosons
and fermions to $m_{b}^{0}$ and $m_{f}^{0}$, and study the effect of
thermal masses on $\Gamma_{b}$ and $\eta_{b}$. Thermal corrections
to the masses of bosons and fermions are computed using standard
Hard Thermal Loop (HTL) method (see e.g. in \cite{kiessig2010}). Let
us notice that, according to the description in \cite{kiessig2009},
this ad-hoc treatment of thermal masses seems intuitive and is
justified, since it equals the HTL treatment with an approximate
fermion propagator. However, it is not equal the full HTL result
\cite{kiessig2010}.
\par
We will then focus on the fermionic part of the shear viscosity, and
derive its dependence on the fermionic spectral width. This build
the central part of the analytical results of the present paper.
Here, in contrast to the approximations made in \cite{lang2013}, we
use the spectral representation of retarded two-point Green's
function presented for the first time in \cite{weldon1989} (see also
\cite{weldon1999}). The latter is used in
\cite{kiessig2010,plasmino-app,blaizot1996,jeon2007,plasmino-NJL,
plasmino-QED,plasmino-QCD,plasmino-yukawa-1,plasmino-yukawa-2,balizot2014}
within the context of Yukawa theory, NJL model, QED and QCD. In
\cite{weldon1989}, it is shown that a fermionic system at finite
temperature has twice as many fermionic modes as at zero
temperature. Besides propagating quark and antiquarks, there are
also propagating quark holes and antiholes. Thus, thermal fermions
have, apart from normal excitation, a collective excitation,
referred to either as a hole or as a plasmino \cite{weldon1999}. The
latter appears as an additional pole in the fermion propagator, and
as a consequence of the preferred frame defined by the heat bath.
Hence, the two poles lead to two different dispersion relations,
both with positive energy. It turns out that in the chiral limit
$m_{f}^{0}\to 0$, the normal excitation has the same chirality and
helicity, while the collective excitation possesses opposite
chirality and helicity \cite{weldon1999}. Denoting the spectral
widths, corresponding to the normal and collective (plasmino)
excitations, with $\Gamma_{+}$ and $\Gamma_{-}$, respectively, we
will use the aforementioned Laurent expansion to derive a novel
analytic relation for $\eta_{f}$ in term of $\Gamma_{\pm}$ up to
${\cal{O}}(\Gamma_{\pm}^{0})$. We will then determine the $p$, $T$,
$\mu$ and $\xi_{0}$ dependence of $\Gamma_{\pm}$ in a one-loop
perturbative expansion in the orders of the Yukawa coupling. Using
$\eta_{f}[\Gamma_{\pm}]$, it is then possible to explore the thermal
properties of $\eta_{f}$ for various $\xi_{0}$. Adding thermal
corrections to the bosonic and fermionic masses, the effect of
thermal masses on $\Gamma_{\pm}$ and $\eta_{f}$ will be also
studied. Let us notice at this stage that in the literature
\cite{nardi2009,jeon2007,plasmino-yukawa-1}, the difference between
$\Gamma_{+}$ and $\Gamma_{-}$, as well as their $p$ dependence are
often neglected, and $\Gamma_{\pm}(p)$ is approximated by
$\Gamma_{\pm}(0)\propto g^{2}T$, where $g$ is the coupling constant
of the theory \cite{nardi2009,plasmino-yukawa-1}. We, however, will
explicitly determine the $p$-dependence of $\Gamma_{+}$ and
$\Gamma_{-}$, and use it in the numerical computation of $\eta_{f}$.
Then, we will assume $\Gamma_{+}= \Gamma_{-}$, and will determine
the difference between $\eta_{f}[\Gamma_{+}\neq\Gamma_{-}]$ and
$\eta_{f}[\Gamma_{+}=\Gamma_{-}]$ in terms of $T$ and $\mu$. It
turns out that, depending on $T$ and/or $\mu$,
$\eta_{f}[\Gamma_{+}=\Gamma_{-}]$ is larger than
$\eta_{f}[\Gamma_{+}\neq\Gamma_{-}]$.
\par
The organization of this paper is a follows: In Sec. \ref{sec2}, we
will review the Green-Kubo formalism, and present the shear
viscosity in terms of retarded correlators of the traceless part of
the viscous energy-momentum tensor. In Sec. \ref{sec3}, we start
with the Lagrangian density of the Yukawa theory, and derive the
bosonic and fermionic contributions to the shear viscosity, in a
one-loop skeleton expansion, in terms of bosonic and fermionic
spectral density functions, $\rho_{b}$ and $\rho_{f}$. Eventually,
using an appropriate Laurent expansion in the orders of bosonic and
fermionic spectral widths, $\eta_{b}[\Gamma_{b}]$ and
$\eta_{f}[\Gamma_{\pm}]$ are determined (see Secs. \ref{sec3a} and
\ref{sec3b} as well as Apps. \ref{appA} and \ref{appC}). In Sec.
\ref{sec4}, the spectral bosonic and fermionic widths, $\Gamma_{b}$
and $\Gamma_{\pm}$ are separately computed in one-loop perturbative
expansion in the orders of the Yukawa coupling (see Sec. \ref{sec4a}
for the bosonic and Sec. \ref{sec4b} for the fermionic spectral
widths). In order to derive the imaginary part of the retarded
two-point Green's functions, corresponding to bosons and fermions,
the standard Schwinger-Keldysh real-time formalism
\cite{schwinger1961} is used. We will mainly use the notations of
\cite{dasbook} and \cite{kobes1985}. In Sec. \ref{sec5}, we will
present our numerical results. Here, the $T,\mu$ and $\xi_{0}$
dependence of $\Gamma_{b}$ and $\Gamma_{\pm}$, as well as the
thermal properties of $\eta_{b}[\Gamma_{b}]$ and
$\eta_{f}[\Gamma_{\pm}]$, will be explored.  As it turns out,
$\Gamma_{b}$ and $\Gamma_{+}$ decreases with increasing $T$ or
$\mu$. In contrast, $\Gamma_{-}$ increases with increasing $T$ or
$\mu$. Whereas this behavior changes when thermal corrections are
added to $m_{b}^{0}$ and $m_{f}^{0}$, $\Gamma_{+}$ and $\Gamma_{-}$
still exhibit different $T$ and $\mu$ dependence. This difference
increases with increasing $T$ or $\mu$. As concerns the shear
viscosities, $\eta_{b}$ ($\eta_{f}$) increases (decreases) with
increasing $T$ or $\mu$. Moreover, it turns out that the
contribution of plasminos on $\eta_{f}$ becomes negligible with
increasing (decreasing) $T$ ($\mu$). A summary of our results is
presented in Sec. \ref{sec6}.
\setcounter{equation}{0}
\section{Shear Viscosity in Relativistic Hydrodynamics}\label{sec2}
\noindent An ideal and locally equilibrated relativistic fluid is
mainly described by the dynamics of the corresponding
energy-momentum tensor
\begin{eqnarray}\label{T1a}
T^{\mu\nu}_{0}=\epsilon\hspace{0.02cm}
u^{\mu}u^{\nu}+P\hspace{0.02cm}\Delta^{\mu\nu},
\end{eqnarray}
where $\epsilon$ is the energy density, $P$ the pressure and
$u_{\mu}(x)=\gamma(x)(1,\mathbf{v}(x))$ is the four velocity of the
fluid, which is defined by the variation of the four-coordinate
$x^{\mu}$ with respect to the proper time $\tau$. Here, the Lorentz
factor $\gamma(x)\equiv (1-\mathbf{v}^{2}(x))^{-1}$. In (\ref{T1a}),
$\Delta^{\mu\nu}$ is defined by $\Delta^{\mu\nu}\equiv
g^{\mu\nu}-u^{\mu}u^{\nu}$, with the metric
$g^{\mu\nu}=\mbox{diag}\left(+,-,-,-\right)$. It satisfies
$u_{\mu}\Delta^{\mu\nu}=0$. Moreover, for the four-velocity
$u_{\mu}$, we have $u_{\mu}u^{\mu}=1$. If there are no external
sources, the energy-momentum tensor (\ref{T1a}) is conserved
\begin{eqnarray}\label{T2a}
\partial_{\mu}T^{\mu\nu}_{0}=0.
\end{eqnarray}
Apart from (\ref{T2a}), an ideal fluid is characterized by the
entropy current conservation law $\partial_{\mu}s^{\mu}=0$, where
the entropy current, $s_{\mu}\equiv s u_{\mu}$, includes the entropy
density $s$. In a system without conserved charges, $\epsilon$ and
$P$ satisfy $\epsilon+P=Ts$, where $T$ is the local temperature of
the fluid.
\par
To include dissipative effects to the fluid, the viscous-stress
tensor $\tau^{\mu\nu}$ is to be added to $T_{0}^{\mu\nu}$ from
(\ref{T1a}). The total energy-momentum tensor then reads
\begin{eqnarray}\label{T3a}
T^{\mu\nu}=T^{\mu\nu}_{0}+\tau^{\mu\nu},
\end{eqnarray}
where $\tau^{\mu\nu}$ satisfies $u_{\mu}\tau^{\mu\nu}=0$. In an
expansion in the orders of derivatives of $u_{\mu}$, the viscous
stress tensor is determined using the second law of thermodynamics,
$T\partial_{\mu}s^{\mu}\geq 0$, that replaces the conservation law
$\partial_{\mu}s^{\mu}=0$ of the ideal fluid. The viscous stress
tensor is often split as
\begin{eqnarray}\label{T4a}
\tau^{\mu\nu}=\pi^{\mu\nu}+\Delta^{\mu\nu}\Pi,
\end{eqnarray}
where $\pi^{\mu\nu}$ is the traceless part ($\pi^{\mu}_{~\mu}=0$)
and $\Pi$ is the remaining part with non-vanishing trace. Each part
of $\tau^{\mu\nu}$ is then parameterized by a number of viscous
coefficients. In the first order derivative expansion,
$\tau^{\mu\nu}$ is characterized by the shear and bulk viscosities,
$\eta$ and $\zeta$, that appear in the traceless part of
$\tau^{\mu\nu}$,
\begin{eqnarray}\label{T5a}
\pi^{\mu\nu}=\eta
\left(\nabla^{\mu}u^{\nu}+\nabla^{\nu}u^{\mu}-\frac{2}{3}\Delta^{\mu\nu}\nabla^{\rho}u_{\rho}\right),
\end{eqnarray}
and in the part of $\tau^{\mu\nu}$ with non-vanishing trace,
\begin{eqnarray}\label{T6a}
\Pi=\zeta\nabla^{\mu}u_{\mu},
\end{eqnarray}
respectively. Here, $\nabla^{\mu}\equiv
\Delta^{\mu\nu}\partial_{\nu}$. Using the properties of
$\Delta^{\mu\nu}$ in $d=4$ dimensional space-time,
$\Delta^{\mu\nu}u_{\nu}=0$ as well as
$\Delta^{\rho}_{\mu}\Delta^{\mu}_{~\nu}=\Delta^{\rho}_{~\nu}$, we
get
\begin{eqnarray}\label{T7a}
\epsilon=u_{\mu}u_{\nu}T^{\mu\nu},\qquad
P=-\frac{1}{3}\Delta_{\mu\nu}T^{\mu\nu},
\end{eqnarray}
as well as
\begin{eqnarray}\label{T8a}
\hspace{-0.3cm}\tilde{\pi}^{\mu\nu}=\left(\Delta^{\rho\mu}\Delta^{\sigma\nu}+\Delta^{\rho\nu}\Delta^{\sigma\mu}-\frac{2}{3}\Delta^{\mu\nu}\Delta^{\rho\sigma}\right)T_{\rho\sigma}.
\end{eqnarray}
Here,  $\tilde{\pi}^{\mu\nu}\equiv \eta^{-1}\pi^{\mu\nu}$ is
introduced. In the rest of this paper, we will focus on the shear
viscosity $\eta$. Following Zubarev's approach \cite{green-kubo} and
within linear response theory, it is determined by the Kubo-type
formula \cite{lang2012}
\begin{eqnarray}\label{T9a}
\eta=\frac{\beta_{s}}{10}\int d^{3}x'\int_{-\infty}^{t}dt'
\left(\tilde{\pi}^{\mu\nu}(0),\tilde{\pi}_{\mu\nu}(\mathbf{x}',t')\right),
\end{eqnarray}
where the inverse proper temperature $\beta_{s}\equiv \gamma \beta$
with $\beta\equiv T^{-1}$, and
\begin{eqnarray}\label{T10a}
\hspace{-0.3cm}\left(X,Y\right)=\frac{1}{\beta}\int_{0}^{\beta}d\tau\langle
X[~e^{H\tau} Y e^{-H\tau}-\langle Y\rangle_{0}~]\rangle_{0}.
\end{eqnarray}
Here, $H$ is the free part of the Hamiltonian of a fully interacting
theory, which is given in terms of the energy momentum tensor
$T^{\mu\nu}$, via $\beta H=\int d^{3}x
\beta_{s}(\mathbf{x},\tau)u^{\mu}(\mathbf{x},\tau)T_{0\mu}(\mathbf{x},\tau)$.
Moreover, $\langle \cdots\rangle_{0}$ is the thermal expectation
value with respect to the equilibrium statistical operator
$\rho_{0}$, and is defined by $\langle
\cdot\rangle_{0}=\mbox{tr}(\cdot \rho_{0})$ \cite{lang2012}. The
correlator appearing in (\ref{T9a}) can be expressed as a real-time
integral over a retarded correlator
\begin{eqnarray}\label{T11a}
\hspace{-0.3cm}\left(X(t),Y(t')\right)\sim
-\frac{1}{\beta}\int_{-\infty}^{t'}dt''\langle
X(t),Y(t'')\rangle_{R},
\end{eqnarray}
with
\begin{eqnarray}\label{T12a}
\hspace{-0.3cm}\langle
X(t),Y(t')\rangle_{R}=-i\theta(t-t')\langle\big[X(t),Y(t')\big]\rangle_{0}.
\end{eqnarray}
In the large-time limit $t'\to \infty$, when the system approaches
global equilibrium, the approximation appearing in (\ref{T11a})
becomes exact. Combining at this stage (\ref{T9a}) and (\ref{T11a}),
and evaluating the resulting expression in the local rest-frame,
where $\beta_{s}=\beta$, the Kubo-formula for the shear viscosity
reads
\begin{eqnarray}\label{T13a}
\eta=-\frac{1}{10}\int_{-\infty}^{0}dt\int_{-\infty}^{t}dt'~
\Pi_{R}(t'),
\end{eqnarray}
with retarded Green's function
\begin{eqnarray}\label{T14a}
\hspace{-0.5cm}\Pi_{R}(t)\equiv -i\theta(-t)\int
d^{3}x~\langle[\tilde{\pi}^{\mu\nu}(0),\tilde{\pi}_{\mu\nu}(\mathbf{x},t)]\rangle_{0},
\end{eqnarray}
and $\tilde{\pi}^{\mu\nu}$ given in (\ref{T8a}). Equivalently $\eta$
is given by
\begin{eqnarray}\label{T15a}
\eta=\frac{i}{10}\frac{d}{dp_{0}}\Pi_{R}(p_{0})\bigg|_{p_{0}=0}.
\end{eqnarray}
It arises by replacing the Fourier transformation of
$\Pi_{R}(t)=\int\frac{dp_{0}}{2\pi}e^{-ip_{0}t}\Pi_{R}(p_{0})$ in
(\ref{T13a}), and integrating over $t$ and $t'$ using the functional
identity \cite{lang2012}
\begin{eqnarray}\label{T16a}
\int_{-\infty}^{0}dt'\int_{t}^{0}dt~e^{-ip_{0}t'}\to -2\pi
i\delta(p_{0})\frac{d}{dp_{0}}.
\end{eqnarray}
It is the purpose of this paper to determine the thermal properties
of the shear viscosity of a Yukawa theory by computing $\Pi_{R}$
from (\ref{T14a}) in a weak coupling expansion in the orders of the
Yukawa coupling. To this purpose, we will first introduce a Yukawa
theory including a real scalar and a fermionic field, and then,
using an appropriate weak coupling expansion up to one-loop level,
we will determine $\eta$ for these fields separately.
\section{Shear viscosity of a Yukawa Theory: General Considerations}\label{sec3}
\setcounter{equation}{0}\noindent In this section, we will first
review the method presented in \cite{lang2012}, and determine the
bosonic part of the shear viscosity of a Yukawa theory in terms of
the bosonic spectral width. We will then use this method as a
guideline, and derive the fermionic part of the shear viscosity of
the Yukawa theory in terms of fermionic spectral widths. Here, we
will explicitly consider the contributions of the normal and
collective (plasmino) excitations of fermions, with different
spectral widths. This is in contrast with the result recently
presented in \cite{lang2013}, where within a quasi-particle
approximation, a Breit-Wigner type formula is presented for the
fermionic shear viscosity in terms of one and the same fermionic
spectral width.
\par
\begin{figure}[hbt]
\includegraphics[width=8cm,height=1.5cm]{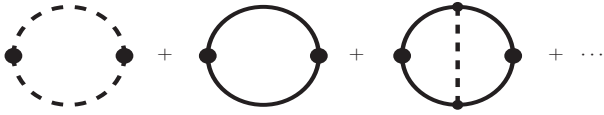}
\caption{The skeleton expansion of $\Pi_{T}(\tau)$ from (\ref{A3b}).
Dashed and solid lines denote the dressed bosonic and fermionic
two-point function $G_{T}(\mathbf{p},\tau)$ from (\ref{A10b}) and
$S_{T}(\mathbf{p},\tau)$ from (\ref{A11b}), respectively. In our
computation up to one-loop skeleton expansion, only the first two
diagrams in the above series are considered.}\label{fig1}
\end{figure}

Let us start with the Lagrangian density of a Yukawa theory
\begin{eqnarray}\label{A1b}
{\cal{L}}=\bar{\psi}\left(i\gamma\cdot
\partial-m_{f}\right)\psi+\frac{1}{2}\partial_{\mu}\varphi\partial^{\mu}\varphi-\frac{1}{2}m_{b}^{2}\varphi^{2}+g\bar{\psi}\psi\varphi,\hspace{-0.3cm}\nonumber\\
\end{eqnarray}
where, $\varphi$ is a real scalar field and $\bar{\psi},\psi$ are
fermionic fields. Moreover, $m_{b}$ and $m_{f}$ corresponds to the
masses of bosons and fermions, respectively. According to
(\ref{T13a}), the shear viscosity $\eta$ for this theory is given by
a two-point Green's function of the tensor field
$\tilde{\pi}^{\mu\nu}$, which is defined in (\ref{T8a}) in terms of
the energy-momentum tensor $T_{\mu\nu}$. The energy-momentum tensor
of the Yukawa theory is given by
\begin{eqnarray}\label{A2b}
T_{\mu\nu}=i\bar{\psi}\gamma_{\mu}\partial_{\nu}\psi+\partial_{\mu}\varphi
\partial_{\nu}\varphi-{\cal{L}}g_{\mu\nu},
\end{eqnarray}
where ${\cal{L}}$ is given in (\ref{A1b}). As it turns out,
$T_{\mu\nu}$, and consequently the shear viscosity include a bosonic
and a fermionic part. In what follows, we will denoted them by
$\eta_{b}$ and $\eta_{f}$, where the subscripts correspond to bosons
($b$) and fermions ($f$), respectively. To compute these two parts
separately, we will use (\ref{T15a}). Introducing the imaginary time
$\tau\equiv it$ in (\ref{T14a}), the thermal Green's function,
$\Pi_{T}(\tau)$, reads
\begin{eqnarray}\label{A3b}
\hspace{-0.3cm}\Pi_{T}(\tau)\equiv \int d^{3}x~\langle
{\cal{T}}_{\tau}[\tilde{\pi}^{\mu\nu}(0)\tilde{\pi}_{\mu\nu}(\mathbf{x},\tau)]\rangle_{0},
\end{eqnarray}
where ${\cal{T}}_{\tau}$ stands for the time-ordering prescription.
According to the above descriptions, it is given by
\begin{eqnarray}\label{A4b}
\Pi_{T}(\tau)=\Pi_{T}^{b}(\tau)+\Pi_{T}^{f}(\tau),
\end{eqnarray}
with the bosonic part
\begin{eqnarray}\label{A5b}
\lefteqn{\hspace{-0.5cm}\Pi_{T}^{b}(\tau)=2\int d^{3}x~
\eta^{\alpha\beta\rho\sigma}
}\nonumber\\
&&\times \langle\partial_{\beta}\varphi(0)\partial_{\rho}\varphi(0)
\partial_{\alpha}\varphi(\mathbf{x},\tau)\partial_{\sigma}\varphi(\mathbf{x},\tau)
\rangle_{0},
\end{eqnarray}
and the fermionic part
\begin{eqnarray}\label{A6b}
\lefteqn{\hspace{-0.5cm}\Pi_{T}^{f}(\tau)=-2\int d^{3}x~
\eta^{\alpha\beta\rho\sigma}
}\nonumber\\
&&\times\langle
\bar{\psi}(0)\gamma_{\beta}\partial_{\rho}\psi(0)\bar{\psi}(\mathbf{x},\tau)\gamma_{\alpha}\partial_{\sigma}\psi(\mathbf{x},\tau)\rangle_{0}.
\end{eqnarray}
In the above relations, $\eta^{\alpha\beta\rho\sigma}$ is defined by
\begin{eqnarray}\label{A7b}
\eta^{\alpha\beta\rho\sigma}\equiv
\Delta^{\alpha\beta}\Delta^{\rho\sigma}+\Delta^{\beta\sigma}\Delta^{\rho\alpha}-\frac{2}{3}\Delta^{\alpha\sigma}\Delta^{\beta\rho}.
\end{eqnarray}
Performing a Fourier transformation into the momentum space, using
$\tilde\varphi(\mathbf{p},\tau)=\int d^{3}x~
e^{i\mathbf{p}\cdot\mathbf{x}}\varphi(\mathbf{x},\tau)$ and
$\tilde{\psi}(\mathbf{p},\tau)=\int d^{3}x~
e^{i\mathbf{p}\cdot\mathbf{x}}\psi(\mathbf{x},\tau)$, evaluating the
resulting four-point functions arising in (\ref{A5b}) and
(\ref{A6b}) using an appropriate expansion up to one-loop skeleton
expansion, as is described in \cite{lang2012}, and eventually
neglecting the disconnected parts of the Green's functions, the
bosonic part of $\Pi_{T}(\tau)$ reads
\begin{eqnarray}\label{A8b}
\lefteqn{\hspace{-0.5cm}\Pi_{T}^{b}(\omega_{n})=4\int_{0}^{\beta}d\tau e^{i\omega_{n}\tau}}\nonumber\\
&&\times \int \frac{d^{3}p}{(2\pi)^{3}}\eta^{\alpha\beta\rho\sigma}
p_{\alpha}p_{\beta}p_{\rho}p_{\sigma} G_{T}^{2}(\mathbf{p},\tau),
\end{eqnarray}
and the fermionic part of $\Pi_{T}(\tau)$ is given by
\begin{eqnarray}\label{A9b}
\Pi_{T}^{f}(\omega_{n})&=&2\int_{0}^{\beta}d\tau
e^{i\omega_{n}\tau}\int
\frac{d^{3}p}{(2\pi)^{3}}\eta^{\alpha\beta\rho\sigma}p_{\rho}p_{\sigma} \nonumber\\
&&\hspace{-0.3cm}\times \mbox{tr}
\big[S_{T}(\mathbf{p},\tau)\gamma_{\alpha}S_{T}({\mathbf{p}},-\tau)\gamma_{\beta}\big].
\end{eqnarray}
Let us notice that in the above relations $G_{T}(\mathbf{p},\tau)$
and $S_{T}(\mathbf{p},\tau)$ are exact (dressed) bosonic and
fermionic two-point functions, respectively. They are defined by
\begin{eqnarray}\label{A10b}
G_{T}(\mathbf{p},\tau)\equiv V^{-1}\langle
T_{\tau}[\tilde{\varphi}(0)\tilde{\varphi}(\mathbf{p},\tau)]\rangle_{0},
\end{eqnarray}
and
\begin{eqnarray}\label{A11b}
S_{T}(\mathbf{p},\tau)\equiv V^{-1}\langle
T_{\tau}[\tilde{\psi}(0)\tilde{\bar{\psi}}(\mathbf{p},\tau)]\rangle_{0}.
\end{eqnarray}
Moreover, in (\ref{A8b}) and (\ref{A9b}), the bosonic and fermionic
Matsubara frequencies are given by $\omega_{n}=2n\pi T$ and
$\omega_{n}=(2n+1)\pi T$, respectively. As aforementioned, the
expressions presented in (\ref{A8b}) and (\ref{A9b}) are the
one-loop contributions in the skeleton expansion. The latter is
diagrammatically presented in Fig. \ref{fig1}. In what follows, we
will separately evaluate the bosonic and fermionic thermal two-point
functions (\ref{A8b}) and (\ref{A9b}). The results will be then used
to determine the bosonic and fermionic parts of the shear viscosity
$\eta$ in term of bosonic and fermionic spectral widths.
\par
\subsection{The bosonic contribution to $\eta$ in the one-loop skeleton expansion}\label{sec3a}
\noindent To evaluate the bosonic part of the shear viscosity
$\eta_{b}$, we will use the method described in \cite{lang2012},
whose main steps will be reviewed in what follows.
\par
Let us first consider (\ref{A8b}). According to the standard
K\"allen-Lehmann representation, the two-point Green's function
$G_{T}(\mathbf{p},\omega_{n})$ is given in terms of bosonic spectral
density function $\rho_{b}(\mathbf{p},\omega)$ as
\begin{eqnarray}\label{A12b}
G_{T}({\mathbf{p}},\omega_{n})=\frac{1}{2\pi}\int_{-\infty}^{+\infty}d\omega~\frac{\rho_{b}(\mathbf{p},\omega)}{\omega+i\omega_{n}}.
\end{eqnarray}
Plugging this relation in
\begin{eqnarray}\label{A13b}
G_{T}(\mathbf{p},\tau)=\sum_{n=-\infty}^{+\infty}e^{-i\omega_{n}\tau}G_{T}(\mathbf{p},\omega_{n}),
\end{eqnarray}
and adding over bosonic Matsubara frequencies $\omega_{n}=2n\pi T$,
we arrive at
\begin{eqnarray}\label{A14b}
\lefteqn{\hspace{-0.5cm}G_{T}({\mathbf{p}},\tau)}\nonumber\\
&&\hspace{-0.7cm}=\frac{1}{2\pi}\int_{-\infty}^{+\infty}
d\omega~e^{-\omega
|\tau|}\rho_{b}({\mathbf{p}},\omega)[1+n_{b}(\omega)],
\end{eqnarray}
where, the bosonic distribution function $n_{b}(\omega)$ reads
\begin{eqnarray}\label{A15b}
n_{b}(\omega)\equiv \frac{1}{e^{\beta\omega}-1}.
\end{eqnarray}
To derive (\ref{A14b}), we have used the symmetry property
$\rho_{b}(\mathbf{p},-\omega)=-\rho_{b}(\mathbf{p},\omega)$, that
yields, in particular, $|\tau|$ on the right hand side (r.h.s.) of
(\ref{A14b}). Plugging further $G_{T}(\mathbf{p},\tau)$ from
(\ref{A14b}) in (\ref{A8b}), and integrating over $\tau$, we arrive
after analytical continuation, $i\omega_{n}\to p_{0}+i\epsilon$, at
\begin{eqnarray}\label{A16b}
\lefteqn{\hspace{-0.5cm}\Pi_{R}^{b}(p_{0})=4\int\frac{d^{3}p}{(2\pi)^{3}}\eta^{\alpha\beta\rho\sigma}p_{\alpha}p_{\beta}p_{\rho}p_{\sigma}
}\nonumber\\
&&\times\int_{-\infty}^{+\infty}\frac{d\omega_{1}d\omega_{2}}{(2\pi)^{2}}\rho_{b}({\mathbf{p}},\omega_{1})
\rho_{b}({\mathbf{p}},\omega_{2})\nonumber\\
&&\times~
n_{b}(\omega_{1})n_{b}(\omega_{2})W_{\epsilon}(\omega_{12},p_{0}),
\end{eqnarray}
where $\eta^{\alpha\beta\rho\sigma}$ is defined in (\ref{A7b}),
$\omega_{12}\equiv \omega_{1}+\omega_{2}$, and
$W_{\epsilon}(\omega_{12},p_{0})$ is given by
\begin{eqnarray}\label{A17b}
\hspace{-0.5cm}W_{\epsilon}(\omega_{12},p_{0})\equiv
\frac{1}{p_{0}+i\epsilon-\omega_{12}}-\frac{1}{p_{0}+i\epsilon+\omega_{12}}.
\end{eqnarray}
At this stage, we use the definition of the bosonic spectral density
function $\rho_{b}$ in terms of retarded two-point Green's function,
$G_{R}(p)$,
\begin{eqnarray}\label{A18b}
\rho_{b}(p)\equiv -2~\mathfrak{Im}[G_{R}(p)],
\end{eqnarray}
to formulate $\rho_{b}$ in terms of the bosonic renormalized energy
\begin{eqnarray}\label{A19b}
E_{b}(p)\equiv
\sqrt{\omega_{b}^{2}+\mathfrak{Re}[\Sigma_{R}^{b}(p)]},
\end{eqnarray}
with $\omega_{b}^{2}\equiv \mathbf{p}^{2}+m_{b}^{2}$, and the
bosonic spectral width
\begin{eqnarray}\label{A20b}
\Gamma_{b}(p)\equiv
-\frac{1}{2p_{0}}\mathfrak{Im}[\Sigma_{R}^{b}(p)].
\end{eqnarray}
Using
\begin{eqnarray}\label{A21b}
G_{R}^{-1}(p)&=&p^{2}-m_{b}^{2}-\Sigma_{R}^{b}(p)\nonumber\\
&\simeq& [p_{0}+i\Gamma_{b}(p)]^{2}-E_{b}^{2}(p),
\end{eqnarray}
the bosonic spectral density function (\ref{A18b}) is given by
\begin{eqnarray}\label{A22b}
\lefteqn{\rho_{b}(\mathbf{p},\omega)
}\nonumber\\
&=&\frac{4\omega\Gamma_{b}(\mathbf{p},\omega_{b})}{[\omega^{2}-
E_{b}^{2}(\mathbf{p},\omega_{b})-\Gamma_{b}^{2}(\mathbf{p},\omega_{b})]^{2}+4\omega^{2}\Gamma_{b}^{2}(\mathbf{p},\omega_{b})},\nonumber\\
\nonumber\\
\end{eqnarray}
where $E_{b}=E_{b}(\mathbf{p},\omega_{b})$ and
$\Gamma_{b}=\Gamma_{b}(\mathbf{p},\omega_{b})$ are to be evaluated
on mass-shell. Plugging now $\rho_{b}(\mathbf{p},\omega)$ from
(\ref{A22b}) in (\ref{A16b}), and using \cite{lang2012}
\begin{eqnarray}\label{A23b}
\frac{i}{10}\frac{d}{dp_{0}}W_{\epsilon}(\omega_{12},p_{0})=-\frac{\pi}{5}\delta'(\omega_{12}),
\end{eqnarray}
we arrive first at
\begin{eqnarray}\label{A24b}
\eta_{b}=\frac{4\beta}{5\pi}\int\frac{d^{3}p}{(2\pi)^{3}}\eta^{\alpha\beta\rho\sigma}p_{\alpha}p_{\beta}p_{\rho}p_{\sigma}\int_{-\infty}^{+\infty}d\omega
F_{b}(\mathbf{p},\omega),\nonumber\\
\end{eqnarray}
with $\omega\equiv \frac{1}{2}\bar{\omega}_{12}\equiv
\frac{1}{2}(\omega_{1}-\omega_{2})$ and $F_{b}(\mathbf{p},\omega)$
given by
\begin{eqnarray}\label{A25b}
\lefteqn{F_{b}(\mathbf{p},\omega)
}\nonumber\\
&=&\frac{2\omega^{2}e^{\beta\omega}}{(e^{\beta\omega}-1)^{2}}\frac{\Gamma_{b}^{2}}{[E_{b}^{2}-(\omega-i\Gamma_{b})^{2}]^{2}[E_{b}^{2}-(\omega+i\Gamma_{b})^{2}]^{2}}.\nonumber\\
\end{eqnarray}
Plugging further (\ref{A25b}) in (\ref{A24b}) and integrating over
$\omega$, using of the same procedure as in \cite{lang2012}, and
which will be described below, we arrive at the bosonic part of the
shear viscosity of the Yukawa theory in terms of the renormalized
energy $E_{b}$ from (\ref{A19b}) and the bosonic spectral width
$\Gamma_{b}$ from (\ref{A20b}),
\begin{eqnarray}\label{A26b}
\eta_{b}=\frac{\beta}{30\pi^{2}}\int_{0}^{\infty}dp~
\frac{\mathbf{p}^{6}}{E_{b}^{2}}~\frac{e^{\beta E_{b}}}{\left(e^{\beta E_{b}}-1\right)^{2}}\frac{1}{\Gamma_{b}}+{\cal{O}}(\Gamma_{b}^{0}).\nonumber\\
\end{eqnarray}
To derive (\ref{A26b}), the pole structure of
$F_{b}(\mathbf{p},\omega)$ from (\ref{A25b}) is to be considered.
Following \cite{lang2012}, the integral over $\omega$ in
(\ref{A24b}) is to be performed by closing the contour in the upper
half-plane, i.e. by considering only two poles $\omega^{\pm}\equiv
\pm E_{b}+i\Gamma_{b}$ from four poles $\omega^{\pm}$ and
$-\omega^{\pm}$, and eventually expanding the resulting analytical
expression in the orders of small $\Gamma_{b}$. This results in
\begin{eqnarray}\label{A27b}
2\pi i\sum_{\omega=\omega_{b}^{\pm}}F_{b}(\omega)=\frac{e^{\beta
E_{b}}}{(e^{\beta
E_{b}}-1)^{2}}\frac{\pi}{16E_{b}^{2}\Gamma_{b}}+{\cal{O}}(\Gamma_{b}^{0}).\nonumber\\
\end{eqnarray}
Let us  notice that apart from the aforementioned poles
$\omega^{\pm}$ and $-\omega^{\pm}$, there are also an infinite
number of poles arising from the denominator $e^{\beta\omega}-1$ in
(\ref{A25b}). But as it is shown in \cite{lang2012}, the
contributions of their residues are proportional to
$\Gamma_{b}^{2}$, and, if we assume that $\Gamma_{b}$ is small
enough, they are suppressed relative to the leading
$\Gamma_{b}^{-1}$ term in (\ref{A27b}). Plugging therefore
(\ref{A27b}) in (\ref{A24b}), and considering the local rest frame
of the fluid, we arrive at the bosonic part of the shear viscosity
from (\ref{A26b}). To perform the $p$-integration in (\ref{A26b})
and study eventually the $T$-dependence of $\eta_{b}$, the $p$ and
$T$-dependence of $E_{b}$ and $\Gamma_{b}$ are to be determined
perturbatively in an appropriate loop-expansion in the orders of the
Yukawa coupling. In this paper, we will approximate $E_{b}\simeq
\omega_{b}$ and will determine in Sec. \ref{sec4} only $\Gamma_{b}$
at one-loop level. The result will eventually used to determine the
$T$ dependence of $\eta_{b}$. As concerns the $\mu$ dependence of
$\eta_{b}$, we will use the same relation (\ref{A26b}). In this
case, the $\mu$ dependence of $\eta_{b}$ arises only from
$\Gamma_{b}$ on the r.h.s. of (\ref{A26b}).
\subsection{The fermionic contribution to $\eta$ in the one-loop skeleton expansion}\label{sec3b}
\noindent To determine the fermionic part of the shear viscosity
$\eta_{f}$, we will follow the same steps as in the previous
section, and will present $\eta_{f}$ in terms of fermionic spectral
widths $\Gamma_{\pm}$, corresponding to normal and collective
excitations of fermions. The resulting expression builds the central
analytical result of the present paper.
\par
To start, let us first consider (\ref{A9b}). Using the standard
K\"allen-Lehmann representation, the fermionic two-point Green's
function $S_{T}(\mathbf{p},\omega_{n})$ can be given in terms of
fermionic spectral density function $\rho_{f}(\mathbf{p},\omega)$ as
\begin{eqnarray}\label{A28b}
S_{T}({\mathbf{p}},\omega_{n})&=&\frac{1}{2\pi}\int_{-\infty}^{+\infty}d\omega~\frac{\rho_{f}(\mathbf{p},\omega)}{\omega+i\omega_{n}}.
\end{eqnarray}
Plugging this relation in
\begin{eqnarray}\label{A29b}
S_{T}({\mathbf{p}},\tau)=T\sum_{n=-\infty}^{+\infty}e^{-i\omega_{n}\tau}S_{T}({\mathbf{p}},\omega_{n}),
\end{eqnarray}
and adding over fermionic Matsubara frequencies
$\omega_{n}=(2n+1)\pi T$, we arrive at
\begin{eqnarray}\label{A30b}
\lefteqn{\hspace{-0.5cm}S_{T}({\mathbf{p}},\tau)=\frac{1}{2\pi}\int_{-\infty}^{+\infty}
d\omega~e^{-\omega \tau}\rho_{f}({\mathbf{p}},\omega)
}\nonumber\\
&&\hspace{-0.5cm}\times
\big[\theta(\tau)(1-n_{f}(\omega))-\theta(-\tau)n_{f}(\omega)\big],
\end{eqnarray}
with the fermionic distribution function
\begin{eqnarray}\label{A31b}
n_{f}(\omega)=\frac{1}{e^{\beta\omega}+1}.
\end{eqnarray}
Plugging $S_{T}(\mathbf{p},\tau)$ from (\ref{A30b}) in (\ref{A9b}),
and integrating over $\tau$, using
\begin{eqnarray}\label{A32b}
\lefteqn{\int_{0}^{\beta}d\tau
e^{(i\omega_{n}-\omega_{1}+\omega_{2})\tau}
}\nonumber\\
&&\times
[\theta(\tau)(1-n_{f}(\omega_{1}))-\theta(-\tau)n_{f}(\omega_{1})]\nonumber\\
&&\times[\theta(-\tau)(1-n_{f}(\omega_{2}))-\theta(\tau)n_{f}(\omega_{2})]\nonumber\\
&&=\frac{(1-n_{f}(\omega_{1}))n_{f}(\omega_{2})-(1-n_{f}(-\omega_{1}))n_{f}(-\omega_{2})}{i\omega_{n}-\omega_{1}+\omega_{2}},\nonumber\\
\end{eqnarray}
we arrive first at
\begin{eqnarray}\label{A33b}
\lefteqn{\hspace{-0.5cm}\Pi_{T}^{f}(\omega_{n})=\frac{1}{2\pi^{2}}\int\frac{d^{3}p}{(2\pi)^{3}}\eta^{\alpha\beta\rho\sigma}p_{\rho}p_{\sigma}
}\nonumber\\
&&\hspace{-0.5cm}\times\int
d\omega_{1}d\omega_{2}(1-n_{f}(\omega_{1}))n_{f}(\omega_{2})\nonumber\\
&&\hspace{-0.5cm}\times \bigg\{
\frac{\mbox{tr}\left(\rho_{f}(\omega_{1},\mathbf{p})\gamma_{\alpha}\rho_{f}(\omega_{2},\mathbf{p})\gamma_{\beta}\right)}{i\omega_{n}-\omega_{1}+\omega_{2}}\nonumber\\
&&\hspace{-0.5cm}-
\frac{\mbox{tr}\left(\rho_{f}(-\omega_{1},-\mathbf{p})\gamma_{\alpha}\rho_{f}(-\omega_{2},-\mathbf{p})\gamma_{\beta}\right)}{i\omega_{n}+\omega_{1}-\omega_{2}}\bigg\},
\end{eqnarray}
where $\eta^{\alpha\beta\rho\sigma}$ is defined below (\ref{A6b}).
To evaluate $\Pi_{T}(\omega_{n})$, let us use at this stage, in
analogy to the bosonic case, the definition of the fermionic
spectral density function $\rho_{f}$ in term of the retarded
two-point Green's function $S_{R}$,
\begin{eqnarray}\label{A34b}
\rho_{f}(p)=-2~\mathfrak{Im}[S_{R}(p)],
\end{eqnarray}
and the decomposition of $S_{R}(p)$ in term the fermion self-energy
$\Sigma_{R}^{f}$,
\begin{eqnarray}\label{A35b}
S_{R}^{-1}(p)=\gamma\cdot p-m_{f}+\Sigma_{R}^{f}(p).
\end{eqnarray}
Using the method, described in details in App. A, the spectral
density function of fermions is given by
\begin{eqnarray}\label{A36b}
\lefteqn{\hspace{-0.8cm}\rho_{f}(\mathbf{p},\omega)=\frac{2\Gamma_{+}(\mathbf{p},\omega_{f})}{[\omega-E_{+}(\mathbf{p},\omega_{f}]^{2}+\Gamma_{+}^{2}(\mathbf{p},\omega_{f})}\hat{g}_{+}(\mathbf{p},\omega_{f})
}\nonumber\\
&&\hspace{-1cm}-\frac{2\Gamma_{-}(\mathbf{p},\omega_{f})}{[\omega+E_{-}(\mathbf{p},\omega_{f})]^{2}+\Gamma_{-}^{2}(\mathbf{p},\omega_{f})}\hat{g}_{-}(\mathbf{p},\omega_{f}),
\end{eqnarray}
where $\omega_{f}^{2}=\mathbf{p}^{2}+m_{f}^{2}$, and
\begin{eqnarray}\label{A37b}
\hat{g}_{\pm}(\mathbf{p},\omega_{f})=\frac{1}{2\omega_{f}}\big[\gamma_{0}\omega_{f}\mp\left(\gamma.\mathbf{p}-m_{f}\right)\big].
\end{eqnarray}
In (\ref{A36b}), $E_{\pm}$ and $\Gamma_{\pm}$ are defined by [see
App. \ref{appA} for more details]
\begin{eqnarray}\label{A38b}
E_{\pm}(\mathbf{p},\omega_{f})&\equiv&
\omega_{f}\pm\frac{1}{2}\mbox{tr}\left(\hat{g}_{\pm}(\mathbf{p},\omega_{f})\mathfrak{Re}[\Sigma_{R}^{f}(\mathbf{p},\omega_{f})]\right),\nonumber\\
\Gamma_{\pm}(\mathbf{p},\omega_{f})&\equiv&\pm\frac{1}{2}\mbox{tr}\left(\hat{g}_{\pm}(\mathbf{p},\omega_{f})\mathfrak{Im}[\Sigma_{R}^{f}(\mathbf{p},\omega_{f})]\right).\nonumber\\
\end{eqnarray}
In \cite{jeon2007}, almost the same expression for $\rho_{f}$ as in
(\ref{A36b}) is introduced. However, in contrast to (\ref{A36b}),
only one spectral width for the fermion appears in the relation
presented in \cite{jeon2007}. Apparently, $\Gamma_{+}\simeq
\Gamma_{-}$ is assumed. In what follows, we do not make this
approximation, and after deriving $\eta_{f}$ in terms of
$\Gamma_{\pm}$, we will explore the effect of $\Gamma_{+}$ and
$\Gamma_{-}$ on the thermal properties of $\eta_{f}$. Let us notice
at this stage, that the plus and minus signs appearing on $E_{\pm}$
and $\Gamma_{\pm}$ correspond to the normal and collective
(plasmino) modes of the fermions \cite{weldon1989}. In the chiral
limit $m_{f}\to 0$, they correspond to the same and opposite
helicity and chirality of massless fermions, respectively
\cite{weldon1999}.
\par
Let us now consider (\ref{A33b}), which will be simplified in what
follows. Using the symmetry properties of $E_{\pm}(p)$ and
$\Gamma_{\pm}(p)$,
\begin{eqnarray}\label{A39b}
E_{\pm}(\mathbf{p},-\omega_{f})&=&-E_{\pm}(\mathbf{p},\omega_{f}),\nonumber\\
\Gamma_{\pm}(\mathbf{p},-\omega_{f})&=&+\Gamma_{\pm}(\mathbf{p},\omega_{f}),
\end{eqnarray}
which we could verify only at one-loop level, we obtain
\begin{eqnarray}\label{A40b}
\lefteqn{\hspace{-0.5cm}\rho_{f}(-\mathbf{p},-\omega)=\rho_{f}(\mathbf{p},\omega)}\nonumber\\
&&\hspace{-0.5cm}-\frac{2m_{f}}{\omega_{f}}\bigg\{\frac{\Gamma_{+}(\mathbf{p},\omega_{f})}{[\omega-E_{+}(\mathbf{p},\omega_{f})]^{2}+\Gamma_{+}^{2}(\mathbf{p},\omega_{f})}\nonumber\\
&&\hspace{-0.5cm} +
\frac{\Gamma_{-}(\mathbf{p},\omega_{f})}{[\omega+E_{-}(\mathbf{p},\omega_{f})]^{2}+\Gamma_{-}^{2}(\mathbf{p},\omega_{f})}\bigg\}.
\end{eqnarray}
Using (\ref{A40b}) together with the properties of the traces of
Dirac $\gamma$-matrices, we have
\begin{eqnarray}\label{A41b}
\lefteqn{\hspace{-0.5cm}\mbox{tr}\left(\rho_{f}(-\mathbf{p},-\omega_{1})\gamma_{\alpha}\rho_{f}(-\mathbf{p},-\omega_{2})\gamma_{\beta}\right)}\nonumber\\
&&=\mbox{tr}\left(\rho_{f}(\mathbf{p},\omega_{1})\gamma_{\alpha}\rho_{f}(\mathbf{p},\omega_{2})\gamma_{\beta}\right).
\end{eqnarray}
Implementing now this relation in (\ref{A33b}), we arrive after
analytical continuation, $i\omega_{n}\to p_{0}+i\epsilon$, at
\begin{eqnarray}\label{A42b}
\lefteqn{\Pi_{R}^{f}(p_{0})=\frac{1}{2\pi^{2}}\int\frac{d^{3}p}{(2\pi)^{3}}\eta^{\alpha\beta\rho\sigma}p_{\rho}p_{\sigma}}\nonumber\\
&&\times\int_{-\infty}^{+\infty}d\omega_{1}d\omega_{2}\mbox{tr}\left(\rho_{f}(\omega_{1},\mathbf{p})\gamma_{\alpha}\rho_{f}(\omega_{2},\mathbf{p})\gamma_{\beta}\right)\nonumber\\
&&\times~
(1-n_{f}(\omega_{1}))n_{f}(\omega_{2})W_{\epsilon}(\bar{\omega}_{12},p_{0}),
\end{eqnarray}
where $\bar{\omega}_{12}\equiv \omega_{1}-\omega_{2}$ and
$W_{\epsilon}$ is defined in (\ref{A17b}). To derive the fermionic
part of the shear viscosity $\eta_{f}$ from (\ref{T15a}), we follow
the same steps as is presented in the previous section for the
bosonic case. Plugging (\ref{A36b}) in (\ref{A42b}), and after
performing a straightforward mathematical computation, where mainly
the relations
\begin{eqnarray}\label{A43b}
\frac{i}{10}\frac{d}{dp_{0}}W_{\epsilon}(p_{0},\bar{\omega}_{12})\bigg|_{p_{0}=0}=-\frac{\pi}{5}\delta'(\bar{\omega}_{12}),
\end{eqnarray}
and
\begin{eqnarray}\label{A44b}
\lefteqn{\mbox{tr}\left(\hat{g}_{\pm} \gamma _ {\alpha}
\hat{g}_{\mp} \gamma_{\rho}\right)
}\nonumber\\&=&\frac{1}{\omega_{f}^2}\bigg\{2\omega_{f}^2
(g_{0\alpha}g_{0\rho}-g_{\alpha \rho})-p_ip_j(g_{\alpha}^i
g_{\rho}^j +g_{\rho}^i g_{\alpha}^j)\bigg\}.\nonumber\\
\lefteqn{\mbox{tr}\left(\hat{g}_{\pm} \gamma _ {\alpha}
\hat{g}_{\pm}
\gamma_{\rho}\right)=\frac{1}{\omega_{f}^2}\bigg\{2\omega_{f}^{2}g_{0\alpha}g_{0\rho}
\mp 2\omega_{f} p_i(g_{\alpha}^0 g_{\rho}^i }\nonumber\\
&& +g_{\rho}^0 g_{\alpha}^i)+p_ip_j(g_{\alpha}^i g_{\rho}^j
+g_{\rho}^i g_{\alpha}^j)\bigg\},
\end{eqnarray}
in the local rest frame of the fluid are used, we arrive at
\begin{eqnarray}\label{A45b}
\eta_{f}=\frac{8\beta}{15\pi}\int\frac{d^{3}p}{(2\pi)^{3}}\int_{-\infty}^{+\infty}
d\omega~F_{f}(\mathbf{p},\omega),
\end{eqnarray}
with
\begin{eqnarray}\label{A46b}
\lefteqn{F_{f}(\mathbf{p},\omega)\equiv \frac{e^{\beta
\omega}}{(e^{\beta\omega}+1)^{2}}\mathbf{p}^{2}
}\nonumber\\
&&\times\bigg\{\frac{\mathbf{p}^{2}}{\omega_{f}^{2}}\left(\frac{\Gamma_{+}}{(\omega-E_{+})^{2}+\Gamma_{+}^{2}}+\frac{\Gamma_{-}}{(\omega+E_{-})^{2}+\Gamma_{-}^{2}}\right)^{2}
\nonumber\\
&&-\frac{2~\Gamma_{+}\Gamma_{-}}{[(\omega-E_{+})^{2}+\Gamma_{+}^{2}][(\omega+E_{-})^{2}+\Gamma_{-}^{2}]}\bigg\},
\end{eqnarray}
where
$\omega\equiv\frac{1}{2}\omega_{12}=\frac{1}{2}(\omega_{1}-\omega_{2})$,
and $E_{\pm}=E_{\pm}(\mathbf{p},\omega_{f})$ as well as
$\Gamma_{\pm}=\Gamma_{\pm}(\mathbf{p},\omega_{f})$ are defined in
(\ref{A38b}). To evaluate the integration over $\omega$ in
(\ref{A46b}), the pole structure of $F_{f}(\mathbf{p},\omega)$ is to
be considered. Similar to the previous case of bosonic fields, the
contributions of the poles arising from the denominator
$e^{\beta\omega}+1$ in (\ref{A46b}) turn out to be proportional to
$\Gamma_{\pm}^{2}$, and, assuming that $\Gamma_{+}$ and $\Gamma_{-}$
are small enough, they can be neglected. As concerns the residue of
the remaining poles, we have to close the contour in the upper
half-plane and consider only two residue $\omega^{\pm}\equiv \pm
E_{\pm}+i\Gamma_{\pm}$. Expanding the resulting expression in the
orders of $\Gamma_{\pm}$ in a Laurent series in the orders of
$\Gamma_{\pm}$, and using
\begin{eqnarray}\label{A47b}
\lefteqn{\int_{-\infty}^{+\infty}d\omega\frac{e^{\beta\omega}}{(e^{\beta\omega}+1)^{2}}\frac{\Gamma_{\pm}^{2}}{[(\omega\mp
E_{\pm})^{2}+\Gamma^{2}_{\pm}]^{2}}
}\nonumber\\
&&\approx \pi\frac{e^{\beta E_{\pm}}}{(e^{\beta
E_{\pm}}+1)^{2}}\frac{1}{2
\Gamma_{\pm}},\nonumber\\
\lefteqn{\int_{-\infty}^{+\infty}d\omega
\frac{\Gamma_{+}\Gamma_{-}}{[(\omega-E_{+})^{2}+\Gamma_{+}^{2}][(\omega+E_{-})^{2}+\Gamma_{-}^{2}]}
}\nonumber\\
&&\approx\pi\sum_{s=\pm}\frac{e^{\beta E_{s}}}{(e^{\beta
E_{s}}+1)^{2}}\frac{\Gamma_{f}^{+}-\Gamma_{s}}{[E_{f}+is~\Gamma_{f}^{+}][E_{f}+i\Gamma_{f}^{-}]},\nonumber\\
\end{eqnarray}
where
\begin{eqnarray}\label{A48b}
E_{f}\equiv E_{+}+E_{-},\qquad \Gamma_{f}^{\pm}\equiv \Gamma_{+}\pm
\Gamma_{-},
\end{eqnarray}
we arrive, after performing the integration over three-dimensional
angles in (\ref{A45b}), at the fermionic part of the shear viscosity
of the Yukawa theory,
\begin{eqnarray}\label{A49b}
\lefteqn{\hspace{0cm}\eta_{f}=\frac{2\beta}{15\pi^{2}}\int_{0}^{\infty}
dp
\frac{\mathbf{p}^{4}}{\omega_{f}^{2}}\sum_{s=\pm}\bigg\{\frac{e^{\beta
E_{s}}}{(e^{\beta E_{s}}+1)^{2}}
}\nonumber\\
&& \times \bigg[\frac{\mathbf{p}^{2}}{\Gamma_{s}}-\frac{4m_{f}^{2}(\Gamma_{f}^{+}-\Gamma_{s})}{[E_{f}+is\Gamma_{f}^{+}]
[E_{f}+i\Gamma_{f}^{-}]}\bigg]\bigg\}+{\cal{O}}(\Gamma_{\pm}^{0}).\nonumber\\
\end{eqnarray}
Here, $E_{f}=E_{f}(\mathbf{p},\omega_{f})$,
$\Gamma_{\pm}=\Gamma_{\pm}(\mathbf{p},\omega_{f})$ and
$\Gamma_{f}^{\pm}=\Gamma_{f}^{\pm}(\mathbf{p},\omega_{f})$ are
defined in (\ref{A38b}) and (\ref{A48b}). Let us notice that the
first term of above relation for $\eta_{f}$ is comparable with the
shear viscosity corresponding to fermions appearing in
\cite{ghosh2013} in a relaxation time approximation. Moreover, it
resembles the $\eta_{f}$ presented recently in \cite{lang2013}.
Here, the authors express $\eta_{f}$ first in terms of fermionic
density function, $\rho_{f}$, which in contrast to (\ref{A36b}),
possesses a generalized Breit-Wigner shape, including only a
quasiparticle mass $M$ and a fermionic width $\Gamma_{f}$. Using
this Ansatz for $\rho_{f}$, they arrive then at $\eta_{f}$ in this
quasiparticle approximation (see Eq. (22) in \cite{lang2013}). We,
however, will work with (\ref{A49b}) and after determining
$\Gamma_{\pm}$ in a one-loop perturbative expansion, in the next
section, will study the thermal properties of
$\eta_{f}[\Gamma_{\pm}]$ for various masses $m_{b}$ and $m_{f}$. We
will then determine the difference between
$\eta_{f}[\Gamma_{+}=\Gamma_{-}]$ and $\eta_{f}[\Gamma_{+}\neq
\Gamma_{-}]$. In App. \ref{appC}, we will generalize the method
presented in this section for the case of non-vanishing chemical
potential. We will show that in this case (\ref{appD1a}), replaces
(\ref{A49b}), and can be used to explore the thermal properties of
$\eta_{f}$ at finite $T$ and $\mu$.
\par\vspace{0.5cm}
\section{Perturbative computation of bosonic and fermionic spectral
widths}\label{sec4}
\setcounter{equation}{0}
\par\noindent In this section, we will perturbatively compute the
bosonic and fermionic spectral widths of the Yukawa theory from
(\ref{A20b}) and (\ref{A38b}) at one-loop level. To do this, the
imaginary part of the one-loop bosonic and fermionic self-energy
diagrams will be evaluated using the standard Schwinger-Keldysh
real-time formalism \cite{schwinger1961}. In what follows, we will
closely follow the notations of \cite{dasbook} and \cite{kobes1985}.
According to this formalism, the free propagator of scalar bosons is
given by
\begin{eqnarray}\label{G1a}
{\cal{G}}=\left(
\begin{array}{cc}
G_{++}&G_{+-}\\
G_{-+}&G_{--}
\end{array}
\right),
\end{eqnarray}
where $G_{ab}, a,b=\pm$ read
\begin{eqnarray}\label{G2a}
G_{++}(p)&=&-\frac{i}{p^2-m_{b}^2+i\epsilon}-2\pi
n_b(|p_{0}|)\delta(p^2-m_{b}^2).\nonumber\\
G_{+-}(p)&=&-2\pi  [\theta(-p_0)+n_b(|p_0|)]\delta(p^2-m_{b}^2),\nonumber\\
G_{-+}(p)&=&-2\pi  [\theta(p_0)+n_b(|p_0|)]\delta(p^2-m_{b}^2),\nonumber\\
G_{--}(p)&=&\frac{i}{p^2-m_{b}^2-i\epsilon}-2\pi
n_b(|p_{0}|)\delta(p^2-m_{b}^2).\nonumber\\
\end{eqnarray}
Here, $m_{b}$ is the boson mass and $n_{b}(p_{0})$ the bosonic
distribution function defined in (\ref{A15b}). Similarly, the free
fermion propagator is given by
\begin{eqnarray}\label{G3a}
{\cal{S}}=\left(
\begin{array}{cc}
S_{++}&S_{+-}\\
S_{-+}&S_{--}
\end{array}
\right),
\end{eqnarray}
with the components
\begin{widetext}
\par\vspace{-0.5cm}
\begin{eqnarray}\label{G4a}
S_{++}(p)&=&(\gamma\cdot p+m_{f})\left(-\frac{i}{p^2-m_{f}^2+i\epsilon}+2\pi n_f(|p_{0}|)\delta(p^2-m_{f}^2)\right),\nonumber\\
S_{+-}(p)&=&-2\pi (\gamma\cdot p+m_{f})[\theta(-p_0)-n_f(|p_0|)]\delta(p^2-m_{f}^2),\nonumber\\
S_{-+}(p)&=&-2\pi (\gamma\cdot p+m_{f})[\theta(p_0)-n_f(|p_0|)]\delta(p^2-m_{f}^2),\nonumber\\
S_{--}(p)&=&(\gamma\cdot
p+m_{f})\left(\frac{i}{p^2-m^2-i\epsilon}+2\pi
n_f(|p_{0}|)\delta(p^2-m_{f}^2)\right).
\end{eqnarray}
\end{widetext}
Here, $m_{f}$ is the fermion mass and $n_{f}(p_{0})$ the fermionic
distribution function defined in (\ref{A31b}). Combining
$G_{ab},a,b=\pm$ and $S_{ab}, a,b=\pm$, the physical retarded (R)
and advanced (A) two-point Green's functions for scalar bosons,
$G_{R/A}$, and fermions, $S_{R/A}$, are given by
\begin{eqnarray}\label{G5a}
\hspace{-0.5cm}G_{R}=G_{++}+G_{+-}, \qquad G_{A}=G_{++}+G_{-+},
\end{eqnarray}
and
\begin{eqnarray}\label{G6a}
\hspace{-0.5cm}S_{R}=S_{++}+S_{+-}, \qquad
\hspace{0.2cm}S_{A}=S_{++}+S_{-+}.
\end{eqnarray}
\begin{figure}[t]
\vspace{-3.5cm}
\includegraphics[width=15.5cm,height=21cm]{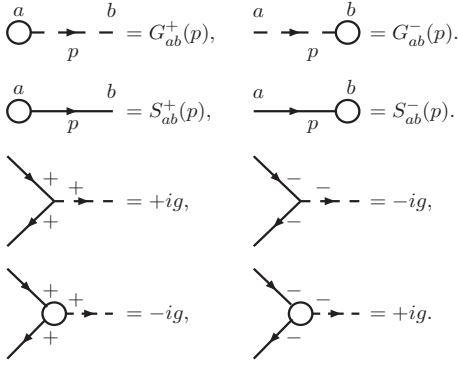}
\vspace{-13cm} \caption{Feynman rules, that are necessary to compute
the imaginary part of bosonic and fermionic one-loop self-energy
diagrams of the Yukawa theory (see Figs. \ref{fig3} and \ref{fig4}).
The definitions of $G_{ab}^{\pm}$ and $S_{ab}^{\pm}$ with $a,b=\pm$
in terms of $G_{ab}$ and $S_{ab}$ from (\ref{G2a}) and (\ref{G4a})
are presented in (\ref{G8a}).}\label{fig2}
\end{figure}
\par\noindent
To determine the spectral widths, $\Gamma_{b}$ and $\Gamma_{\pm}$
from (\ref{A20b}) and (\ref{A38b}), the imaginary parts of the
bosonic and fermionic one-loop self-energies, $\Sigma_{R}^{b}$ and
$\Sigma_{R}^{f}$, are to be computed. In the real-time formalism,
this is done using the finite temperature cutting rules
\cite{dasbook,kobes1985}. The main ingredients of these rules are
specific propagators and vertices, that for the Yukawa theory, are
demonstrated in Fig. \ref{fig2}.
\begin{figure*}[bth]
\vspace{-2cm}
\includegraphics[width=16cm,height=20cm]{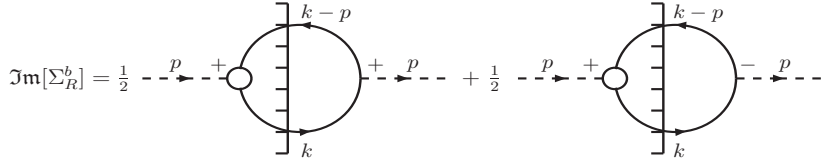}
\vspace{-15cm} \caption{Diagrammatic representation of the cutting
rules leading to the imaginary part of the retarded part of the
one-loop self-energy diagram for scalar bosons, $\Sigma_{R}^{b}$ in
the Yukawa theory. }\label{fig3}
\end{figure*}
Here, $G_{ab}^{\pm}$ and $S_{ab}^{\pm}$ with $a,b=\pm$ are the
retarded ($+$) and advanced ($-$) part of the bosonic and fermionic
Green's functions. They are defined in the following decomposition
for a generic Green's function, ${\cal{D}}_{ab}$, with $a,b=\pm$,
\begin{eqnarray}\label{G7a}
{\cal{D}}_{ab}(x)=\theta(t){\cal{D}}_{ab}^{+}(x)+\theta(-t){\cal{D}}_{ab}^{-}(x).
\end{eqnarray}
Using the definitions ${\cal{D}}_{ab}, a,b=\pm$ and
${\cal{D}}=\{G,S\}$, from (\ref{G2a}) and (\ref{G4a}), we get the
following identities:
\begin{eqnarray}\label{G8a}
&&{\cal{D}}_{++}^{+}={\cal{D}}_{--}^{-}={\cal{D}}_{-+}^{+}={\cal{D}}_{-+}^{-}={\cal{D}}_{-+},\nonumber\\
&&{\cal{D}}_{++}^{-}={\cal{D}}_{--}^{+}={\cal{D}}_{+-}^{+}={\cal{D}}_{+-}^{-}={\cal{D}}_{+-}.
\end{eqnarray}
In what follows, we will separately compute the imaginary part of
the one-loop self-energy corrections to bosonic and fermionic
two-point Green's functions. The results will eventually be used to
determine the bosonic and fermionic spectral widths.
\subsection{Bosonic spectral width in one-loop perturbative expansion}\label{sec4a}
\par\noindent
Let us consider the bosonic spectral width $\Gamma_{b}$ from
(\ref{A20b}), that evaluated at
$\omega_{b}=(\mathbf{p}^{2}+m_{b}^{2})^{1/2}$ reads
\begin{eqnarray}\label{G9a}
\Gamma_{b}(\mathbf{p},\omega_{b})=-\frac{1}{2\omega_{b}}\mathfrak{Im}[\Sigma_{R}^{b}(\mathbf{p},\omega_{b})].
\end{eqnarray}
To determine the imaginary part of $\Sigma_{R}^{b}(p)$ at one-loop
level, we will use the diagrammatic representation of the cutting
rules \cite{dasbook, kobes1985}, demonstrated in Fig. \ref{fig3}.
Using the propagators and vertices presented in Fig. \ref{fig2}, the
imaginary part of $\Sigma_{R}^{b}(p)$ reads
\begin{eqnarray}\label{G10a}
\mathfrak{Im}[\Sigma_{R}^{b}(p)]&=&-\frac{g^{2}}{2}\int
\frac{d^{4}k}{(2\pi)^{4}}\mbox{tr}\left(S_{++}^{-}(k-p)S_{++}^{+}(k)\right.
\nonumber\\
&&\left.-S_{-+}^{-}(k-p)S_{+-}^{+}(k)\right),
\end{eqnarray}
where, according to (\ref{G7a}) with ${\cal{D}}_{ab}=S_{ab}$,
$S_{ab}^{+}$ and $S_{ab}^{-}$ are the retarded and advanced parts of
the fermionic Green's function $S_{ab}, a,b=\pm$ from (\ref{G4a}),
respectively. To derive the spectral width of bosons, we use the
identities (\ref{G8a}) together with (\ref{G4a}), and arrive after
performing the integration over $k_{0}$ and some straightforward
manipulations first at
\begin{eqnarray}\label{G11a}
\lefteqn{\Gamma_b(\mathbf{p},\omega_{b})=\frac{g^{2}}{8\omega_{b}}\int
\frac{d^{3}k}{(2\pi)^{2}}\frac{(4m_{f}^{2}-m_{b}^{2})}{\omega_{1}\omega_{2}}
}\nonumber\\
&&\times\bigg\{\delta(\omega_{b}-\omega_{1}-\omega_{2})[1-n_{f}(\omega_{1})-n_{f}(\omega_{2})]\nonumber\\
&&~~+\delta(\omega_{b}-\omega_{1}+\omega_{2})[n_{f}(\omega_{1})-n_{f}(\omega_{2})]\nonumber\\
&&~~-\delta(\omega_{b}+\omega_{1}-\omega_{2})[n_{f}(\omega_{1})-n_{f}(\omega_{2})]\nonumber\\
&&~~-\delta(\omega_{b}+\omega_{1}+\omega_{2})[1-n_{f}(\omega_{1})-n_{f}(\omega_{2})]\bigg\}.\nonumber\\
\end{eqnarray}
Here, $\omega_{1}^{2}\equiv \mathbf{k}^{2}+m_{f}^{2}$ and
$\omega_{2}^{2}\equiv(\mathbf{k}-\mathbf{p})^{2}+m_{f}^{2}$. The
factor $(4m_{f}^{2}-m_{b}^{2})$ on the r.h.s. of (\ref{G11a}) arises
by considering the on mass-shell relations, $k^{2}=m_{f}^{2}$ and
$(k-p)^{2}=m_{f}^{2}$ from the Dirac-$\delta$-functions, appearing
in $S_{ab}$ from (\ref{G10a}), with $S_{ab}, a,b=\pm$ given in
(\ref{G4a}). Using now the definition of the fermionic distribution
functions $n_{f}(\omega)$ from (\ref{A31b}), we get
\begin{eqnarray}\label{G12a}
\lefteqn{\Gamma_b(\mathbf{p},\omega_{b})=\frac{g^2}{16\omega_{b}}\int
\frac{d^3 k}{(2\pi)^2} \frac{\sinh(\frac{\beta
\omega_{b}}{2})}{\cosh(\frac{\beta\omega_1}{2})\cosh(\frac{\beta\omega_2}{2})}
}\nonumber\\
&&\times\frac{(4m_{f}^{2}-m_{b}^{2})}{\omega_{1} \omega_{2}}
\bigg\{\delta(\omega_{b}-\omega_1 -\omega_2) -\delta(\omega_{b}-\omega_1 +\omega_2)\nonumber\\
&&\qquad -\delta(\omega_{b}+\omega_1 -\omega_2)+
\delta(\omega_{b}+\omega_1 +\omega_2)\bigg\}.
\end{eqnarray}
Note that in the rest-frame of the scalar bosons with
$\mathbf{p}=0$, only the first term on the r.h.s. of (\ref{G11a}),
proportional to $\delta(\omega_{b}-\omega_{1}-\omega_{2})$ will
contribute. It leads to $m_{b}\geq 2m_{f}$, as a constraint on the
relation between bosonic and fermionic masses. Thus, keeping in mind
that $\Gamma_{b}(\mathbf{p},\omega_{b})$ is Lorentz invariant, it is
in general given by
\begin{eqnarray}\label{G13a}
\Gamma_b(\mathbf{p},\omega_{b})&=&\frac{g^{2}}{16\omega_{b}}\int
\frac{d^{3}k}{(2\pi)^{2}}\frac{\sinh(\frac{\beta
\omega_{b}}{2})}{\cosh(\frac{\beta\omega_1}{2})\cosh(\frac{\beta\omega_2}{2})}
\nonumber\\
&&\hspace{-0.2cm}\times
\frac{(4m_{f}^{2}-m_{b}^{2})}{\omega_{1}\omega_{2}}\delta(\omega_{b}-\omega_1
-\omega_2).
\end{eqnarray}
After performing the integration over $k$, using the method
described in App. C, the bosonic part of the spectral width of the
Yukawa theory, evaluated in a one-loop perturbative expansion, reads
\begin{eqnarray}\label{G14a}
\lefteqn{\hspace{-0.9cm}\Gamma_b(\mathbf{p},\omega_{b})=\frac{g^{2}T}{16\pi}\frac{\gamma_{b}^{2}(\xi^{2}-4)}{\xi^{2}\sqrt{1-\gamma_{b}^{2}}}
}\nonumber\\
&&\hspace{-1.3cm}\times\ln\bigg[\frac{1+\cosh\frac{\kappa_{b}}{2}(1+\frac{1}{\xi}\sqrt{(\xi^{2}-4)(1-\gamma_{b}^{2})})}
{1+\cosh\frac{\kappa_{b}}{2}(1-\frac{1}{\xi}\sqrt{(\xi^{2}-4)(1-\gamma_{b}^{2})})}\bigg].
\end{eqnarray}
Here, $\xi\equiv \frac{m_{b}}{m_{f}}$ and  $\gamma_{b}\equiv
\frac{m_{b}}{\omega_{b}}$, with
$\omega_{b}^{2}=\mathbf{p}^{2}+m_{b}^{2}$. Moreover,
$\kappa_{b}\equiv\omega_{b}/T$. In App. \ref{appC}, we have
generalized the result presented in (\ref{G14a}) to the case of
non-vanishing chemical potential, $\mu$. In this case, the one-loop
contribution to the bosonic spectral width is presented in
(\ref{appD14a}). In Sec. \ref{sec5}, we will use (\ref{G14a}) and
(\ref{appD14a}) to study the the thermal properties of $\Gamma_{b}$.
Eventually $\Gamma_{b}$ will be inserted in (\ref{A26b}) and the
thermal properties of $\eta_{b}$ for various $\xi$ will be studied.
\subsection{Fermionic spectral width in one-loop perturbative expansion}\label{sec4b}
\begin{figure*}[bth]
\vspace{-2cm}
\includegraphics[width=16cm,height=20cm]{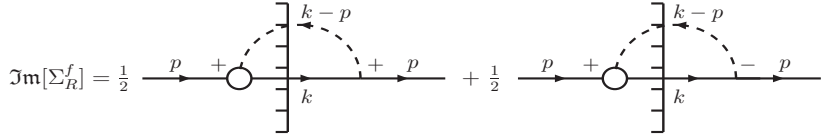}
\vspace{-15cm} \caption{Diagrammatic representation of the cutting
rules leading to the imaginary part of the retarded part of the
one-loop self-energy diagram for fermions, $\Sigma_{R}^{f}$ in the
Yukawa theory.}\label{fig4}
\end{figure*}
\par\noindent
As we have demonstrated in the previous section, fermions possess
two different spectral widths $\Gamma_{\pm}$, defined in
(\ref{A38b}). They can be perturbatively computed by evaluating the
imaginary part of the retarded fermion self-energy $\Sigma_{R}^{f}$
in an appropriate loop expansion. In what follows, in analogy to the
bosonic case, the standard finite temperature cutting rules from
\cite{dasbook, kobes1985} will be used to evaluate the imaginary
part of $\Sigma_{R}^{f}$ at one-loop level. Using the Feynman rules
presented in Fig. \ref{fig2}, and the diagrammatic representation of
$\mathfrak{Im}[\Sigma_{R}^{f}]$ demonstrated in Fig. \ref{fig4}, we
arrive first at
\begin{eqnarray}\label{G15a}
\mathfrak{Im}[\Sigma_{R}^{f}(p)]&=&\frac{g^{2}}{2}\int
\frac{d^{4}k}{(2\pi)^{4}}\big[S_{++}^{+}(k)G_{++}^{-}(k-p)\nonumber\\
&&-S_{+-}^{+}(k)G_{-+}^{-}(k-p)\big],
\end{eqnarray}
where ${\cal{D}}_{ab}^{\pm}, a,b=\pm$ and ${\cal{D}}_{ab}=\{G,S\}$
are defined in (\ref{G7a}). Using the identities (\ref{G8a}), with
${\cal{D}}=\{G,S\}$, together with the definitions of $G_{ab}$ and
$S_{ab}, a,b=\pm$ from (\ref{G2a}) and (\ref{G4a}), we arrive after
performing the integration over $k_{0}$ in (\ref{G15a}) and some
straightforward manipulations, at the fermionic spectral widths
$\Gamma_{\pm}$, defined originally in (\ref{A38b}),
\begin{widetext}
\begin{eqnarray}\label{G16a}
\lefteqn{\hspace{-1cm}\Gamma_{\pm}(\mathbf{p},\omega_{f})=\pm\frac{g^2}{8\omega_{f}}\int
\frac{d^3 k}{(2\pi)^2}\frac{1}{\omega_1 \omega_2} }\nonumber\\
&&\hspace{-0.9cm}\times\bigg[[\omega_{f}\omega_1
\mp\mathbf{p}\cdot\mathbf{k}\pm
m_f^2]\big\{\delta (\omega_{f}-\omega_1 -\omega_2)[1-n_f(\omega_1)+n_b(\omega_2)]+\delta(\omega_{f}-\omega_1 +\omega_2)[n_f(\omega_1)+n_b(\omega_2)]\big\}\nonumber\\
&&\hspace{-0.9cm} +[\omega_{f}\omega_1
\pm\mathbf{p}\cdot\mathbf{k}\mp m_f^2] \big\{\delta
(\omega_{f}+\omega_1 +\omega_2)[1-n_f(\omega_1)+n_b(\omega_2)]
+\delta(\omega_{f}+\omega_1-\omega_2)[n_f(\omega_1)+n_b(\omega_2)]\big\}
\bigg].
\end{eqnarray}
\end{widetext}
According to our notations from Fig. \ref{fig4},
$\omega_{f}^{2}\equiv {\mathbf{p}}^{2}+m_{f}^{2}$ corresponds to the
momentum of the external fermion propagators, and
$\omega_{1}^{2}\equiv k_{0}^{2}=\mathbf{k}^{2}+m_{f}^{2}$ and
$\omega_{2}^{2}\equiv
(k_{0}-p_{0})^{2}=(\mathbf{k}-\mathbf{p})^{2}+m_{b}^{2}$ to the
internal fermion and boson propagators, respectively. Here, in
contrast to the bosonic case, only two terms on the r.h.s. of
(\ref{G16a}), proportional to
$\delta(\omega_{f}-\omega_{1}+\omega_{2})$ and
$\delta(\omega_{f}+\omega_{1}-\omega_{2})$, contribute to the final
results of $\Gamma_{+}$ and $\Gamma_{-}$. This is because of the
specific kinematics of $f\to bf$ process in the rest frame of the
particles. Here, $b$ and $f$ correspond to boson and fermion,
respectively. Thus, the fermionic spectral widths are determined
after some algebraic manipulations, where the definitions
(\ref{A15b}) and (\ref{A31b}) of bosonic and fermionic distribution
functions are used. For $\Gamma_{+}$, we obtain
\begin{eqnarray}\label{G17a}
\lefteqn{\hspace{0cm}\Gamma_{+}(\mathbf{p},\omega_{f}) }\nonumber\\
&&=\frac{g^{2}}{32\omega_{f}}\int \frac{d^3 k}{(2
\pi)^2}\frac{(4m_{f}^{2}-m_{b}^{2})}{\omega_{1}\omega_{2}}\frac{\cosh(\frac{\beta
\omega_{f}}{2})}{\cosh(\frac{\beta \omega_1}{2})\sinh(\frac{\beta
\omega_2}{2})}
\nonumber\\
&&~\times\bigg\{\delta(\omega_{f}-\omega_{1}+\omega_{2})
-\delta(\omega_{f}+\omega_{1}-\omega_{2})\bigg\}.
\end{eqnarray}
As concerns $\Gamma_{-}$, it is given, according to (\ref{A48b}), by
$\Gamma_{-}=\Gamma_{+}-\Gamma_{f}^{-}$, where $\Gamma_{f}^{-}$ is
given by
\begin{eqnarray}\label{G18a}
\lefteqn{ \hspace{-1cm}
\Gamma_{f}^{-}(\mathbf{p},\omega_{f})=\frac{g^2}{8}\int \frac{d^3
k}{(2\pi)^{2}\omega_{2}} \frac{\cosh(\frac{\beta
\omega_{f}}{2})}{\cosh(\frac{\beta
\omega_1}{2})\sinh(\frac{\beta\omega_{2}}{2})}
}\nonumber\\
&&\hspace{-0.8cm}\times
\bigg\{\delta(\omega_{f}-\omega_{1}+\omega_{2})
+\delta(\omega_{f}+\omega_{1}-\omega_{2})\bigg\}.
\end{eqnarray}
Performing the integration over $k$ in (\ref{G17a}),  by making use
of the method presented in App. \ref{appB}, $\Gamma_{+}$ reads
\begin{eqnarray}\label{G19a}
\lefteqn{\Gamma_{+}(\xi,
\gamma_{f},\kappa_{f};T)=\frac{g^{2}T}{32\pi}\frac{\gamma_{f}^{2}(\xi^{2}-4)}{\sqrt{1-\gamma_{f}^{2}}}
}\nonumber\\
&&\times
\bigg\{\ln\bigg[\frac{1-\cosh(2~\Xi_{-})}{\cosh(\Upsilon_{-}+\Xi_{+})-
\cosh(\Upsilon_{-}-\Xi_{+})}\bigg]\nonumber\\
&&-\ln\bigg[
\frac{1+\cosh(2~\Xi_{-}-\kappa_{f})}{\cosh(\Upsilon_{-}+\Xi_{+})+
\cosh(\Upsilon_{+}-\Xi_{+})}
\bigg]\bigg\}.\nonumber\\
\end{eqnarray}
Here, $\xi=\frac{m_{b}}{m_{f}}$ and $\gamma_{f}\equiv
\frac{m_{f}}{\omega_{f}}$ with
$\omega_{f}^{2}=\mathbf{p}^{2}+m_{f}^{2}$. Moreover, we have
\begin{eqnarray}\label{G20a}
\Xi_{\pm}&=&\frac{\kappa_{f}}{4}\xi[\xi\pm\sqrt{(\xi^{2}-4)(1-\gamma_{f}^{2})}],\nonumber\\
\Upsilon_{\pm}&=&\frac{\kappa_{f}}{2}(\gamma_{f}\pm 1),
\end{eqnarray}
with $\kappa_{f}\equiv \omega_{f}/T$. Similarly, the integration
over $k$ in (\ref{G18a}) can be performed analytically. This is done
in App. \ref{appB}, where the final result for $\Gamma_{f}^{-}$ is
presented in (\ref{appC14a}). In App. \ref{appC}, the same method is
used for the case of non-vanishing chemical potential and
$\Gamma_{+}$ and $\Gamma_{f}^{-}$ are determined at one-loop level.
The results for $\Gamma_{+}$ and $\Gamma_{f}^{-}$ are presented in
(\ref{appD17a}) and (\ref{appD19a}), respectively.
\par
In Sec. \ref{sec5}, we will study the qualitative behavior of
dimensionless quantities $\Gamma_{+}/g^{2}T$ and
$\Gamma_{f}^{-}/g^{2}T$ in terms of dimensionless variables
$\xi,\gamma_{f}$ and $\kappa_{f}$. We will then study the $T$ and
$\mu$ dependence of $\Gamma_{+}$ and $\Gamma_{-}$, and will show
that in certain regime of parameter space
$\Gamma_{f}^{-}=\Gamma_{+}-\Gamma_{-}$ is not negligible. Plugging
the resulting expressions for $\Gamma_{\pm}$ and $\Gamma_{f}^{-}$ in
(\ref{A49b}) and assuming that $E_{f}\sim
\omega_{f}=(\mathbf{p}^{2}+m_{f}^{2})^{1/2}$, we will eventually
explore the thermal properties of the fermionic part of the shear
viscosity.
\section{Numerical Results}\label{sec5}
\setcounter{equation}{0}
\par\noindent
In this section, we will mainly study the $T$ and $\mu$ dependence
of bosonic and fermionic spectral widths $\Gamma_{b}$ and
$\Gamma_{\pm}$, as well as the thermal properties of bosonic and
fermionic part of the shear viscosity, $\eta_{b}$ and $\eta_{f}$. We
will first determine the $T$ and $\mu$ dependence of these
quantities for constant $\xi_{0}\equiv m_{b}^{0}/m_{f}^{0}$,
including the $T$ and $\mu$ independent bosonic and fermionic
masses, $m_{b}^{0}$ and $m_{f}^{0}$, respectively. We then consider
the standard thermal corrections of bosonic and fermionic masses
\cite{kiessig2010}, arising from standard  HTL approximation,
\begin{eqnarray}\label{H1}
(m_{b}^{\mbox{\tiny{th}}})^{2}&=&\frac{g^{2}}{6}\left(T^{2}+\frac{3\mu^{2}}{\pi^{2}}\right),\nonumber\\
(m_{f}^{\mbox{\tiny{th}}})^{2}&=&\frac{g^{2}}{16}\left(T^{2}+\frac{\mu^{2}}{\pi^{2}}\right),
\end{eqnarray}
and will add these thermal corrections to the original constant
$m_{b}^{0}$ and $m_{f}^{0}$. Using the definition
\begin{eqnarray}\label{H2}
\xi(T,\mu)\equiv\frac{m_{b}(T,\mu)}{m_{f}(T,\mu)},
\end{eqnarray}
with
\begin{eqnarray}\label{H3}
m_{b}(T,\mu)&\equiv&
m_{b}^{0}+m_{b}^{\mbox{\tiny{th}}}(T,\mu),\nonumber\\
m_{f}(T,\mu)&\equiv& m_{f}^{0}+m_{f}^{\mbox{\tiny{th}}}(T,\mu),
\end{eqnarray}
we will then determine the $T$ and $\mu$ dependence of $\Gamma_{b},
\Gamma_{\pm}$ as well as $\eta_{b}$ and $\eta_{f}$, including the
thermal corrections to bosonic and fermionic masses. According to
the descriptions in \cite{kiessig2009,kiessig2010}, and since in the
Yukawa theory the vertices do not receive any HTL corrections, the
above treatment of thermal masses equals the HTL treatment with an
approximate fermion propagator. In this way, the apparent drawback
of our one-loop perturbative treatment of $\eta_{b}[\Gamma_{b}]$ and
$\eta_{f}[\Gamma_{\pm}]$ is partly compensated. For the fermions, we
mainly focus on the difference between $\Gamma_{+}$ and
$\Gamma_{-}$, arising from normal and collective (plasminos)
excitations of fermions at finite $T$ and $\mu$, respectively. In
the literature, the spectral widths $\Gamma_{+}$ and $\Gamma_{-}$
are often assumed to be equal (see e.g. \cite{jeon2007}). We will
show that depending on $T$ and/or $\mu$, their difference is not
negligible. To study the effect of plasminos on $\eta_{f}$, we will
determine $\eta_{f}$ once for $\Gamma_{+}\neq \Gamma_{-}$ and once
for $\Gamma_{+}= \Gamma_{-}$, and compare the corresponding results.
\subsection{Bosonic contributions}
\subsubsection{Bosonic spectral width}
\begin{figure}[hbt]
\includegraphics[width=7.7cm,height=5cm]{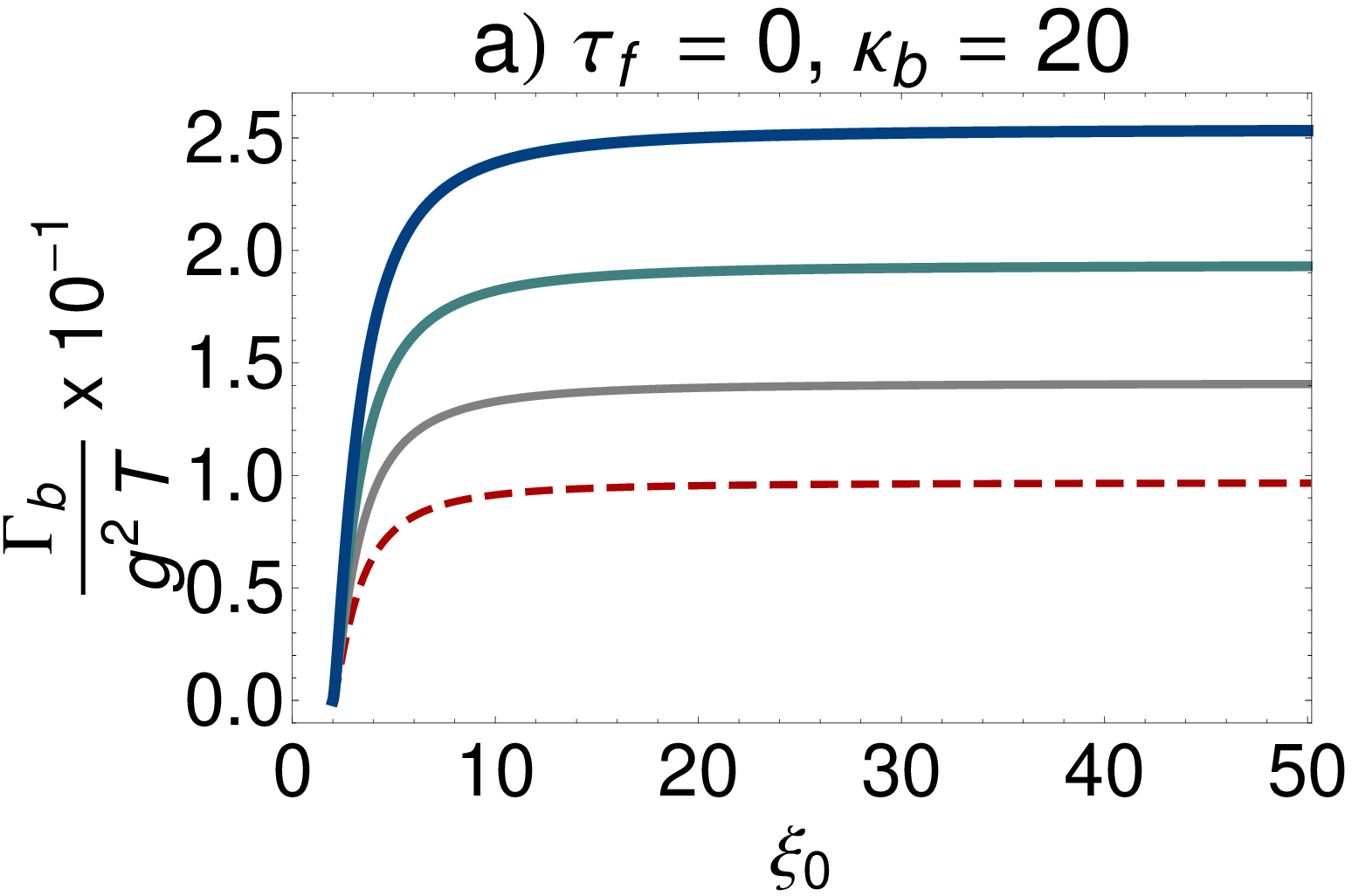}
\includegraphics[width=7.7cm,height=5cm]{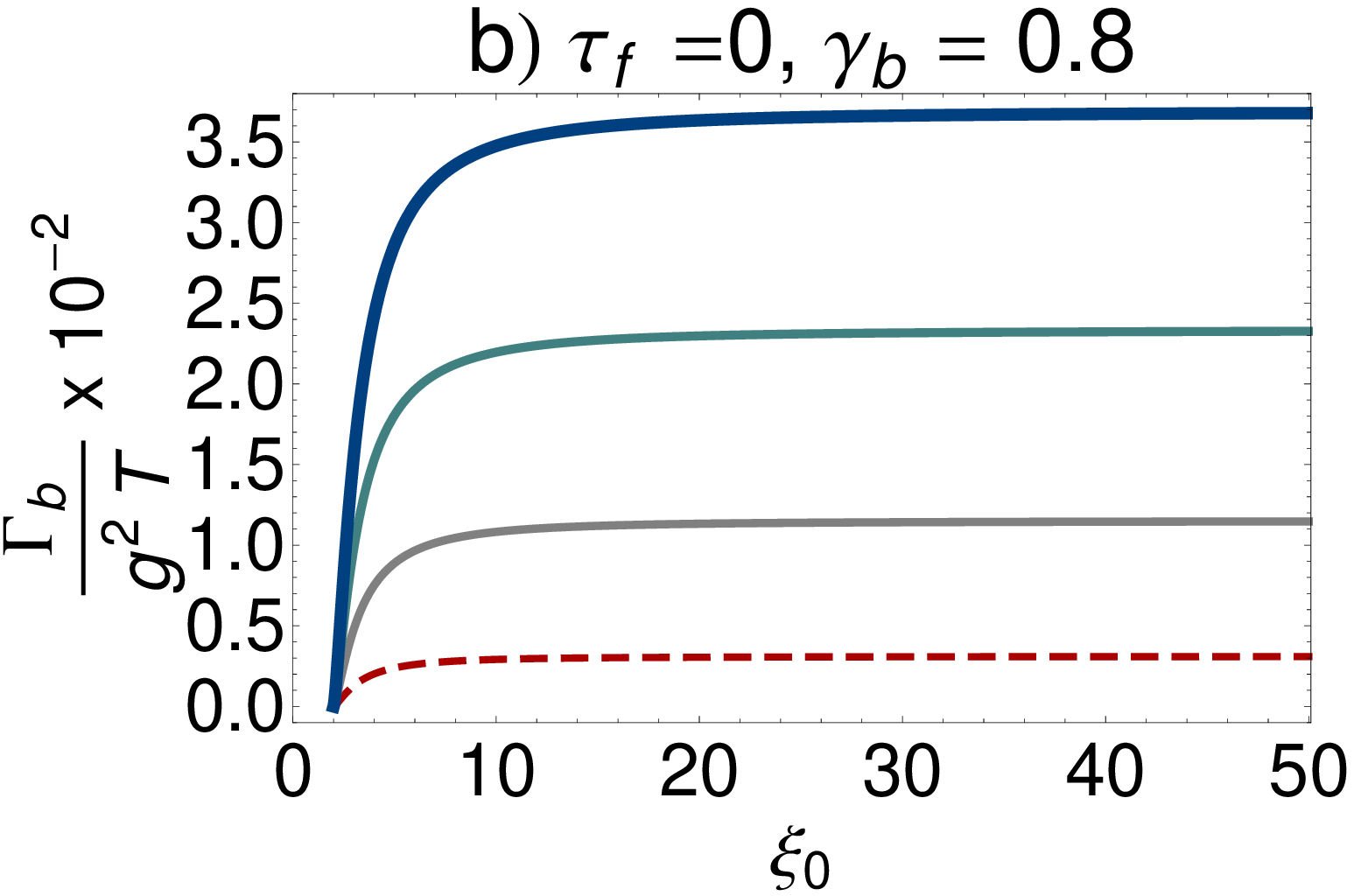}
\caption{(color online).  The $\xi_{0}$ dependence of
$\frac{\Gamma_{b}}{g^{2} T}$ for $\tau_{f}=0$ and (a)
$\kappa_{b}=20$ as well as $\gamma_{b}=0.5,0.6,0.7,0.8$ (from below
to above) and (b) $\gamma_{b}=0.8$ as well as $\kappa_{b}=1,2,3,4$
(from below to above). As it turns out, $\frac{\Gamma_{b}}{g^{2} T}$
remains constant for $\xi_{0}\gtrsim 10$. For fixed values of
$\xi_{0}$ and $\kappa_{b}$ ($\gamma_{b}$), $\frac{\Gamma_{b}}{g^{2}
T}$ increases with increasing $\gamma_{b}$ ($\kappa_{b}$) [see panel
a (b)].}\label{fig5}
\end{figure}
\begin{figure*}[hbt]
\includegraphics[width=5.5cm,height=4cm]{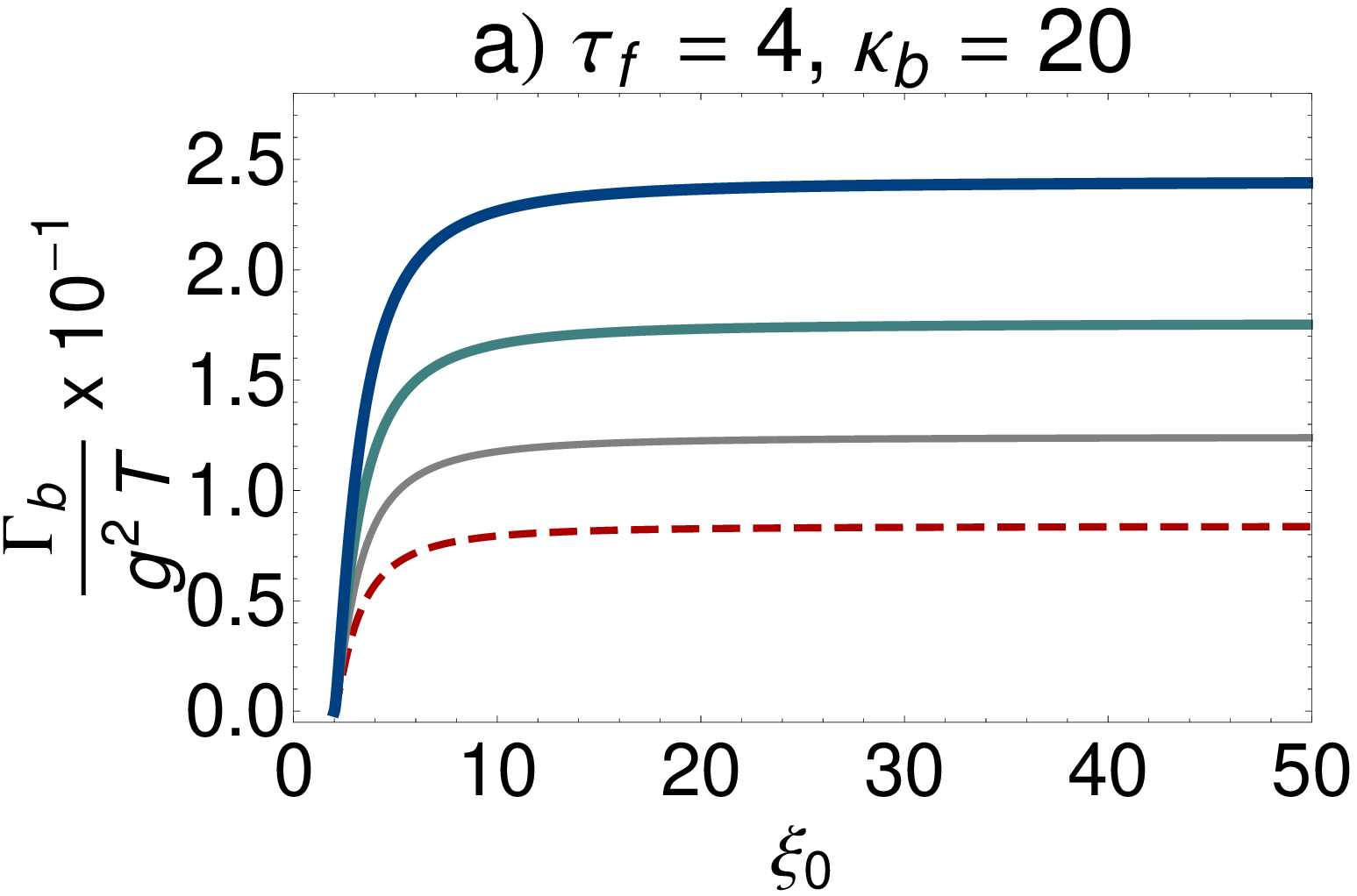}
\includegraphics[width=5.5cm,height=4cm]{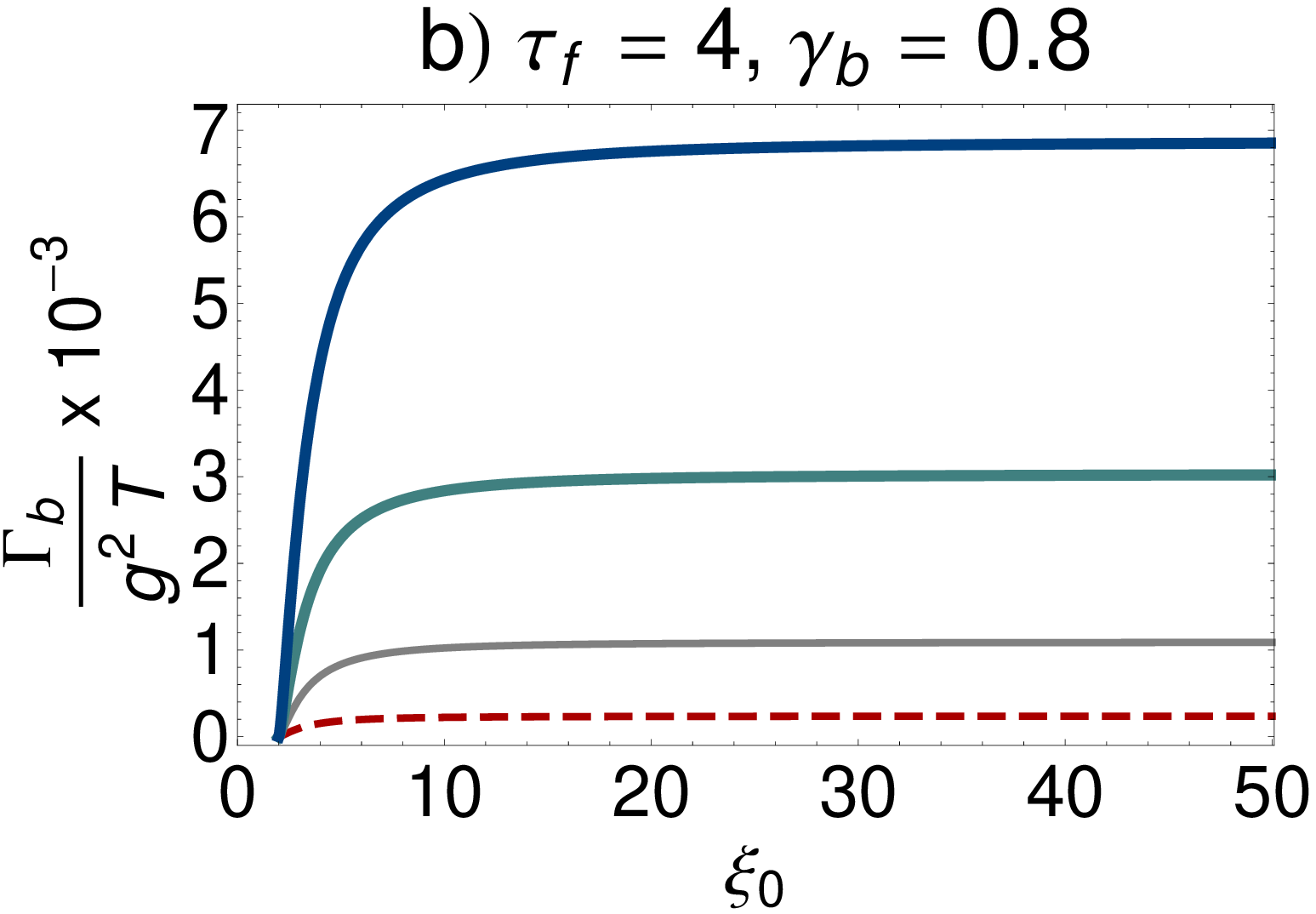}
\includegraphics[width=5.5cm,height=4cm]{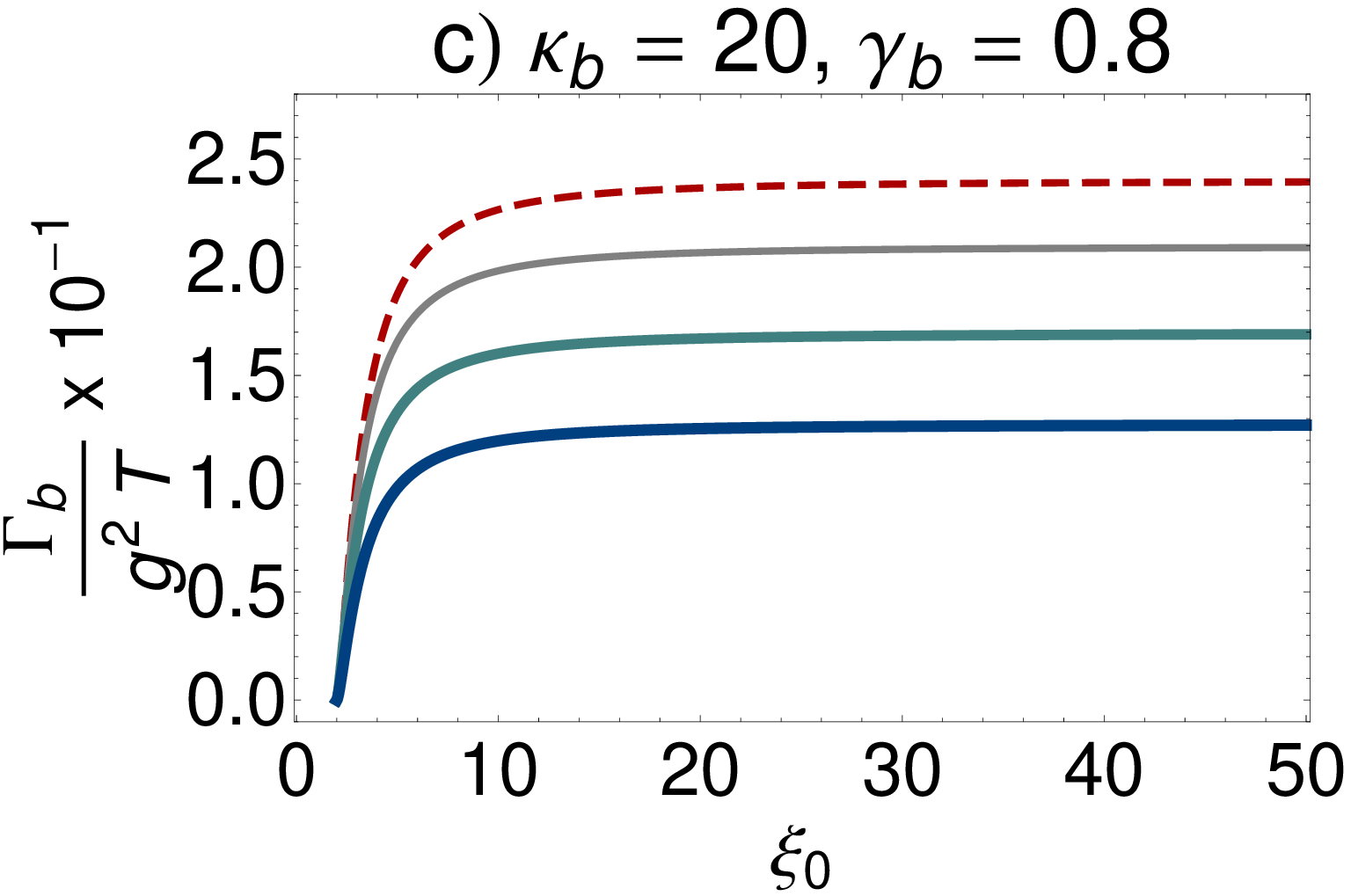}
\caption{(color online).  The $\xi_{0}$ dependence of
$\frac{\Gamma_{b}}{g^{2} T}$ for $\tau_{f}=4$ and (a)
$\kappa_{b}=20$ as well as $\gamma_{b}=0.5,0.6,0.7,0.8$ (from below
to above), and (b) $\gamma_{b}=0.8$ as well as $\kappa_{b}=1,2,3,4$
(from below to above). As it turns out, $\frac{\Gamma_{b}}{g^{2} T}$
remains constant for $\xi_{0}\gtrsim 10$. For a fixed value of
$\xi_{0}$, $\frac{\Gamma_{b}}{g^{2} T}$ increases with increasing
$\gamma_{b}$ for all values of $\kappa_{b}$ (panel a) and with
increasing $\kappa_{b}$ for all values of $\gamma_{b}$ (panel b).
(c) The $\xi_{0}$ dependence of $\frac{\Gamma_{b}}{g^{2}T}$ for
fixed $\kappa_{b}=20, \gamma_{b}=0.8$ and $\tau_{f}=4,6,8,10$ (from
above to below).  As in the previous cases, $\frac{\Gamma_{b}}{g^{2}
T}$ remains constant for $\xi_{0}\gtrsim 10$ for all values of
$\kappa_{b},\gamma_{b}$ and $\tau_{f}$. For a fixed value of
$\kappa_{b},\gamma_{b}$ and $\xi_{0}$, $\frac{\Gamma_{b}}{g^{2} T}$
decreases with increasing $\tau_{f}$.}\label{fig6}
\end{figure*}
\begin{figure}[hbt]
\includegraphics[width=7.7cm,height=5cm]{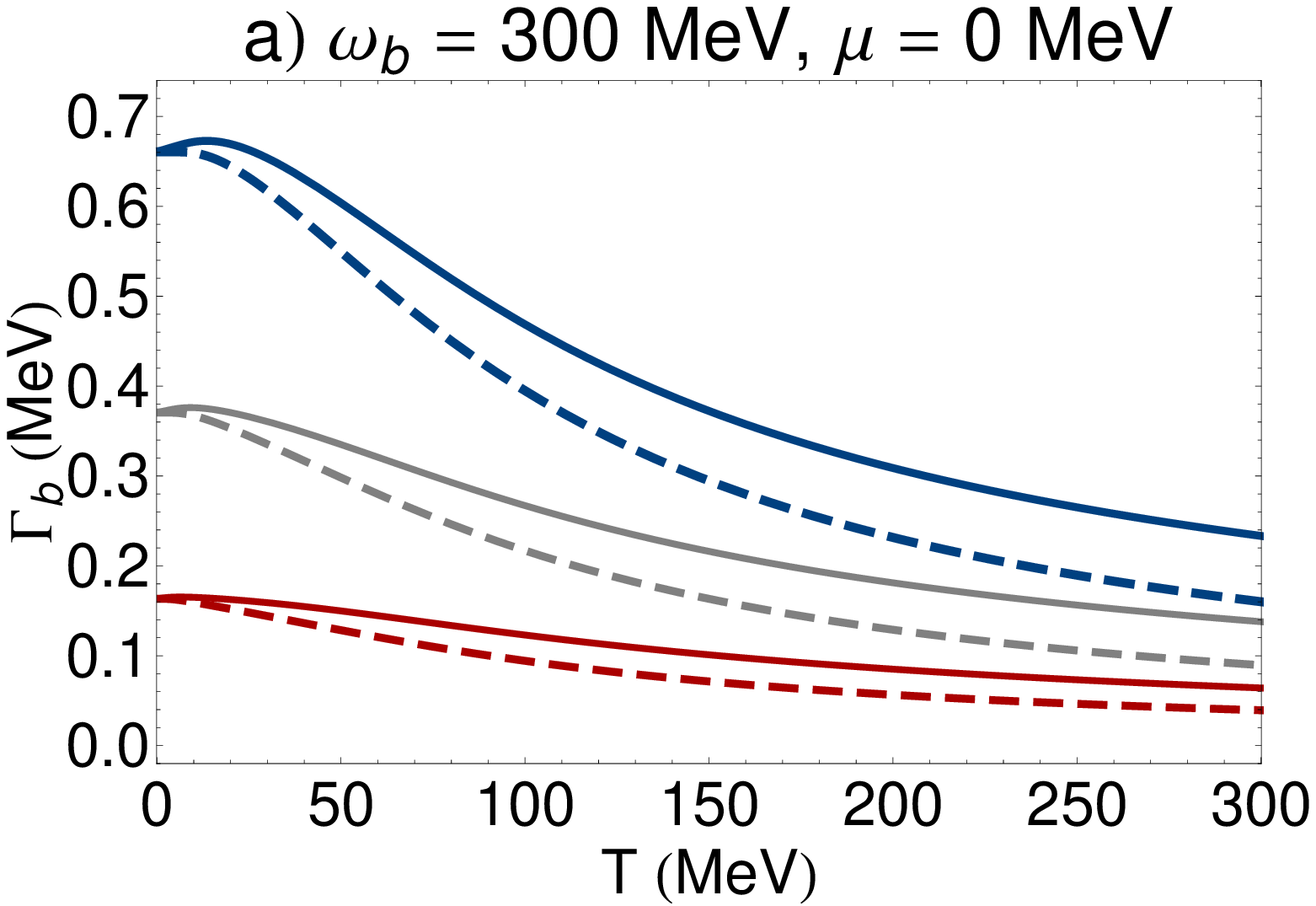}
\includegraphics[width=7.7cm,height=5cm]{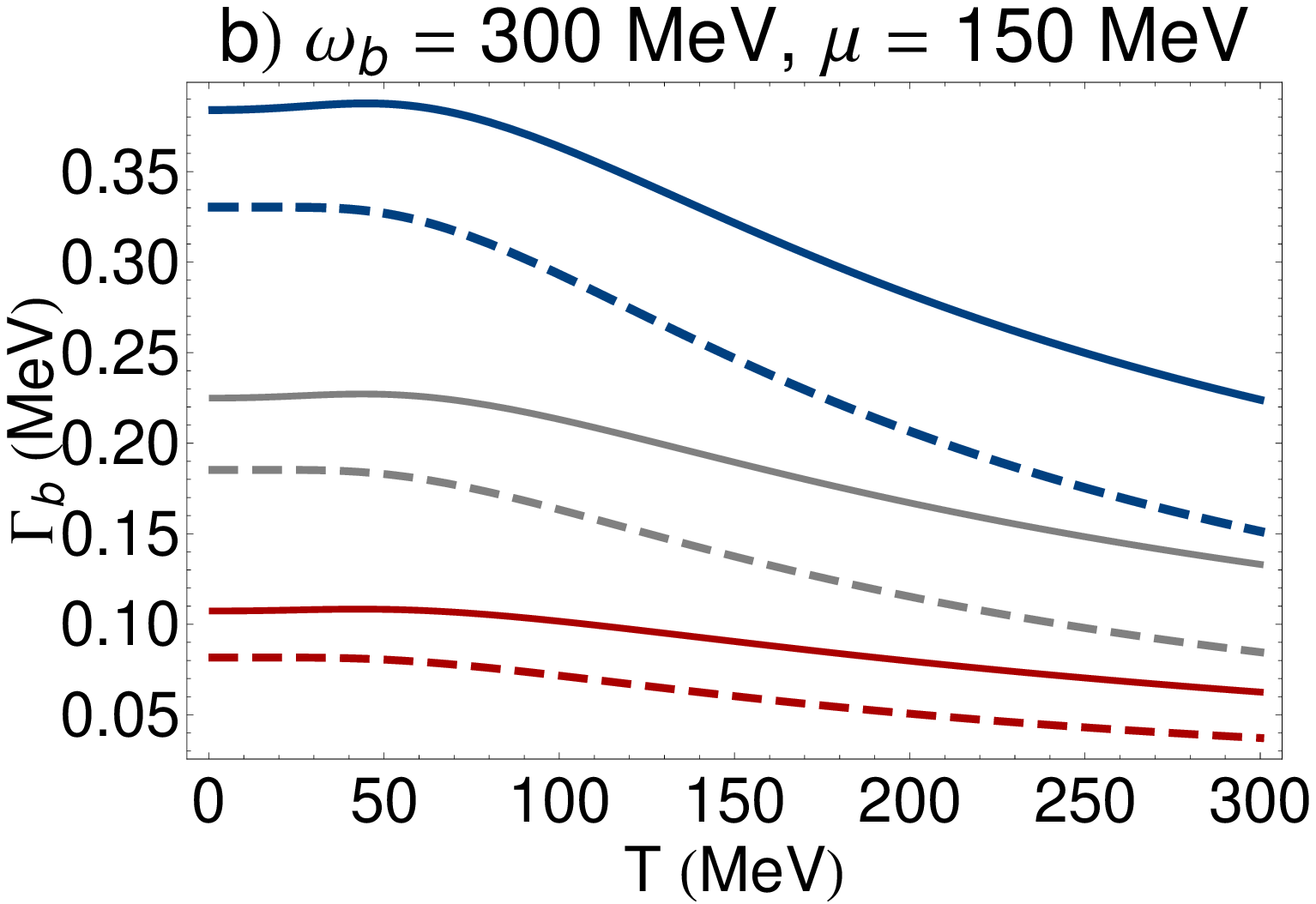}
\caption{ (color online). The $T$ dependence of $\Gamma_{b}$ for
$\omega_{b}=300$ MeV and (a) $\mu=0$ MeV as well as (b) $\mu=150$
MeV. The red, gray and blue lines correspond to $m_{b}^{0}=100, 150,
200$ MeV and $m_{f}^{0}=5$ MeV, respectively. The dashed lines
include only the constant mass contributions of bosons,
$m_{b}^{0}=100,150,200$ MeV, and fermions $m_{f}^{0}=5$ MeV. The
solid lines include,  in addition to the constant mass
contributions, the thermal corrections of the boson and fermion
masses as functions of $T$ and $\mu$ [see (\ref{H1}) to (\ref{H3})].
Here, the Yukawa coupling $g=0.5$ is used. }\label{fig7}
\end{figure}
\begin{figure}[hbt]
\includegraphics[width=7.7cm,height=5cm]{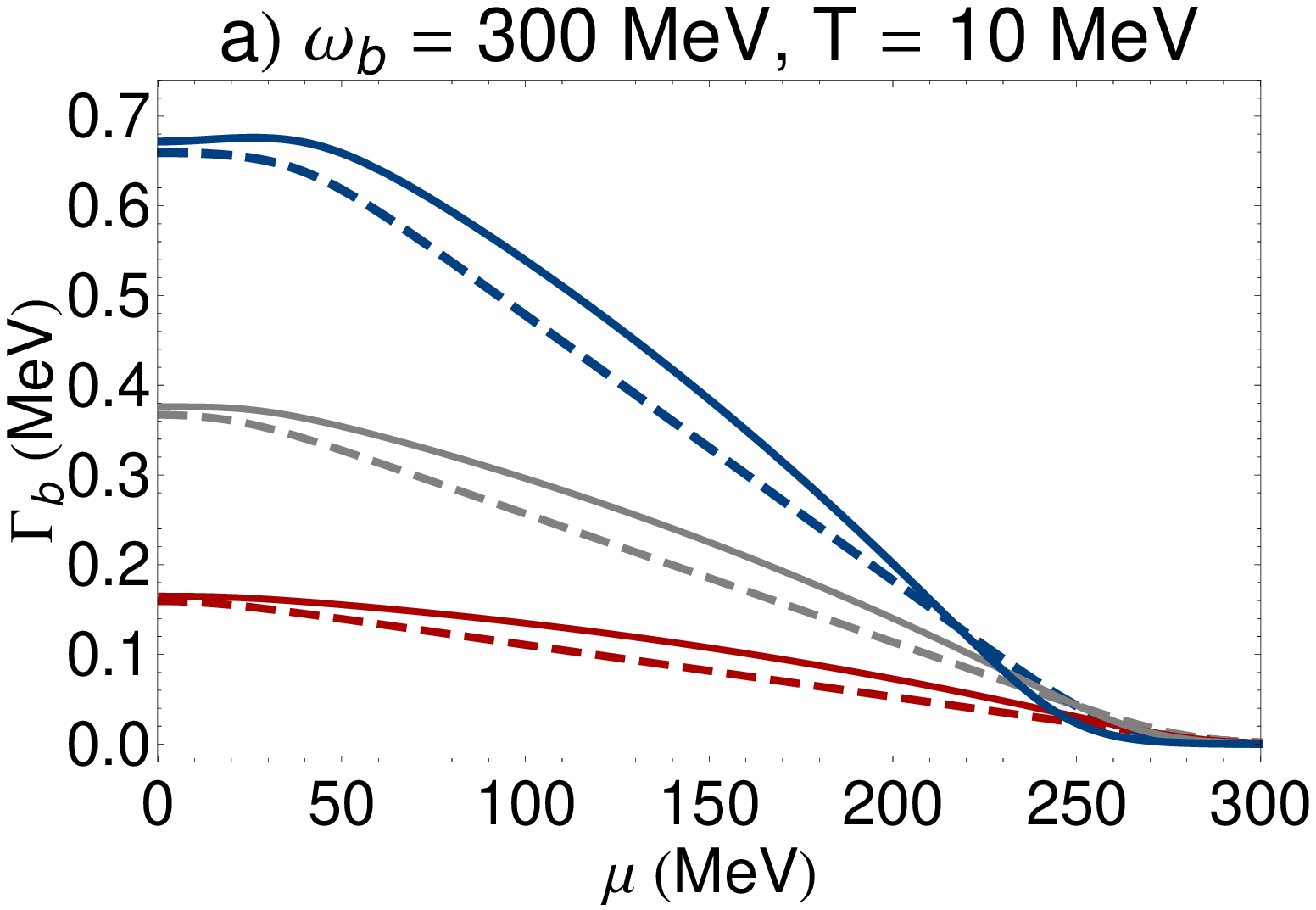}
\includegraphics[width=7.7cm,height=5cm]{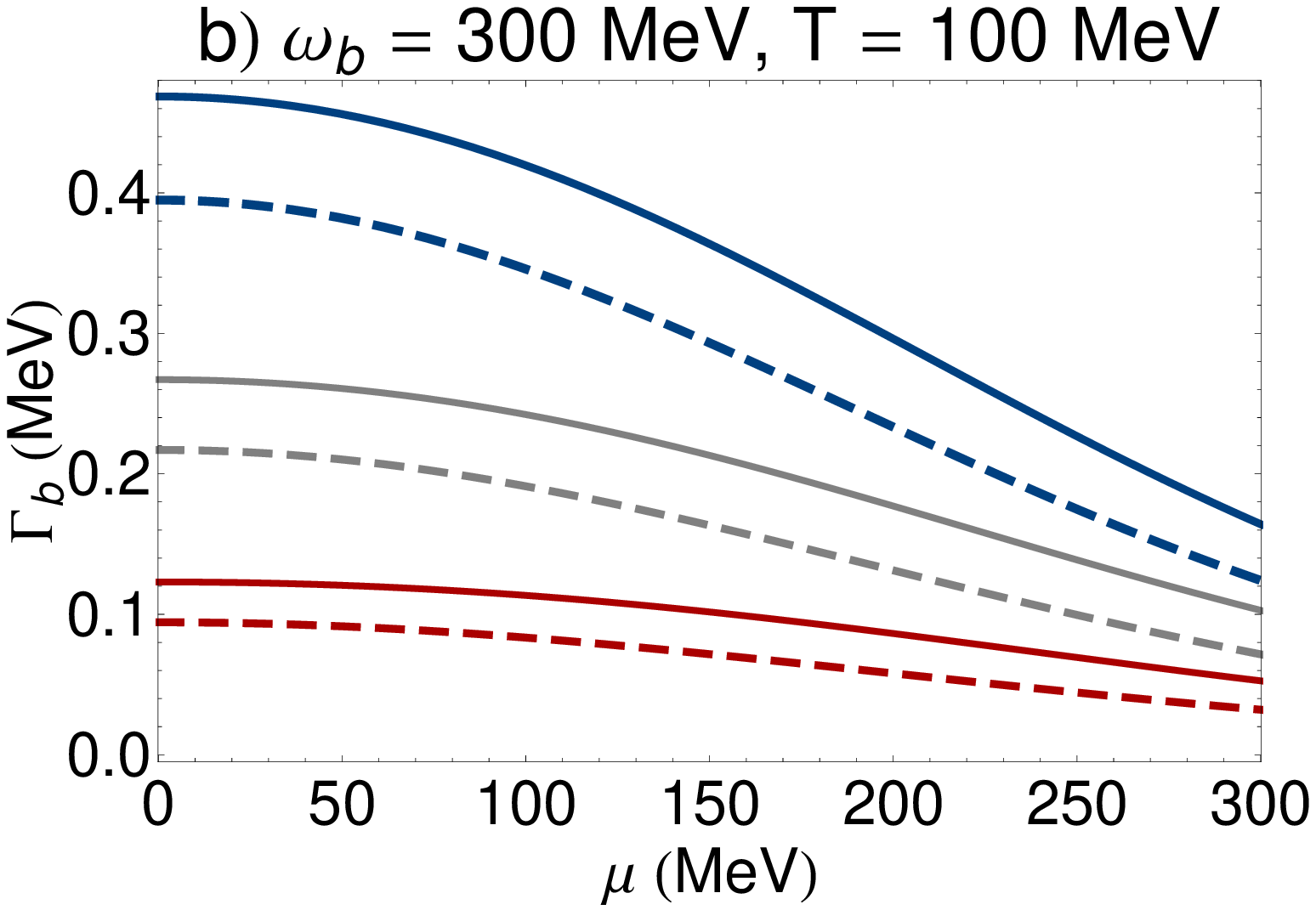}
\caption{ (color online). The $\mu$ dependence of $\Gamma_{b}$ for
$\omega_{b}=300$ MeV and (a) $T=10$ MeV as well as (b) $T=100$ MeV.
The red, gray and blue lines correspond to $m_{b}^{0}=100, 150, 200$
MeV and $m_{f}^{0}=5$ MeV, respectively. The dashed lines include
only the constant mass contributions of bosons,
$m_{b}^{0}=100,150,200$ MeV, and fermions, $m_{f}^{0}=5$ MeV. The
solid lines include, in addition to the constant mass contributions,
the thermal corrections of the boson and fermion masses as functions
of $T$ and $\mu$ [see (\ref{H1}) to (\ref{H3})]. Here, the Yukawa
coupling $g=0.5$ is used.}\label{fig8}
\end{figure}
\begin{figure}[hbt]
\includegraphics[width=7.7cm,height=5cm]{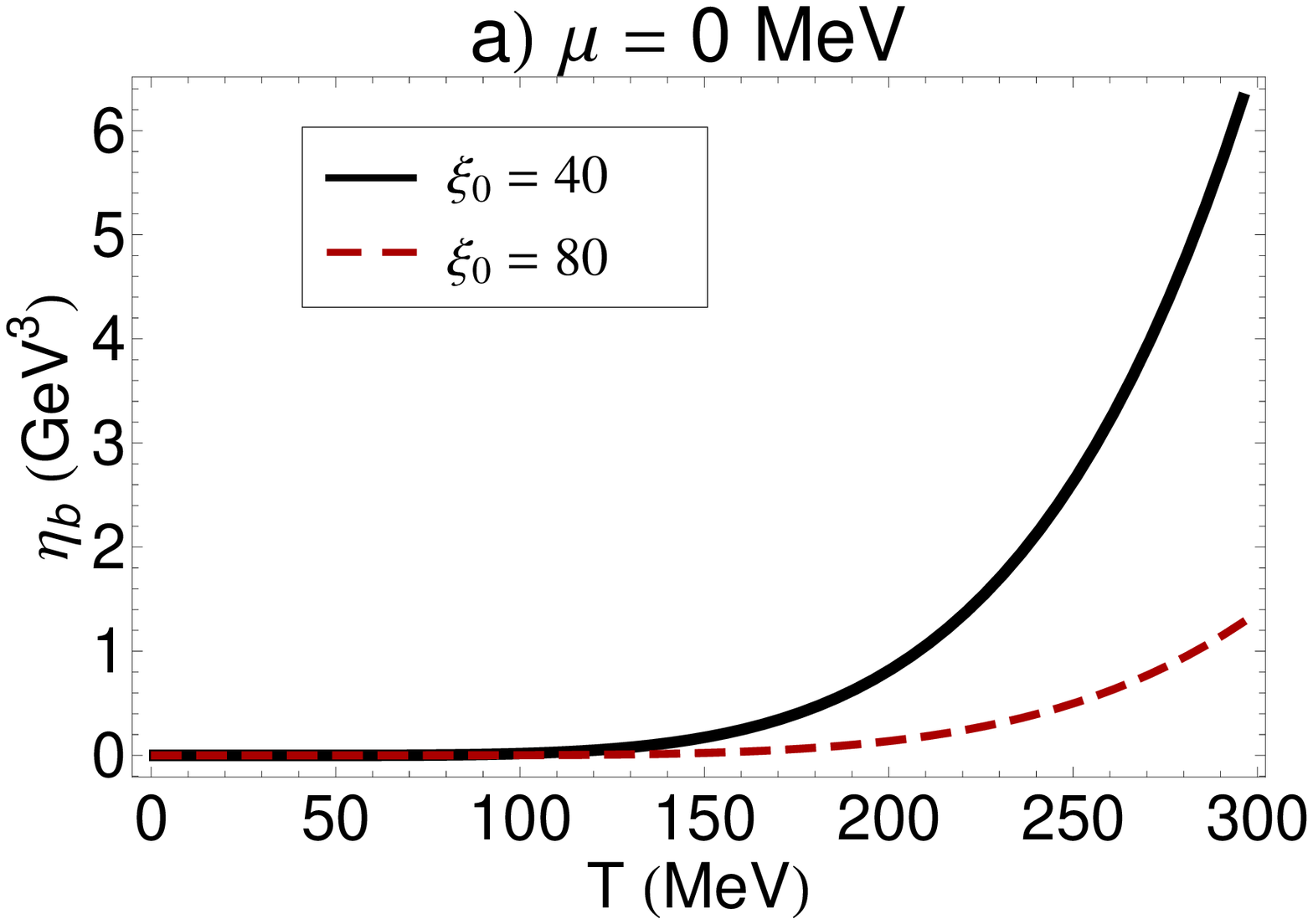}
\includegraphics[width=7.7cm,height=5cm]{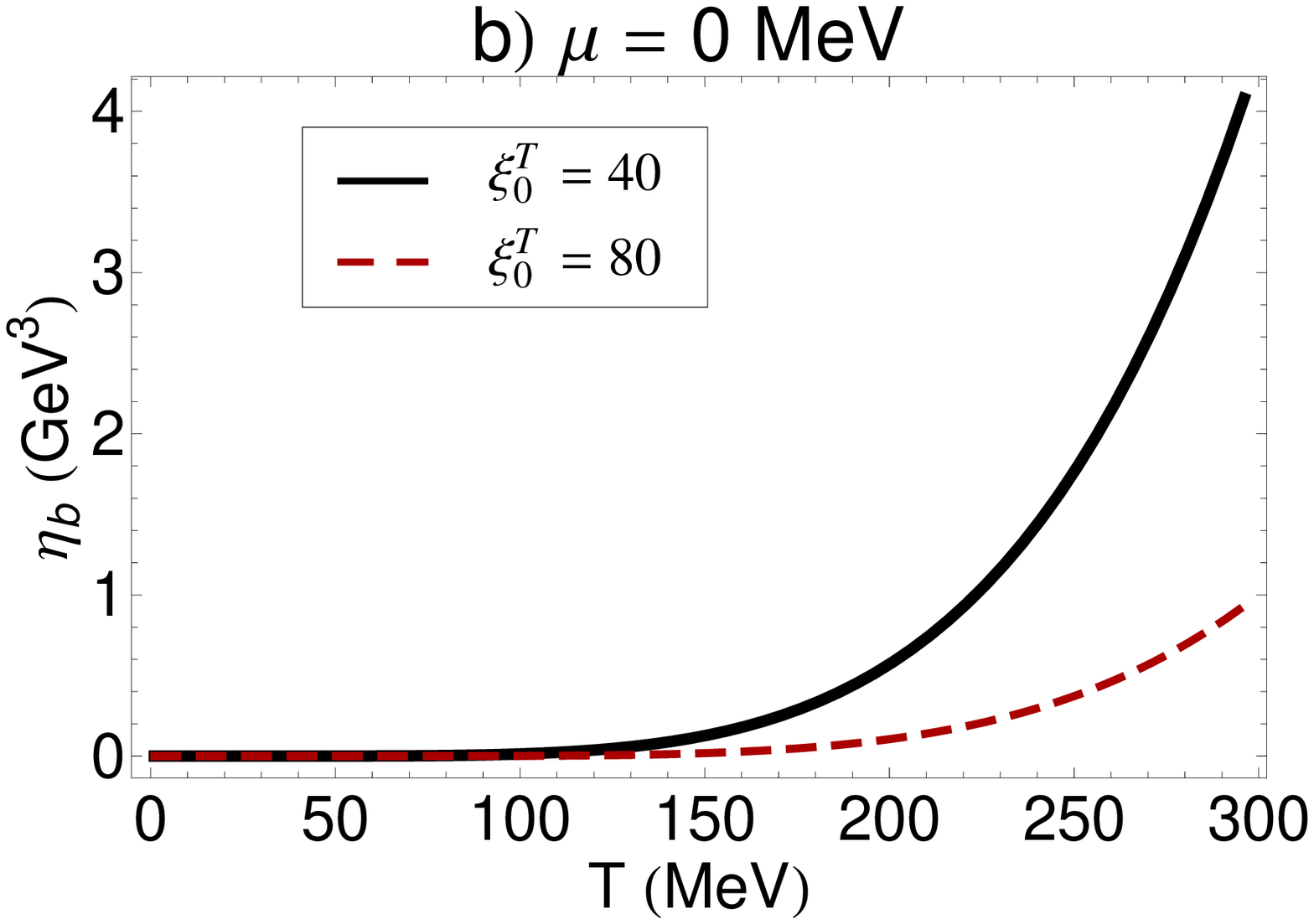}
\caption{ (color online). (a) The $T$ dependence of $\eta_{b}$ is
plotted for $\mu=0$ and $T$ independent $\xi_{0}=40,80$ arising from
$m_{b}^{0}=200,400$ MeV and $m_{f}^{0}=5$ MeV. (b) The $T$
dependence of $\eta_{b}$, including the $T$ dependent thermal
corrections to bosonic and fermionic masses, is plotted for
$m_{b}^{0}=200,400$ MeV and $m_{f}^{0}=5$ MeV. Here, $\xi_{0}^{T}$
denotes the ratio $m_{b}^{0}/m_{f}^{0}$ in $\xi(T,\mu)$ from
(\ref{H2}) and (\ref{H3}).}\label{fig9}
\end{figure}
\begin{figure}[hbt]
\includegraphics[width=7.7cm,height=5cm]{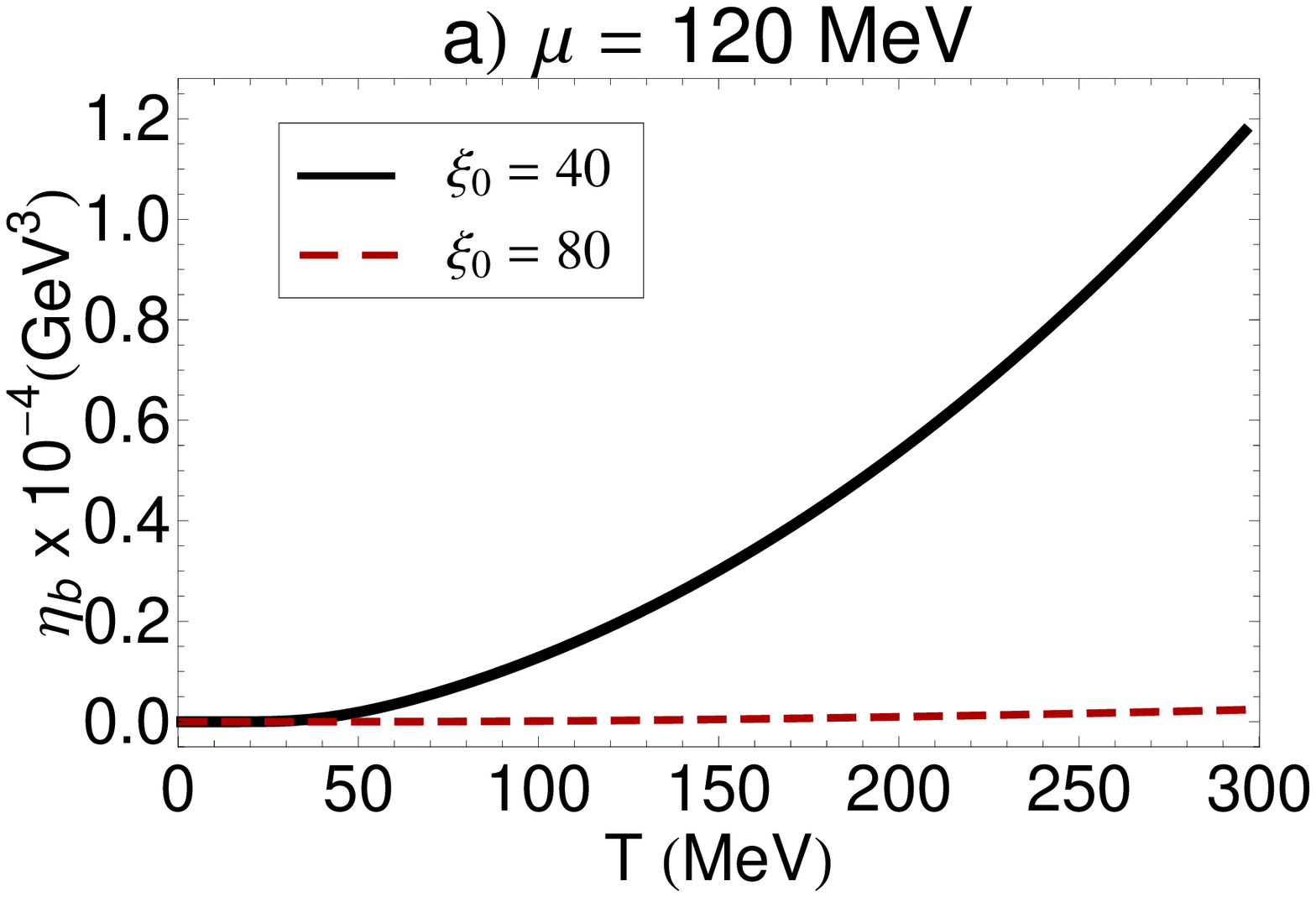}
\includegraphics[width=7.7cm,height=5cm]{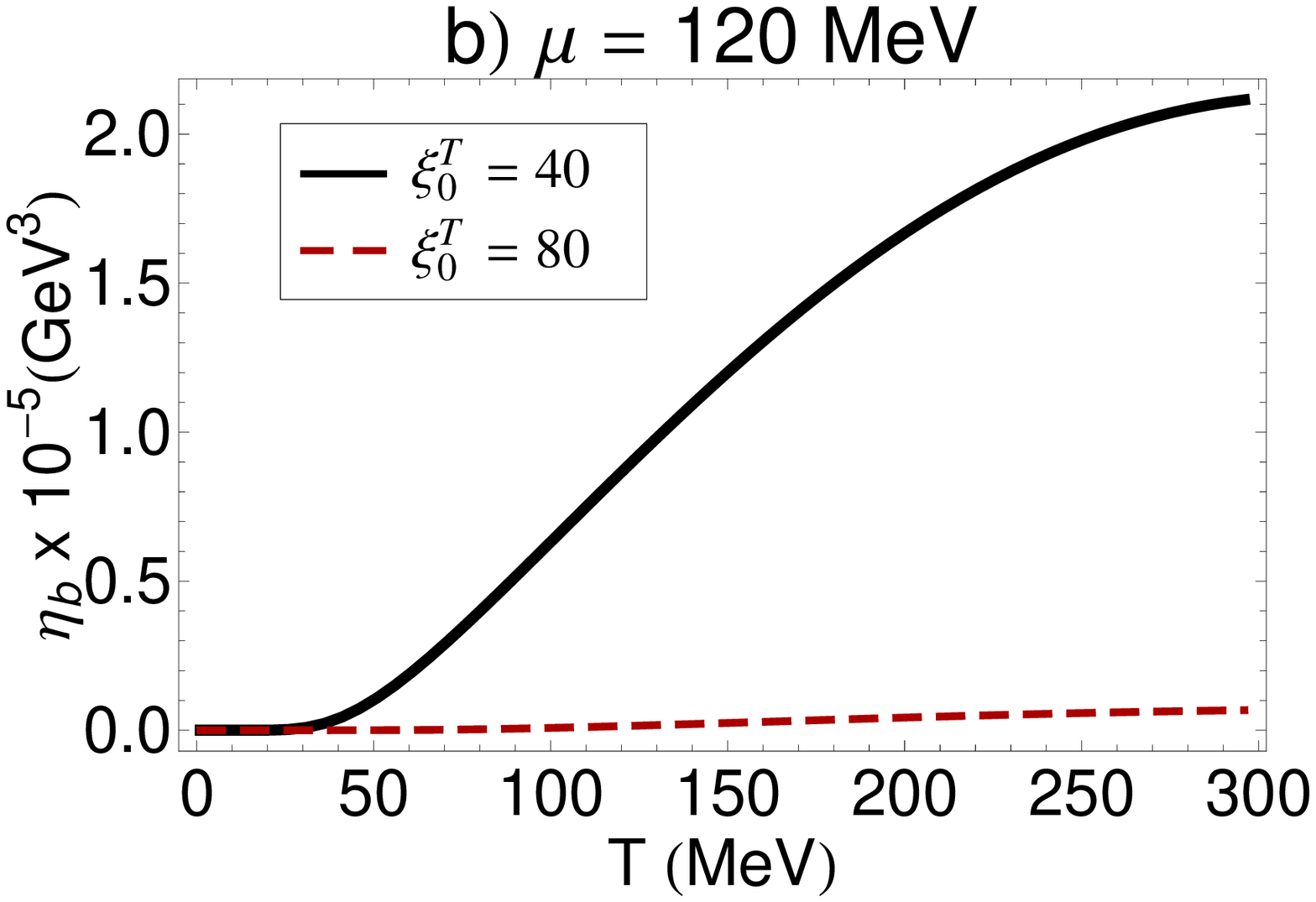}
\caption{ (color online). (a) The $T$ dependence of $\eta_{b}$ is
plotted for $\mu=120$ MeV and $(T,\mu)$ independent $\xi_{0}=40,80$
arising from $m_{b}^{0}=200,400$ MeV and $m_{f}^{0}=5$ MeV. (b) The
$T$ dependence of $\eta_{b}$, including the $T$ and $\mu$ dependent
thermal corrections to bosonic and fermionic masses, is plotted for
$m_{b}^{0}=200,400$ MeV and $m_{f}^{0}=5$ MeV, leading to
$\xi_{0}^{T}=40,80$. }\label{fig10}
\end{figure}
\begin{figure}[hbt]
\includegraphics[width=7.7cm,height=5cm]{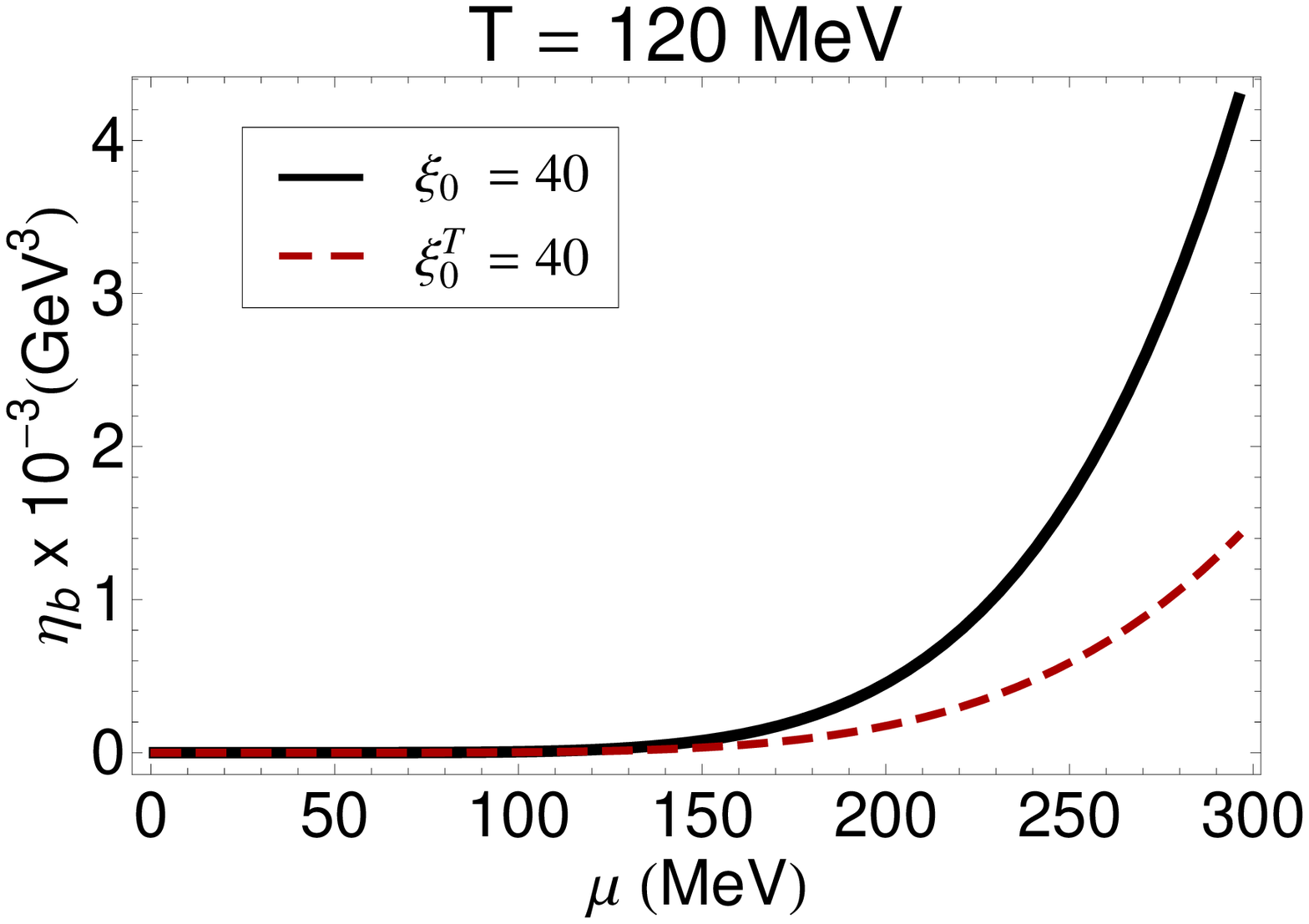}
\caption{ (color online). The $\mu$ dependence of $\eta_{b}$ is
plotted for $T=120$ MeV and $\xi_{0}=\xi_{0}^{T}=40$.}\label{fig11}
\end{figure}
\begin{figure*}[hbt]
\includegraphics[width=5.5cm,height=4cm]{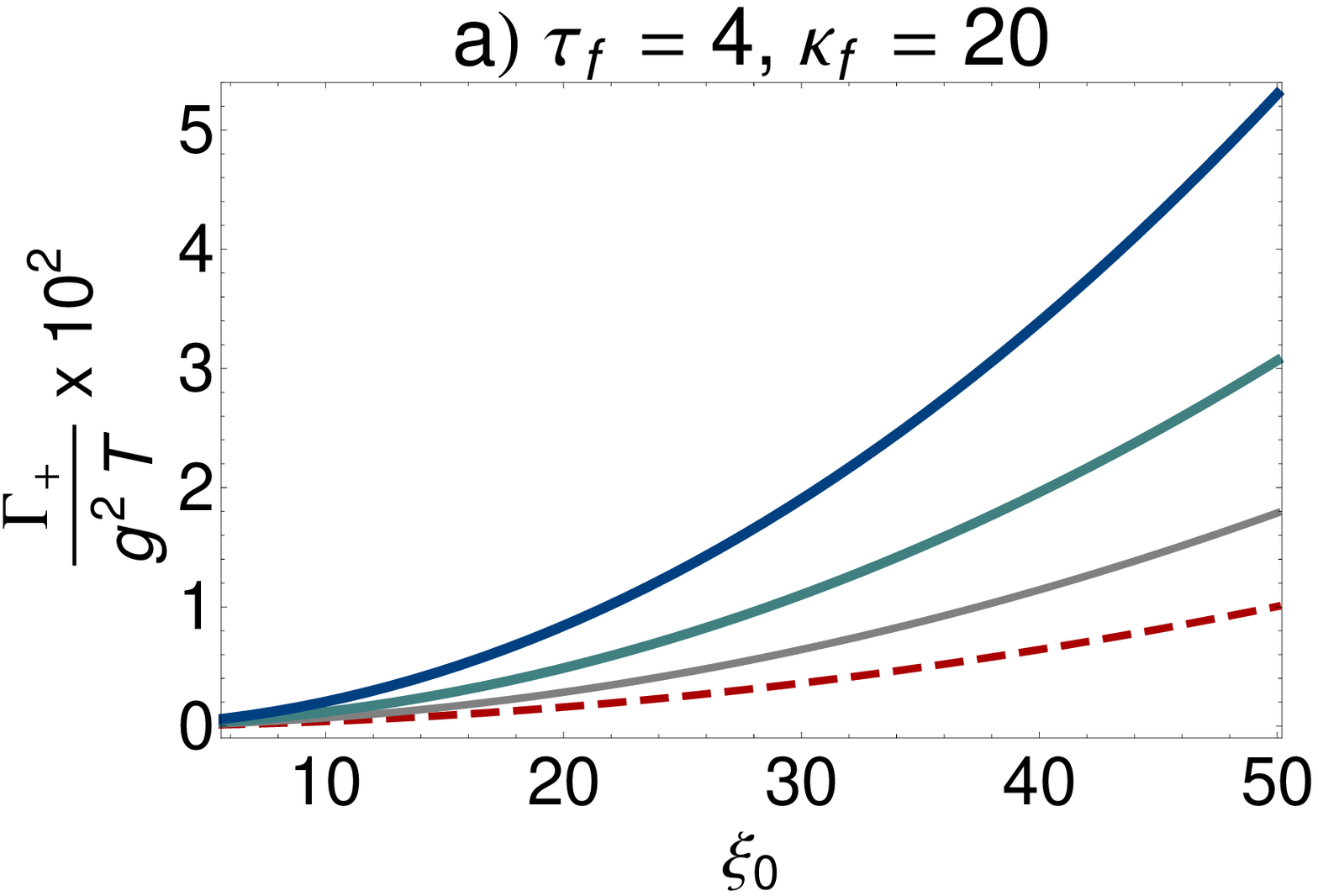}
\includegraphics[width=5.5cm,height=4cm]{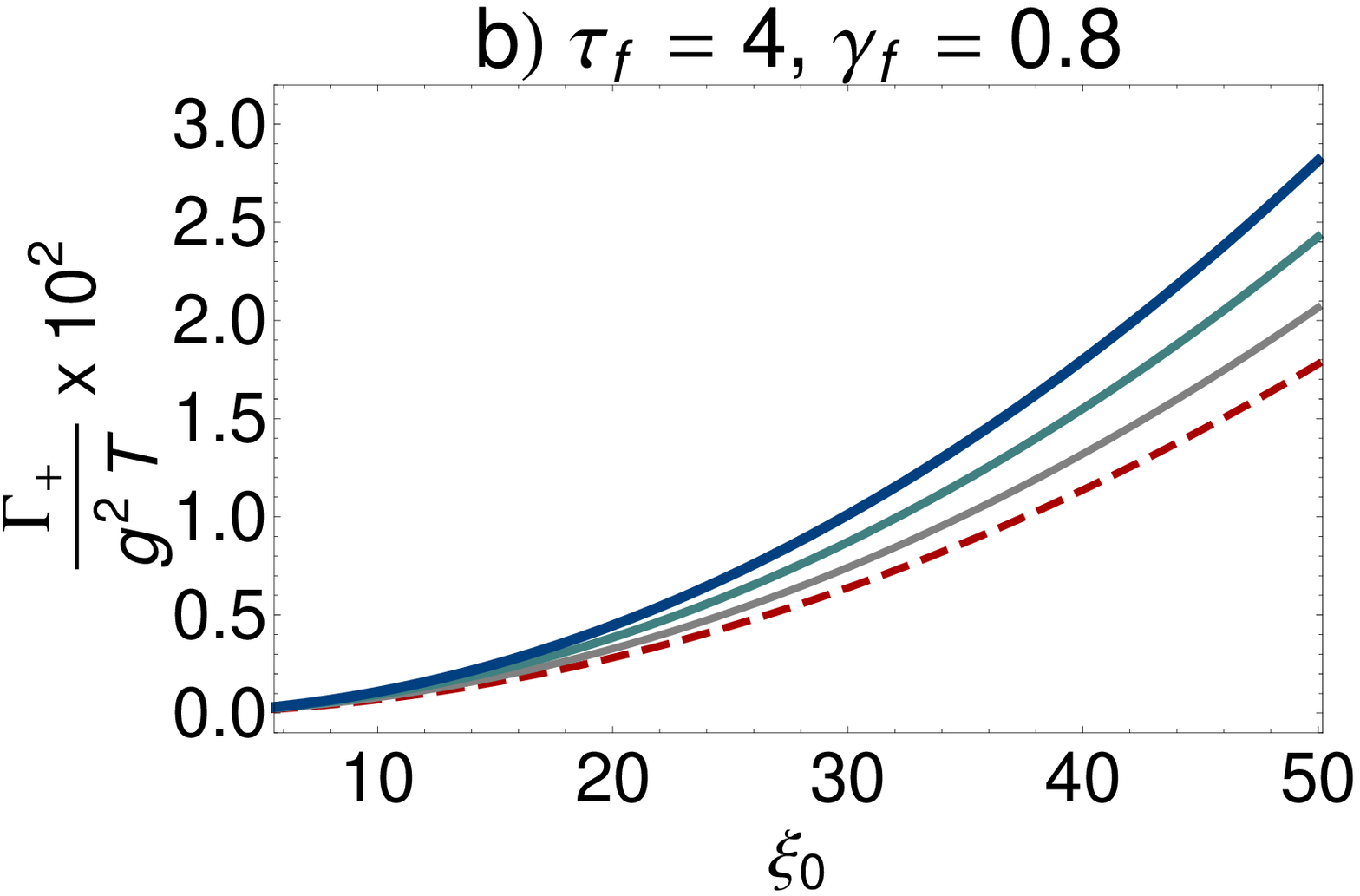}
\includegraphics[width=5.5cm,height=4cm]{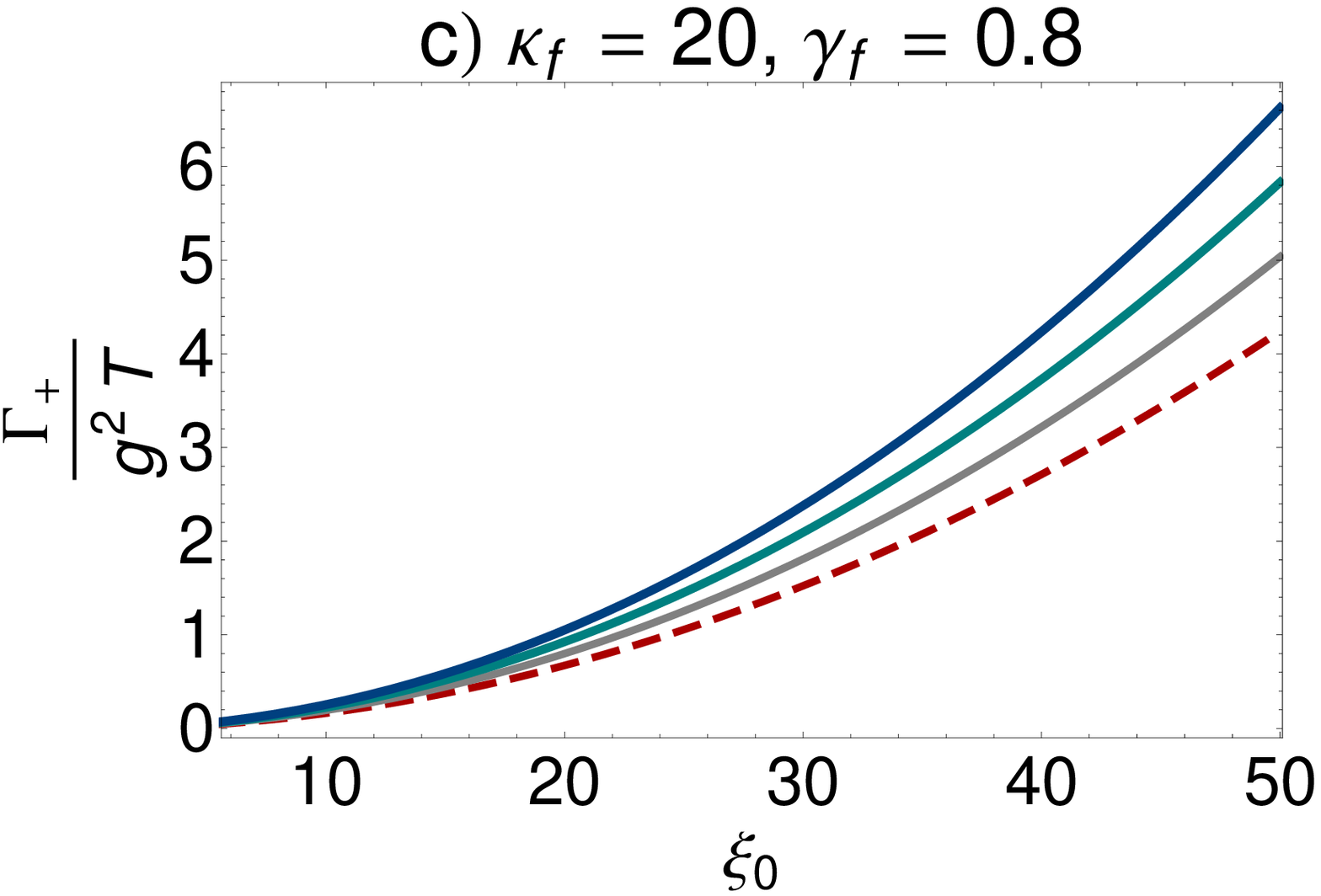}
\caption{ (color online). The $\xi_{0}$ dependence of
$\frac{\Gamma_{+}}{g^{2}T}$ for $\tau_{f}=4$ and (a) $\kappa_{f}=20$
as well as $\gamma_{f}=0.5,0.6,0.7,0.8$ (from below to above), and
(b) $\gamma_{f}=0.8$ as well as $\kappa_{f}=2,4,6,8$ (from below to
above).(c) The $\xi_{0}$ dependence of $\frac{\Gamma_{+}}{g^{2}T}$
for $\kappa_{f}=20,\gamma_{f}=0.8$ and $\tau_{f}=0,3,6,9$ (from
below to above). As it turns out, for a fixed $\xi_{0}$,
$\frac{\Gamma_{+}}{g^{2}T}$ increases whenever one of the parameters
$\gamma_{f},\kappa_{f}$ or $\tau_{f}$ increases and the other two
parameters are held fixed. It can be shown that the same is also
true for $\frac{\Gamma_{-}}{g^{2}T}$.}\label{fig12}
\end{figure*}
\begin{figure*}[hbt]
\includegraphics[width=5.5cm,height=4cm]{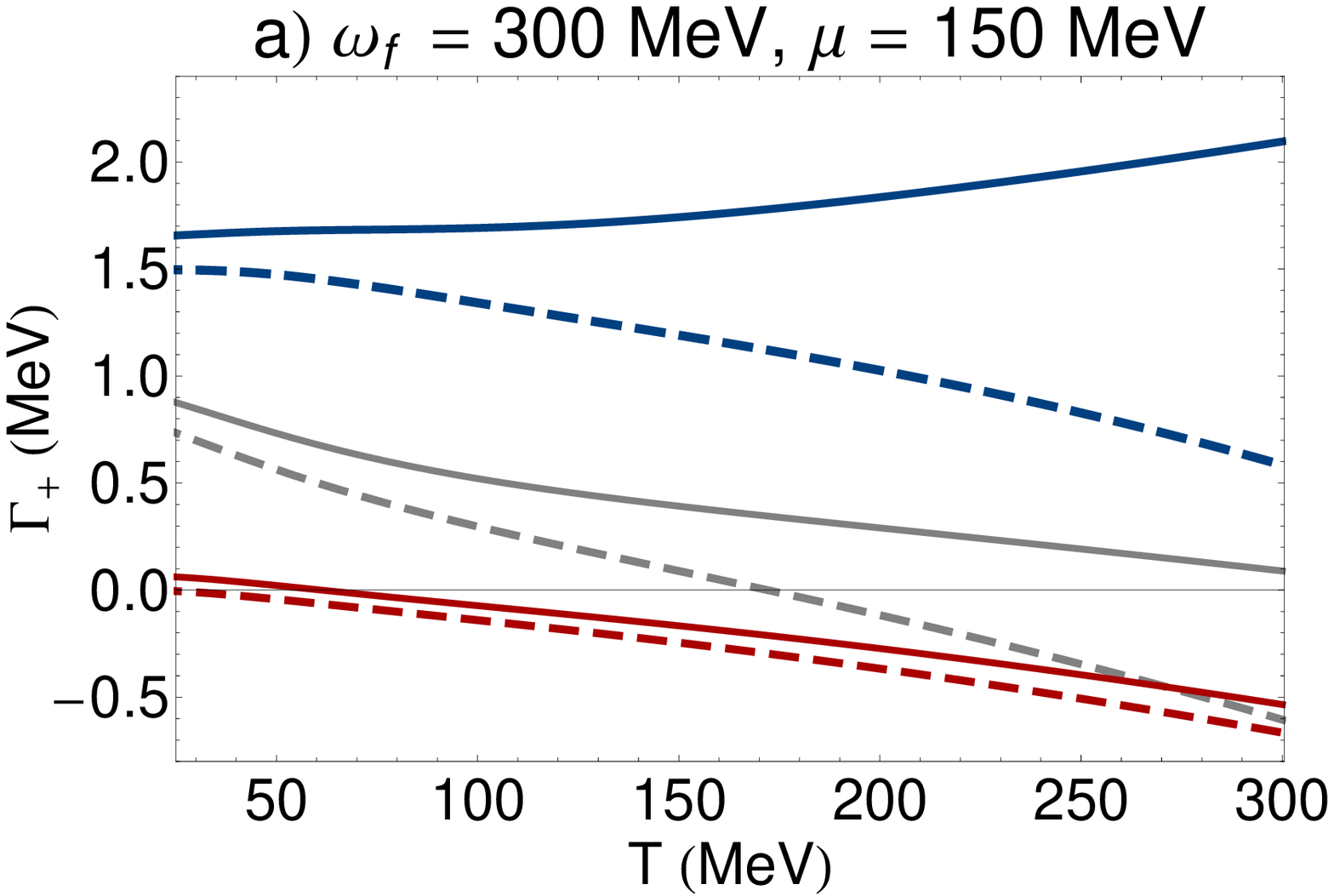}
\includegraphics[width=5.5cm,height=4cm]{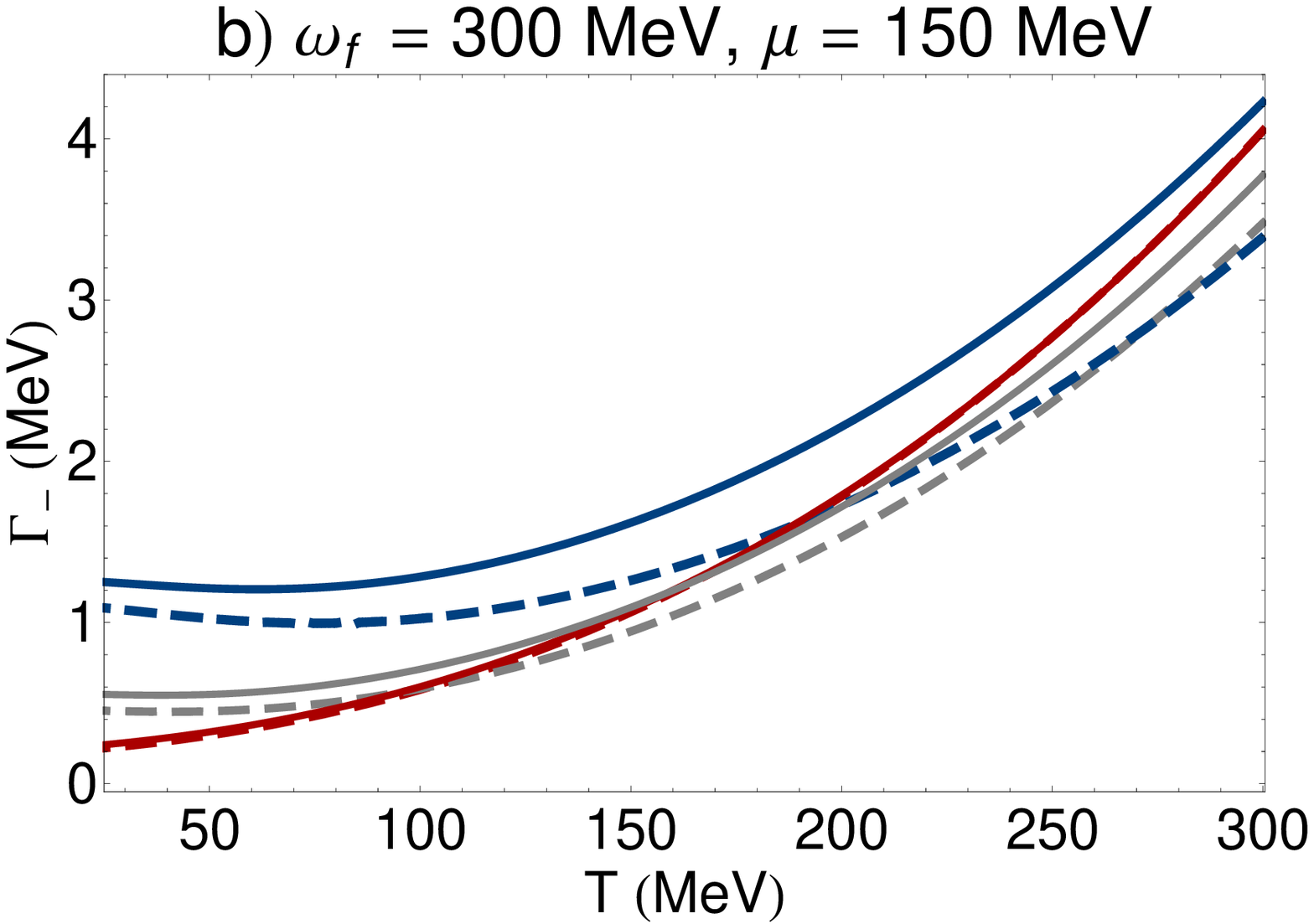}
\includegraphics[width=5.5cm,height=4cm]{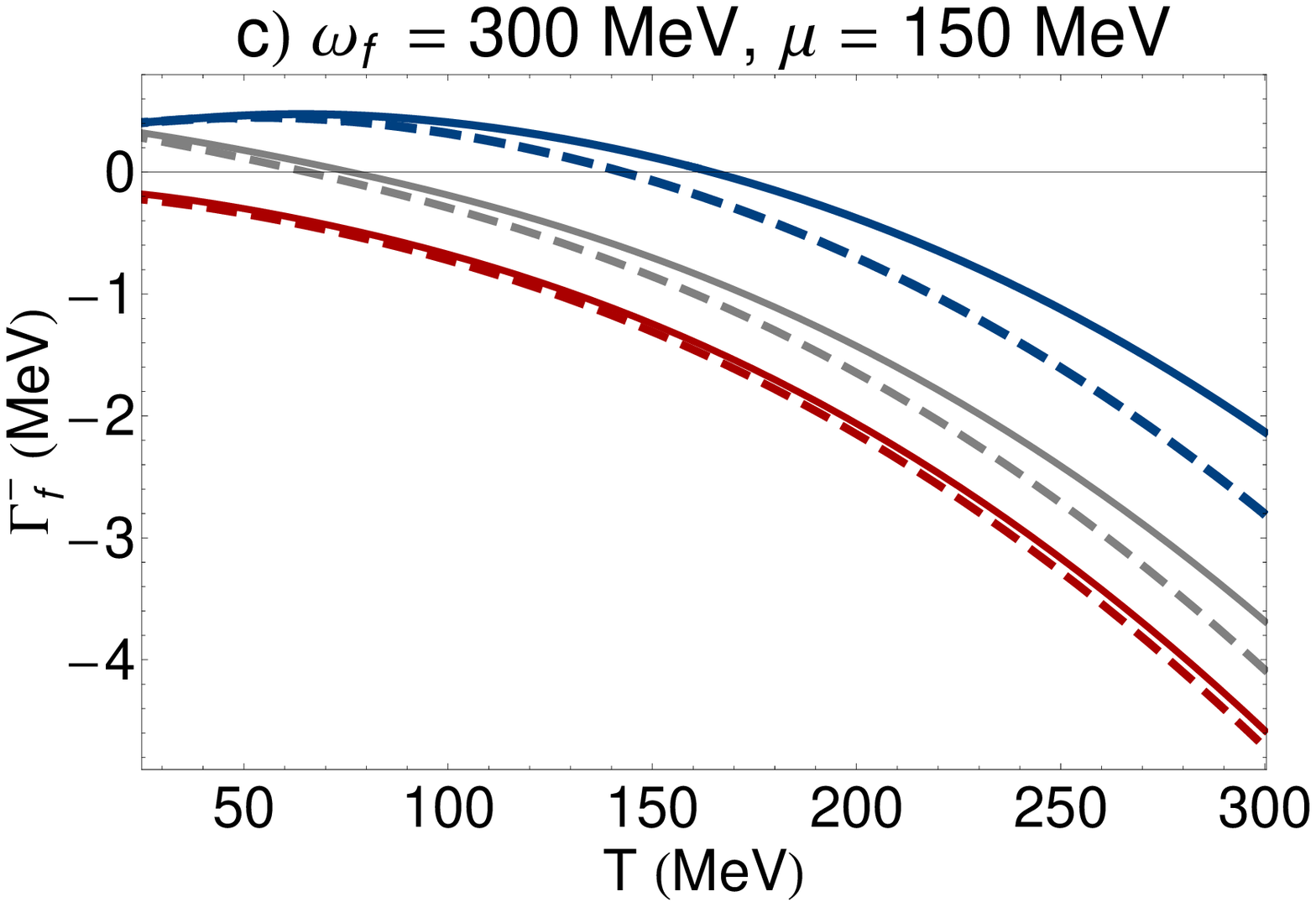}
\caption{ (color online). The $T$ dependence of (a) $\Gamma_{+}$ (b)
$\Gamma_{-}$ and (c) $\Gamma_{f}^{-}=\Gamma_{+}-\Gamma_{-}$ is
plotted for constant $\omega_{f}=300$ MeV and $\mu=150$ MeV. The
red, gray and blue solid and dashed lines correspond to
$m_{b}^{0}=300,450,600$ MeV and $m_{f}^{0}=5$ MeV.  Whereas the
dashed lines correspond to $\Gamma_{\pm}$ and $\Gamma_{f}^{-}$ as
functions of $(T,\mu)$ independent $\xi_{0}=60,90,120$, the solid
lines correspond to the same quantities including the thermal
corrections to bosonic and fermionic masses with $\xi_{0}^{T}=60,90,
120$. It turns out that the absolute value of the difference between
$\Gamma_{+}$ and $\Gamma_{-}$, i.e. $|\Gamma_{f}^{-}|$, increases
with increasing $T$, and decreases with increasing $\xi_{0}$ and
$\xi_{0}^{T}$. Moreover, for small $\xi_{0}$ or $\xi_{0}^{T}$ and
fixed $(T,\mu)$, $\Gamma_{-}$ is always larger than $\Gamma_{+}$.
}\label{fig13}
\end{figure*}
\begin{figure*}[htb]
\includegraphics[width=5.5cm,height=4cm]{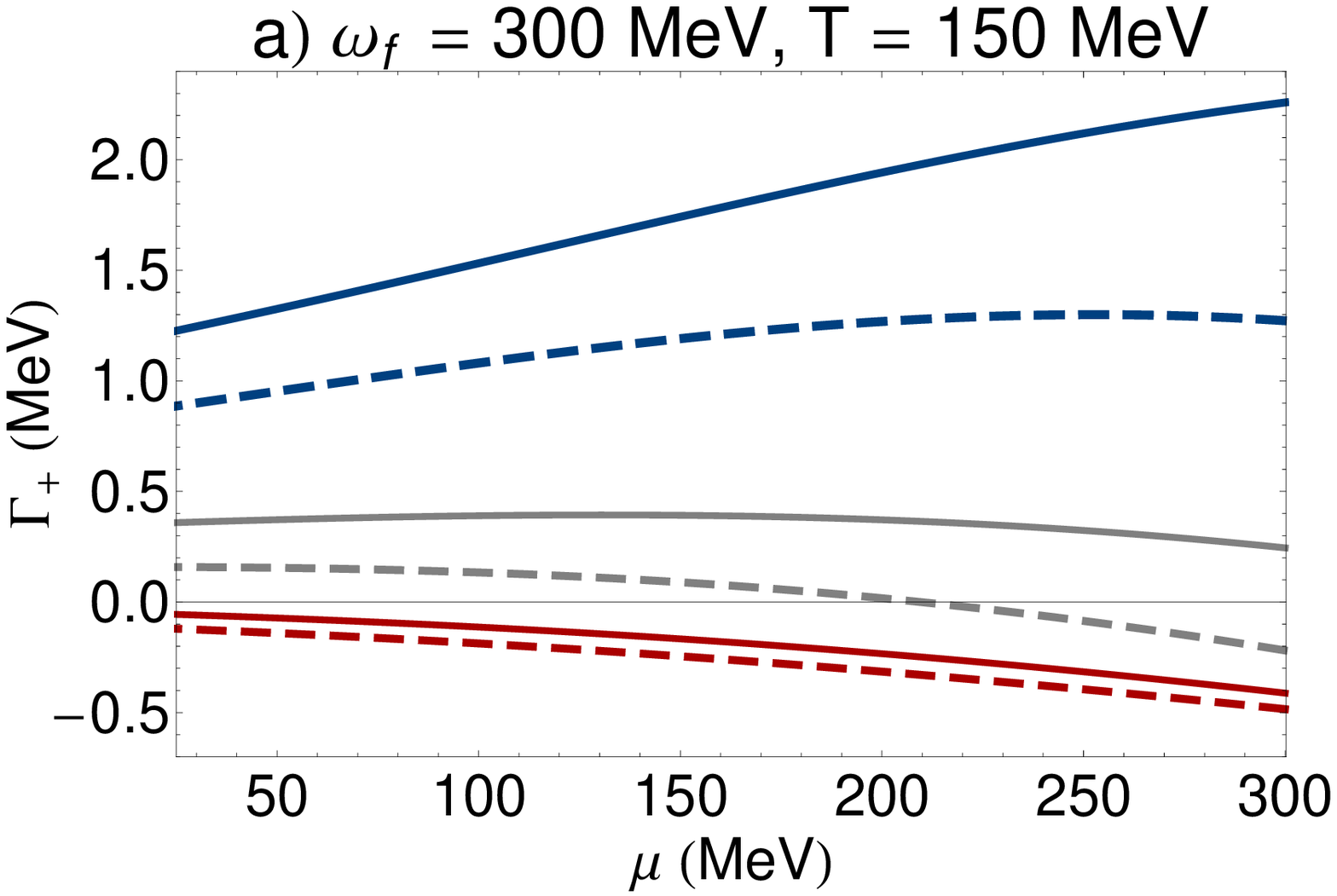}
\includegraphics[width=5.5cm,height=4cm]{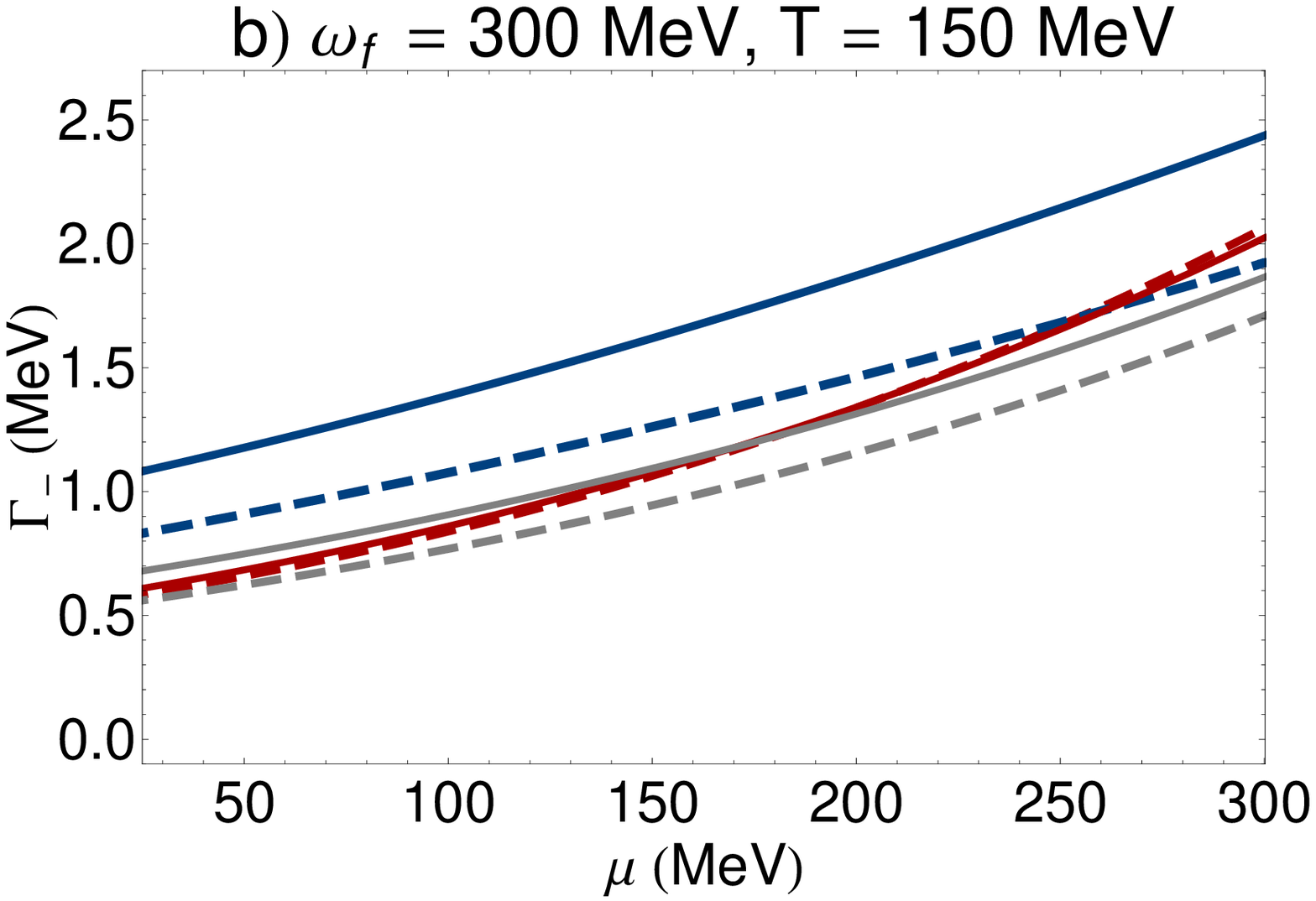}
\includegraphics[width=5.5cm,height=4cm]{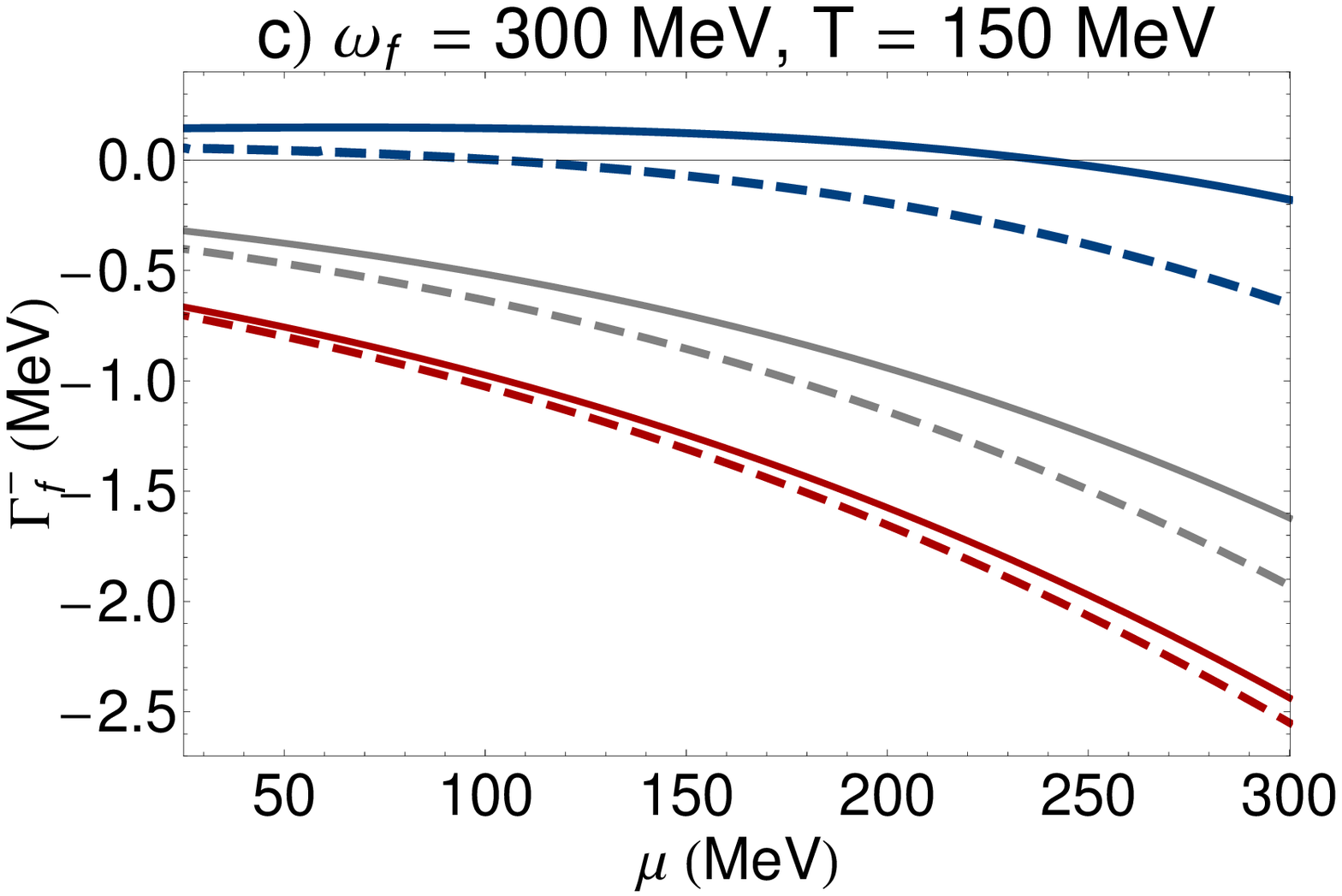}
\caption{ (color online). The $\mu$ dependence of (a) $\Gamma_{+}$
(b) $\Gamma_{-}$ and (c) $\Gamma_{f}^{-}=\Gamma_{+}-\Gamma_{-}$ is
plotted for constant $\omega_{f}=300$ MeV and $T=150$ MeV. The red,
gray and blue solid and dashed lines correspond to
$m_{b}^{0}=300,450,600$ MeV and $m_{f}^{0}=5$ MeV. Whereas the
dashed lines correspond to $\Gamma_{\pm}$ and $\Gamma_{f}^{-}$ as
functions of $(T,\mu)$ independent $\xi_{0}=60,90,120$, the solid
lines correspond to the same quantities including the thermal
corrections to bosonic and fermionic masses with $\xi_{0}^{T}=60,90,
120$. Similar to their $T$ dependence, demonstrated in Fig.
\ref{fig13}, it turns out that $|\Gamma_{f}^{-}|$ increases with
increasing $\mu$, and decreases with increasing $\xi_{0}$ as well as
$\xi_{0}^{T}$. Moreover, for small $\xi_{0}$ or $\xi_{0}^{T}$ and
fixed $(T,\mu)$, $\Gamma_{-}$ is always larger than $\Gamma_{+}$.
}\label{fig14}
\end{figure*}
\begin{figure*}[hbt]
\includegraphics[width=7.cm,height=5cm]{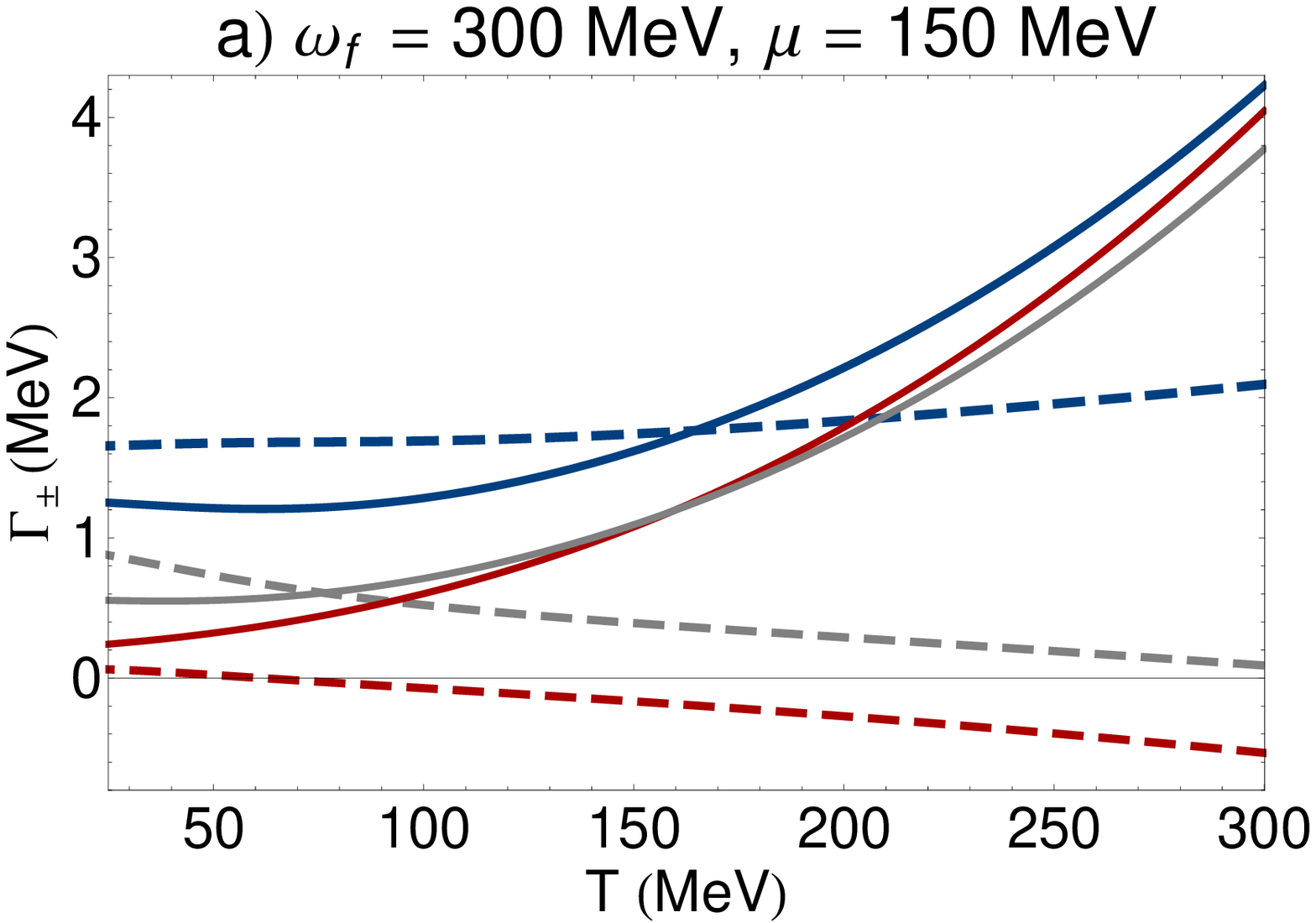}
\includegraphics[width=7.cm,height=5cm]{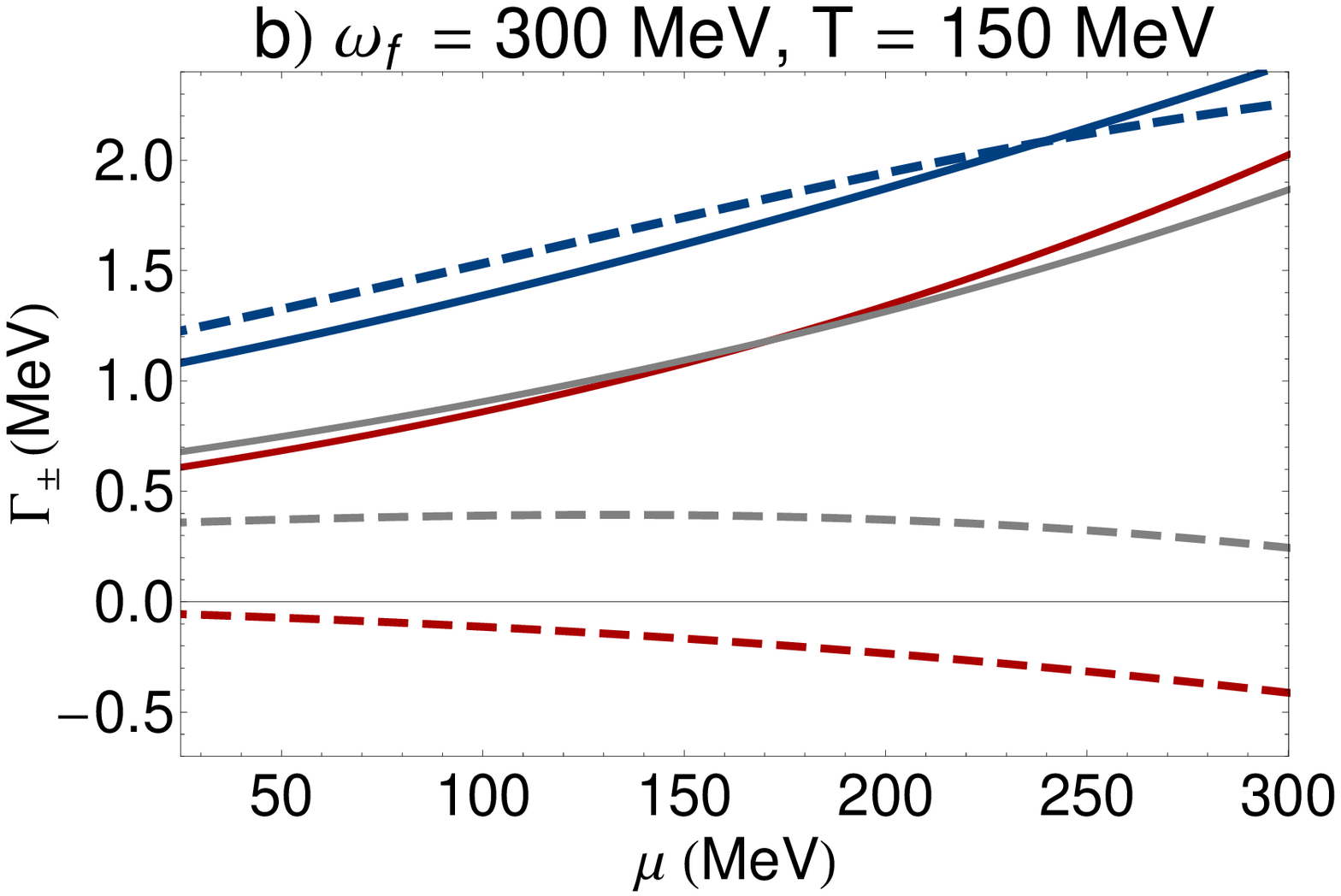}
\caption{ (color online). (a) The $T$ dependence of $\Gamma_{\pm}$
for $\omega_{f}=300$ MeV and $\mu=150$ MeV, including the thermal
corrections to bosonic and fermionic masses. (b) The $\mu$
dependence of $\Gamma_{\pm}$ for $\omega_{f}$ MeV and $T=150$ MeV,
including the thermal corrections to bosonic and fermionic masses.
The dashed (solid) lines correspond to $\Gamma_{+}$ ($\Gamma_{-}$).
The red, gray and blue dashed and solid lines correspond to
$\xi_{0}^{T}=60,90$ and $\xi_{0}^{T}=120$, respectively.
}\label{fig15}
\end{figure*}
\begin{figure}[hbt]
\includegraphics[width=7.7cm,height=5cm]{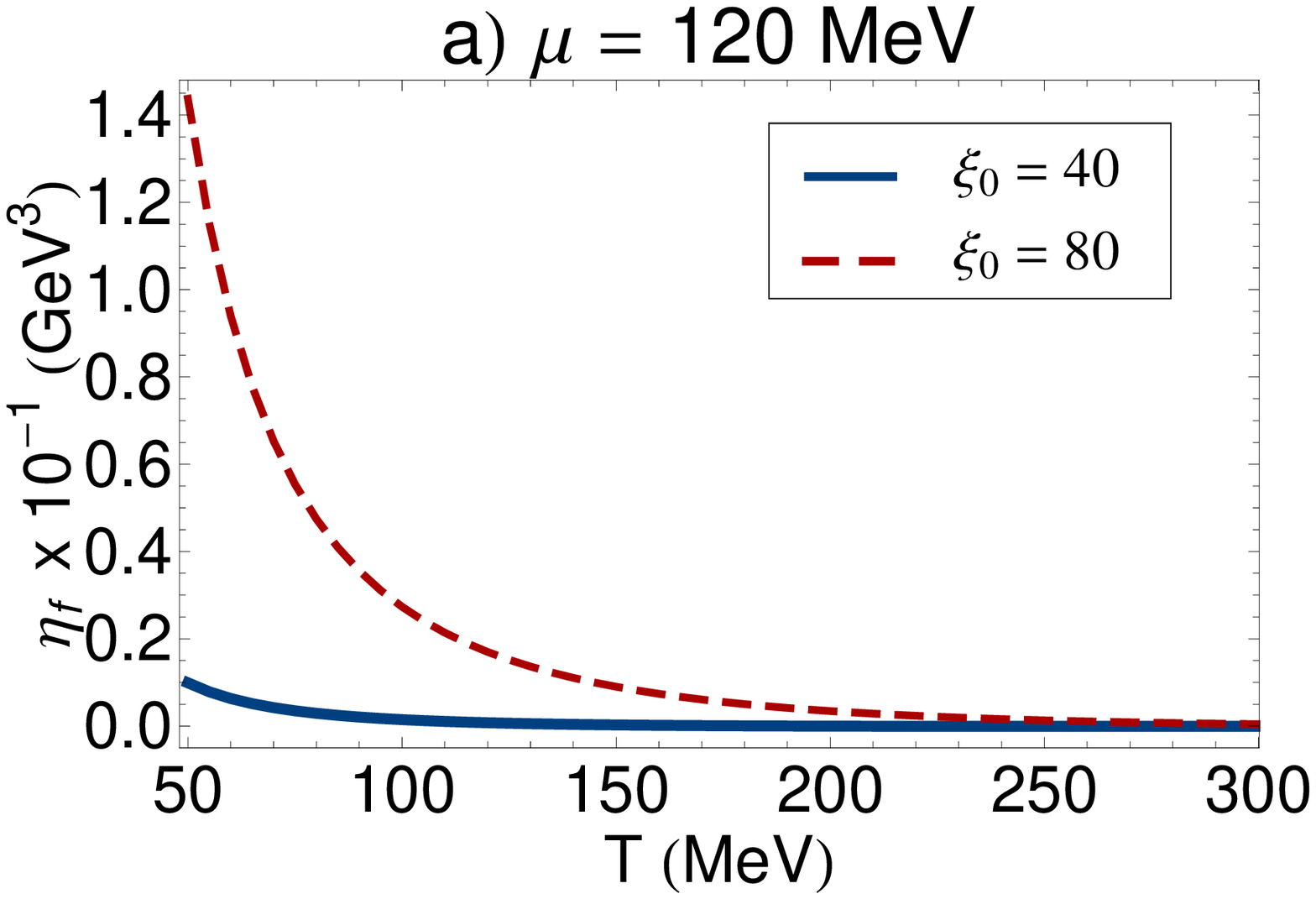}
\includegraphics[width=7.7cm,height=5cm]{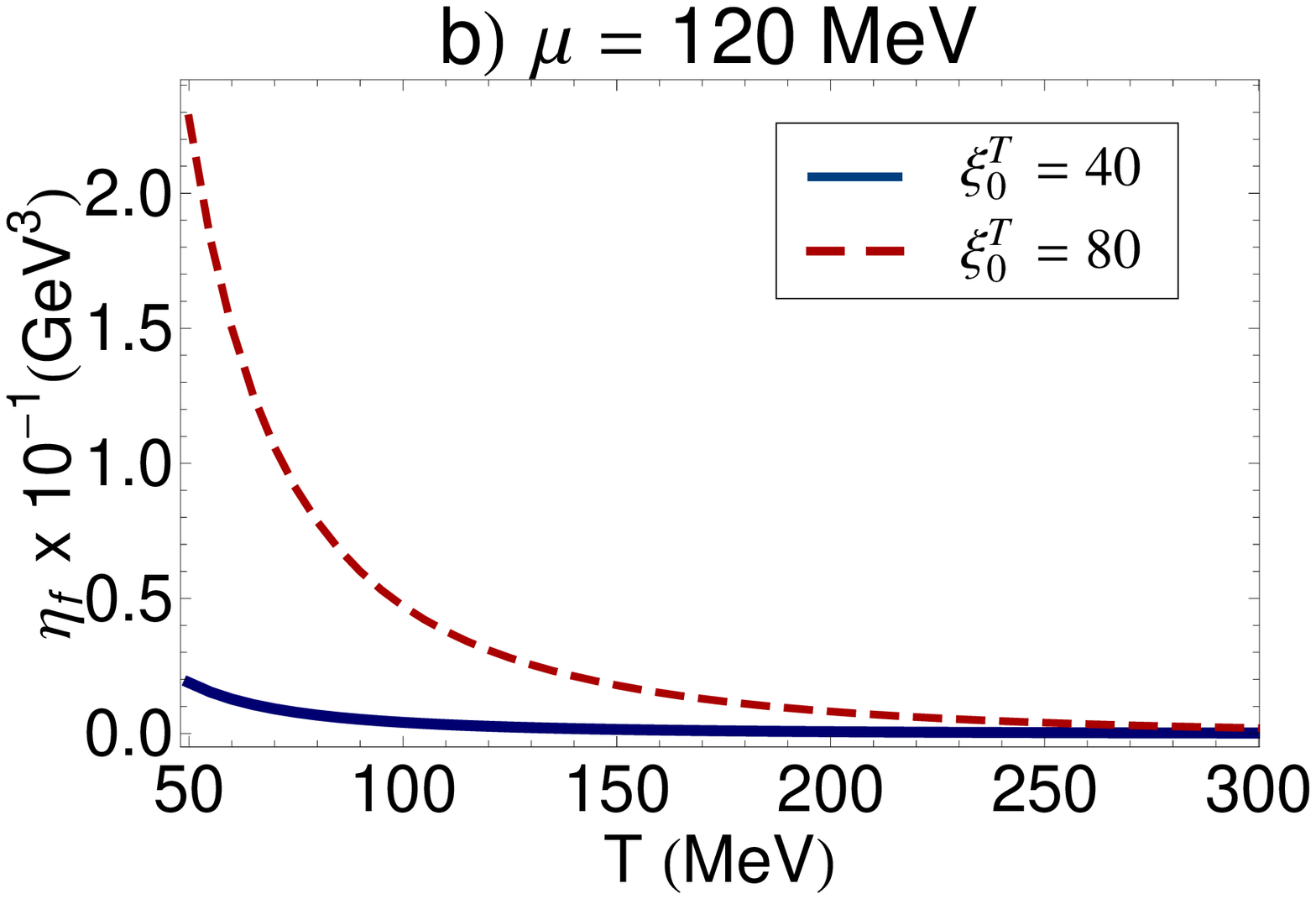}
\caption{ (color online). The $T$ dependence of $\eta_{f}$  is
plotted for $\mu=120$ MeV and $T$ independent $\xi_{0}=40,80$
arising from $m_{b}^{0}=200,400$ MeV and $m_{f}^{0}=5$ MeV. (b) The
$T$ dependence of $\eta_{f}$, including the $T$ and $\mu$ dependent
thermal corrections to bosonic and fermionic masses, is plotted for
$m_{b}^{0}=200,400$ MeV and $m_{f}^{0}=5$ MeV, leading to
$\xi_{0}^{T}=40,80$. }\label{fig16}
\end{figure}
\begin{figure}[hbt]
\includegraphics[width=7.7cm,height=5cm]{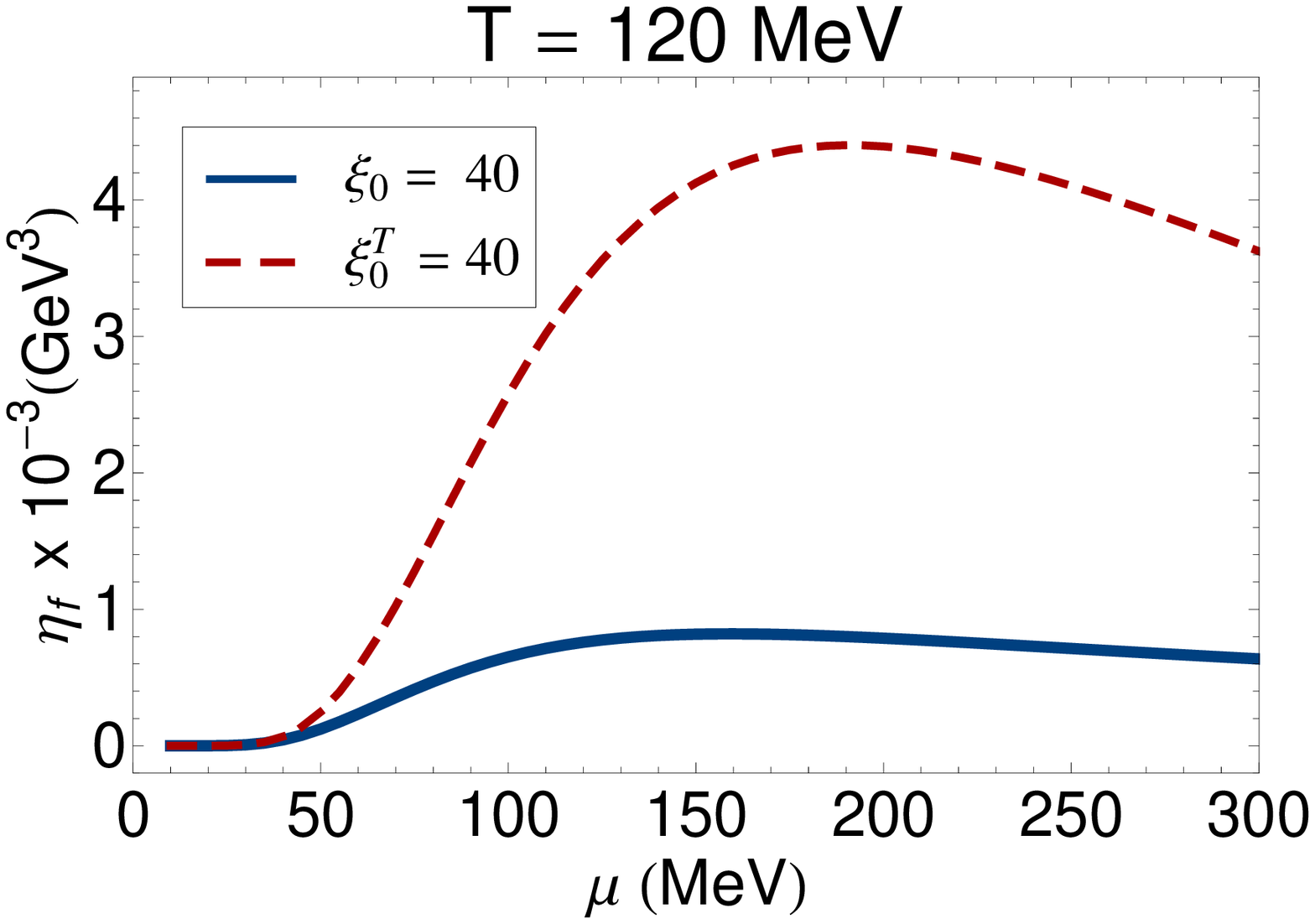}
\caption{ (color online). The $\mu$ dependence of $\eta_{f}$ is
plotted for $T=120$ MeV and $\xi_{0}=\xi_{0}^{T}=40$.}\label{fig17}
\end{figure}
\begin{figure}[hbt]
\includegraphics[width=7.5cm,height=5cm]{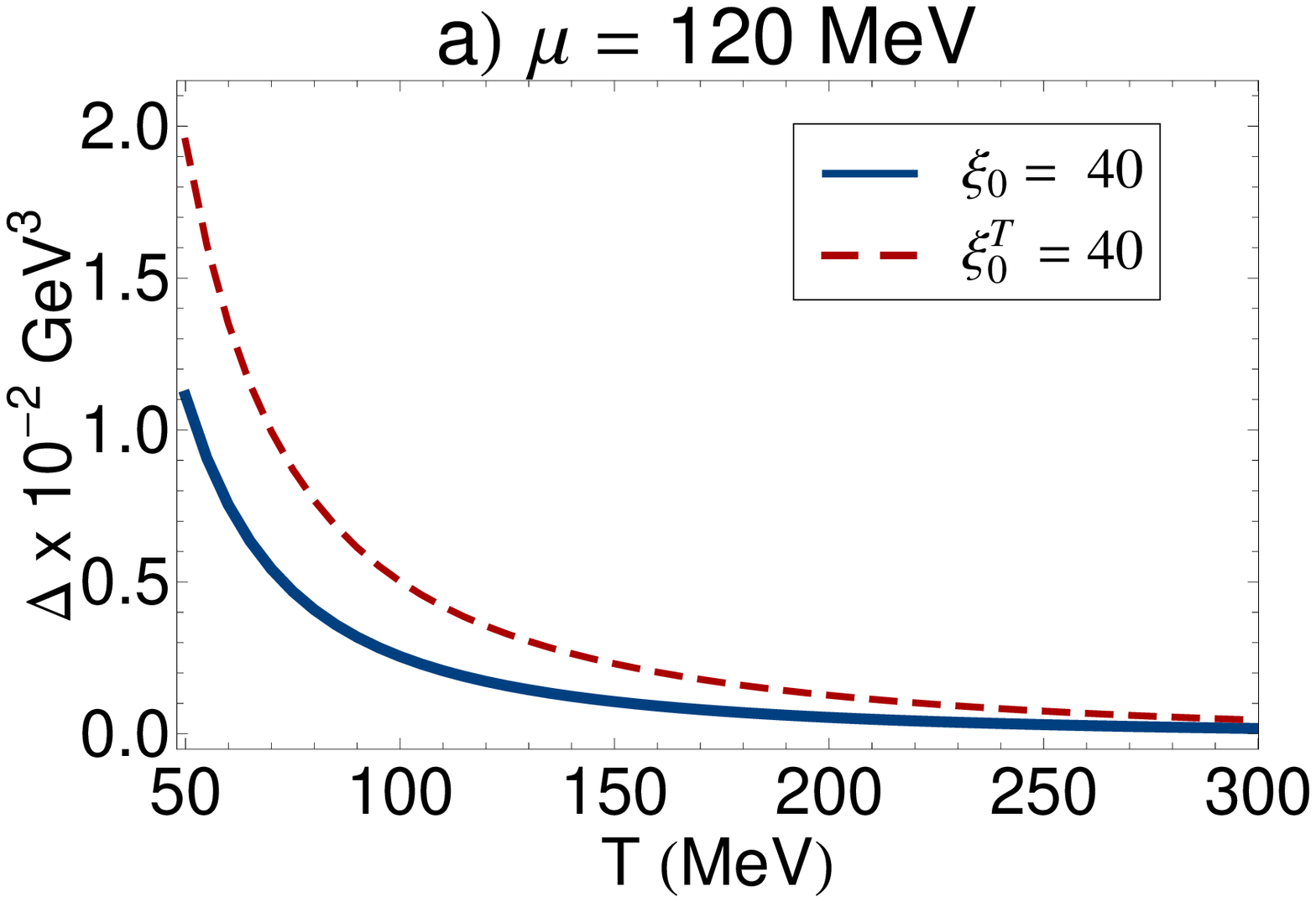}
\includegraphics[width=8.2cm,height=5cm]{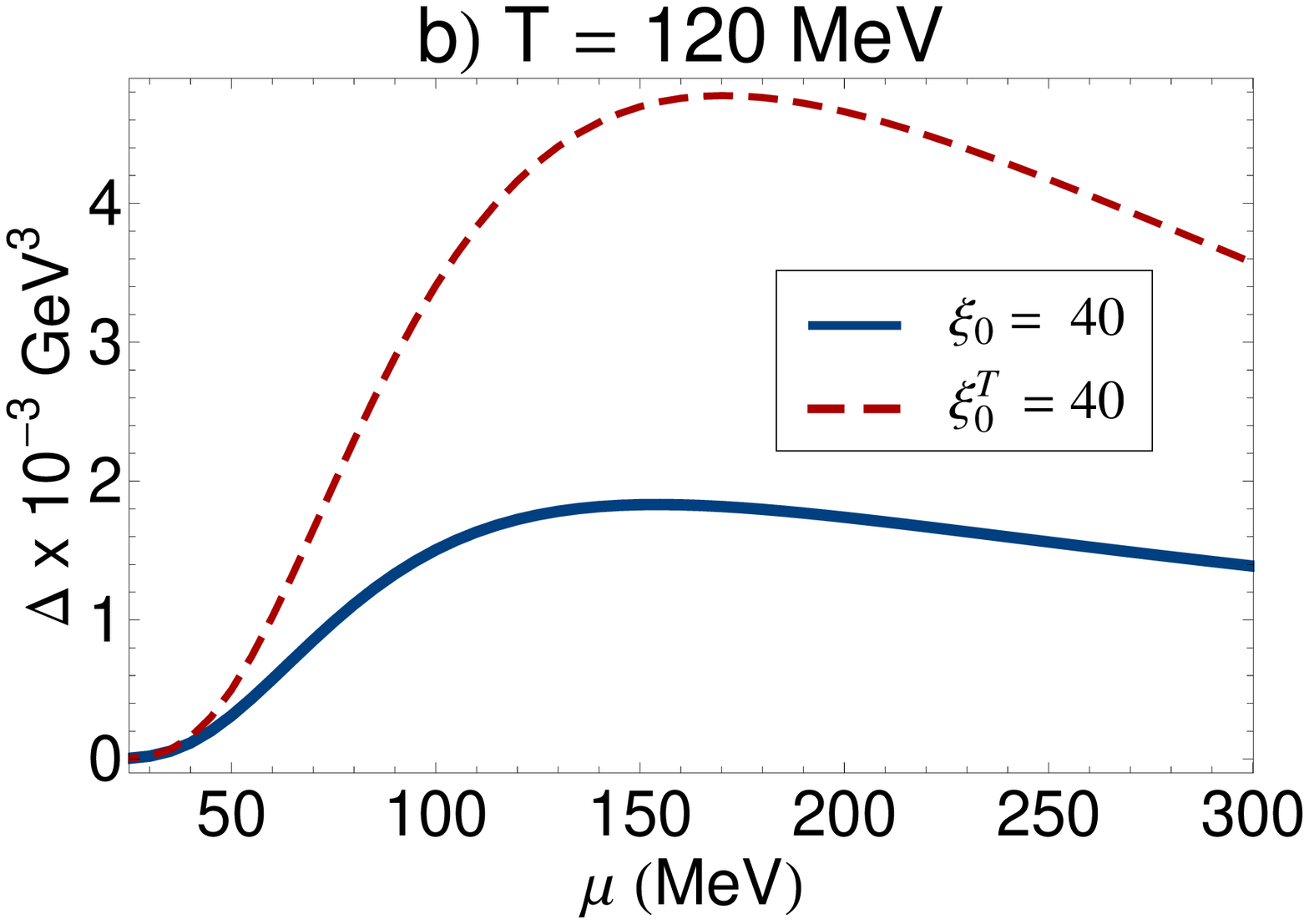}
\caption{ (color online). (a) The $T$ dependence of $\Delta$,
defined in (\ref{H4}), is plotted for $\mu=120$ MeV and
$\xi_{0}=\xi_{0}^{T}=40$. (b) The $\mu$ dependence of $\Delta$ is
plotted for $T=120$ MeV and $\xi_{0}=\xi_{0}^{T}=40$. As it turns
out, $\Delta$ decreases (increases) with increasing $T$ ($\mu$) and
constant mass ratio $\xi_{0}$ as well as
$\xi_{0}^{T}$.}\label{fig18}
\end{figure}
\par\noindent
Let us first consider (\ref{G14a}) and (\ref{appD14a}), where the
bosonic spectral width $\Gamma_{b}$ is presented as a function of
dimensionless parameters, $\gamma_{b}=\frac{m_{b}}{\omega_{b}},
\kappa_{b}=\omega_{b}/T$ with
$\omega_{b}^{2}=\mathbf{p}^{2}+m_{b}^{2}$ and
$\xi=\frac{m_{b}}{m_{f}}$ as well as  $\tau_{f}=\mu/T$ for $\mu=0$
(\ref{G14a}) and $\mu\neq 0$ (\ref{appD14a}). We consider first the
constant mass approach, and replace all $m_{b}$ and $m_{f}$ with
$m_{b}^{0}$ and $m_{f}^{0}$, respectively. We then focus on the
$\xi_{0}$ dependence of $\Gamma_{b}$ for fixed $\kappa_{b},
\gamma_{b}$ and $\tau_{f}$. In Fig. \ref{fig5}(a), the $\xi_{0}$
dependence of dimensionless quantity $\frac{\Gamma_{b}}{g^{2}T}$ is
plotted for $\tau_{f}=0$ and $\kappa_{b}=20$ as well as
$\gamma_{b}=0.5,0.6,0.7,0.8$ [from below (red dashed line) to above
(blue solid line)]. In Fig. \ref{fig5}(b), the $\xi_{0}$ dependence
of $\frac{\Gamma_{b}}{g^{2}T}$ is plotted for $\tau_{f}=0$ and
$\gamma_{b}=0.8$ as well as $\kappa_{b}=1,2,3,4$ [from below (red
dashed line) to above (blue solid line)]. We observe that
$\frac{\Gamma_{b}}{g^{2}T}$ remains constant for $\xi_{0}\gtrsim 10$
in both cases. Moreover, for fixed values of $\xi_{0}$ and
$\kappa_{b}$ ($\gamma_{b}$), the ratio $\frac{\Gamma_{b}}{g^{2} T}$
increases with increasing $\gamma_{b}$ ($\kappa_{b}$) [see panels
(a) and (b) of Fig. \ref{fig5}].
\par
In Fig. \ref{fig6}(a), the $\xi_{0}$ dependence of
$\frac{\Gamma_{b}}{g^{2}T}$ is plotted for $\tau_{f}=4$ and
$\kappa_{b}=20$ as well as $\gamma_{b}=0.5,0.6,0.7,0.8$ [from below
to above]. In Fig. \ref{fig6}(b), the same dimensionless quantity is
plotted for $\tau_{f}=4$ and $\gamma_{b}=0.8$ as well as
$\kappa_{b}=1,2,3,4$ [from below to above]. Similar to the case of
$\tau_{f}=0$, $\frac{\Gamma_{b}}{g^{2}T}$ remains constant for
$\xi_{0}\gtrsim 10$, and increases with increasing $\gamma_{b}
(\kappa_{b})$ for fixed values of $\xi_{0}$ and $\kappa_{b}$
($\gamma_{b}$). In Fig. \ref{fig6}(c), the $\xi_{0}$ dependence of
$\frac{\Gamma_{b}}{g^{2}T}$ is plotted for fixed $\kappa_{b}=20$ and
$\gamma_{b}=0.8$ as well as $\tau_{f}=4,6,8,10$ [from above (red
dashed line) to below (blue solid line)]. In contrast to the
previous cases, $\frac{\Gamma_{b}}{g^{2}T}$ decreases with
increasing $\tau_{f}$ and fixed $\kappa_{b},\gamma_{b}$ and
$\xi_{0}$. These results indicate that $\Gamma_{b}$ decreases with
increasing $T$ and/or $\mu$. This conclusion is compatible with the
observed result demonstrated in Figs. \ref{fig7} and \ref{fig8},
where the $T$ and $\mu$ dependence of $\Gamma_{b}$ is studied for
various fixed parameters.
\par
 In Fig. \ref{fig7}, the $T$ dependence of $\Gamma_{b}$ is plotted
 for $\omega_{b}=300$ MeV, $m_{f}^{0}=5$ MeV and $\mu=0$ MeV
 [Fig. \ref{fig7}(a)] as well as $\mu=150$ MeV [Fig. \ref{fig7}(b)].
 The Yukawa coupling is chosen to be $g=0.5$. Similarly, in Fig.
\ref{fig8}, the $\mu$ dependence of $\Gamma_{b}$ is plotted for
$\omega_{b}=300, m_{f}^{0}=5$ MeV and $T=10$ MeV [Fig.
\ref{fig8}(a)] as well as $T=100$ MeV [Fig. \ref{fig8}(b)].
 The red, gray and blue
 lines in Figs. \ref{fig7} and \ref{fig8} correspond to
 $m_{b}^{0}=100, 150$ and $200$ MeV, respectively. The dashed lines include
 the contributions of constant masses $m_{b}^{0}$ and $m_{f}^{0}$ for bosons and
 fermions, respectively, and the solid lines include the contributions
 of thermal corrections of fermion and boson masses, $m_{b}(T,\mu)$ and
 $m_{f}(T,\mu)$ from (\ref{H3}). As it turns out, $\Gamma_{b}$
 decreases with increasing $T$ and $\mu$. 
Having in mind that $\Gamma_{b}^{-1}$ is essentially proportional to the mean free path of the bosons, $\lambda_{b}$, \cite{lang2012}, the fact that $\Gamma_{b}$ decreases with increasing $T$ and $\mu$ means that $\lambda_{b}$ increases with increasing $T$ and $\mu$. However, for constant $T$ and $\mu$, heavier bosons seems to have smaller $\lambda_{b}$, as expected. Although, according to Figs. \ref{fig7} and \ref{fig8}, adding $T$ and
 $\mu$ dependent (thermal) masses of bosons and fermions to the bare masses $m_{b}^{0}$ and $m_{f}^{0}$ shifts $\Gamma_{b}$
to larger values, but the qualitative interpretation concerning $\lambda_{b}$ remains unchanged. According to (\ref{A26b}), indicating that $\eta_{b}\sim \Gamma_{b}^{-1}$, the thermal behavior of $\lambda_{b}$ is expected to be reflected in thermal behavior of $\eta_{b}$, as it will be shown below.
\subsubsection{Bosonic part of the shear viscosity}
\par\noindent
The bosonic part of the shear viscosity is presented in
(\ref{A26b}), with $\Gamma_{b}$ given in (\ref{G14a}) for $\mu=0$
and (\ref{appD14a}) for $\mu\neq 0$. To determine $\eta_{b}$, we
neglect the contribution of $\mathfrak{Re}[\Sigma_{R}^{b}(p)]$ in
$E_{b}$ from (\ref{A19b}) and set $E_{b}\sim \omega_{b}$. In Fig.
\ref{fig9}, the $T$ dependence of $\eta_{b}$ is plotted for $\mu=0$.
The black solid and red dashed lines in Fig. {\ref{fig9}(a)
correspond to constant ratio $\xi_{0}=40$ MeV and $\xi_{0}=80$ MeV.
The latter arise from $m_{b}^{0}=200, 400$ MeV and $m_{f}^{0}=5$
MeV, respectively. In Fig. \ref{fig9}(b), the $T$ dependence of
$\eta_{b}$ is plotted for $\mu=0$. But, in this case, in contrast to
the plot in Fig. \ref{fig9}(a), $\eta_{b}$ includes thermal masses
$m_{b}(T,\mu)$ and $m_{f}(T,\mu)$ from (\ref{H3}) with
$m_{b}^{0}=200, 400$ MeV and $m_{f}^{0}=5$ MeV. In Fig.
\ref{fig9}(b), $\xi_{0}^{T}$ denotes the ratio $m_{b}^{0}/m_{f}^{0}$
in $\xi(T,\mu)$ from (\ref{H2}). In Figs. \ref{fig10}(a) and (b),
the same quantities are plotted for $\mu=120$ MeV. Comparing the
plots of $\eta_{b}$ for different constant masses in Figs.
\ref{fig9}(a) and \ref{fig10}(a), it turns out that $\eta_{b}$
decreases with increasing $\xi_{0}$. The same is also true for
$\xi(T,\mu)$ [see Figs. \ref{fig9}(b) and \ref{fig10}(b)]. These
results are compatible with our findings in Figs. \ref{fig7}(a) and
\ref{fig7}(b), since for constant $T$ and $\mu$, $\eta_{b}$ is
approximately proportional to $\Gamma_{b}^{-1}$ [see (\ref{A26b})].
Moreover, as expected from Fig. \ref{fig7}, $\eta_{b}$ increases
with increasing $T$. Comparing the results for constant and
$(T,\mu)$ dependent masses in Figs. \ref{fig9} and \ref{fig10}, it
turns out that, as expected from Fig. \ref{fig7}, adding the thermal
corrections to the constant bosonic and fermionic masses decreases
the value of $\eta_{b}$. Moreover, for both constant and $T$ or/and
$\mu$ dependent masses, the difference between $\eta_{b}$ for
different $\xi_{0}$ as well as $\xi_{0}^{T}$ increases with
increasing $T$. However, since the scales in the plots of Fig.
\ref{fig9} and Fig. \ref{fig10} are different, the difference
between $\eta_{b}$ for $\xi_{0}$ and $\xi_{0}^{T}$ seems to be
negligible for the case $\mu\neq 0$ comparing to the case $\mu=0$.
When we compare the plots of Fig. \ref{fig9} with the plots of Fig.
\ref{fig10}, it seems that $\eta_{b}$ decreases with increasing
$\mu$. This conclusion contradicts the result from Figs. \ref{fig7}
and \ref{fig8}, together with the fact that $\eta_{b}\sim
\Gamma_{b}^{-1}$ from (\ref{A26b}). This apparent contradiction may
lie on the fact that for $\mu\neq 0$, the $p$-integration in
(\ref{A26b}) is taken in the interval $p\in
[0,(\mu^{2}-m_{f}^{0~2})^{1/2}]$ for constant fermionic mass
$m_{f}^{0}$, and $p\in [0,[\mu^{2}-m_{f}^{2}(T,\mu)]^{1/2}]$, with
$(T,\mu)$ dependent fermionic mass $m_{f}(T,\mu)$ from (\ref{H3}).
Hence, the $\mu$ dependence of $\Gamma_{b}$ is not the only source
for the $\mu$ dependence of $\eta_{b}$. In Fig. \ref{fig11}, the
$\mu$ dependence of $\eta_{b}$ is demonstrated for constant $T=120$
MeV and $\xi_{0}=40$ as well as $(T,\mu)$ dependent $\xi(T,\mu)$
with $\xi_{0}^{T}=40$. As expected from Figs. \ref{fig7} and
\ref{fig8}, $\eta_{b}$ increases with increasing $\mu$. Recently, in \cite{sarkar2013}, the shear viscosity of a hot pion gas, $\eta_{\pi}$, is determined by solving the relativistic transport equation in Chapman-Enskog and relaxation time approximations. It is shown that for zero pion chemical potential, $\eta_{\pi}$ increases with $T$. 
Although the setup discussed in \cite{sarkar2013} is slightly different from ours -- the self-interaction of pseudoscalar pions is described by the Lagrangian density of chiral perturbation theory -- our results for zero $\mu$ and finite $T$ coincide with the results presented in \cite{sarkar2013}. The fact that, according to our results from Figs. \ref{fig9}-\ref{fig11}, $\eta_{b}$ increases also with $T$ or $\mu$, shows that $T$ and $\mu$ have the same effect on the bosons propagating in a dissipative hot and dense medium. As we have argued in the previous section, the mean free path of bosons, $\lambda_{b}$ increases with increasing $T$ and/or $\mu$. The results of the present section shows that the  thermal properties of $\lambda_{b}$ is directly reflected in the thermal properties of $\eta_{b}$. Moreover, as it turns out heavier bosons have smaller $\eta_{b}$ and $\lambda_{b}$, as expected. 
\subsection{Fermionic contributions}
\subsubsection{Fermionic spectral width}
\par\noindent
In this section, we will focus on the $T$ and $\mu$ dependence of
the fermionic spectral widths $\Gamma_{\pm}$, with the emphasis on
the difference between them. As aforementioned, in the chiral limit
$m_{f}\to 0$ and at finite $(T,\mu)$, $\Gamma_{+}$ and $\Gamma_{-}$
correspond to the normal and collective excitations of fermions,
respectively. The latter is referred to either as a hole or as a
plasmino. Moreover, in the chiral limit, $\Gamma_{+}$ ($\Gamma_{-}$)
corresponds to excitations with the same (opposite) chirality and
helicity. The difference between $\Gamma_{+}$ and $\Gamma_{-}$ is
often neglected in the literature \cite{jeon2007}. We, however,
highlight this difference and study its impact on the fermionic
shear viscosity in different regimes of temperature and chemical
potential.
\par
In (\ref{G19a}) and (\ref{appD17a}), $\Gamma_{+}$ is presented for
vanishing and non-vanishing $\mu$ in terms of dimensionless
parameters $\gamma_{f}=\frac{m_{f}}{\omega_{f}},
\kappa_{f}=\omega_{f}/T$ with
$\omega_{f}^{2}=\mathbf{p}^{2}+m_{f}^{2}$ and
$\xi=\frac{m_{b}}{m_{f}}$ as well as $\tau_{f}=\mu/T$. Similarly,
$\Gamma_{f}^{-}(\gamma_{f},\kappa_{f},\tau_{f};\xi)$ for $\mu=0$ and
$\mu\neq 0$ are presented in (\ref{appC14a}) and (\ref{appD19a}),
respectively. Using $\Gamma_{-}=\Gamma_{+}-\Gamma_{f}^{-}$,
$\Gamma_{-}$ can be determined from the difference between
$\Gamma_{+}$ and $\Gamma_{f}^{-}$. Similar to the bosonic case, let
us replace $m_{b}$ and $m_{f}$ with $(T,\mu)$ independent
$m_{b}^{0}$ and $m_{f}^{0}$, respectively, and focus first on the
$\xi_{0}=m_{b}^{0}/m_{f}^{0}$ dependence of the dimensionless
quantity $\frac{\Gamma_{+}}{g^{2}T}$ as a function of dimensionless
parameters $\gamma_{f},\kappa_{f}$ and $\tau_{f}$.
\par
In Fig. \ref{fig12}(a), the $\xi_{0}$ dependence of
$\frac{\Gamma_{+}}{g^{2}T}$ is plotted for fixed $\tau_{f}=4$ and
$\kappa_{f}=20$ as well as $\gamma_{f}=0.5,0.6,0.7,0.8$ [from below
(red dashed line) to above (blue solid line)]. Similarly, in Fig.
\ref{fig12}(b), the $\xi_{0}$ dependence of
$\frac{\Gamma_{+}}{g^{2}T}$ is plotted for $\tau_{f}=4$ and
$\gamma_{f}=0.8$ as well as $\kappa_{f}=2,4,6.8$ [from below (red
dashed line) to above (blue solid line)]. Finally, in Fig.
\ref{fig12}(c), the $\xi_{0}$ dependence of
$\frac{\Gamma_{+}}{g^{2}T}$ is plotted for fixed $\kappa_{f}=20$ and
$\gamma_{f}=0.8$ as well as $\tau_{f}=0,3,6,9$ [from below (red
dashed line) to above (blue solid line)]. In contrast to the bosonic
case, for a fixed $\xi_{0}$, $\frac{\Gamma_{+}}{g^{2}T}$ increases
whenever one of the parameters $\gamma_{f},\kappa_{f}$ or $\tau_{f}$
increases and the other two parameters are held fixed. Neglecting
the tiny difference between $\frac{\Gamma_{+}}{g^{2}T}$ and
$\frac{\Gamma_{-}}{g^{2}T}$, the same can easily be shown to be true
for $\frac{\Gamma_{-}}{g^{2}T}$. Let us notice at this stage, that to derive the final results for $\Gamma_{\pm}$ for $\mu=0$ and $\mu\neq 0$, the condition $m_{b}^{0}\geq 2m_{f}^{0}$ was necessary. It is easy to show that $\Gamma_{\pm}$ diverges once $m_{b}^{0}=m_{f}^{0}=0$. This is also indicated in \cite{plasmino-yukawa-2}, where it is noted that nonzero boson and fermion mass difference, $\delta m^{2}=m_{b}^{2}-m_{f}^{2}$,  ensures the smoothness of the fermion self-energy, and consequently $\Gamma_{\pm}$, in the far infrared (IR) limit.
\par
Although the $\xi_{0}$ dependence of $\frac{\Gamma_{+}}{g^{2}T}$ and
$\frac{\Gamma_{-}}{g^{2}T}$ as functions of dimensionless parameters
$\gamma_{f},\kappa_{f}$ and $\tau_{f}$ are practically identical,
the $T$ ($\mu$) dependence of $\Gamma_{+}$ and $\Gamma_{-}$ turns
out to be different for fixed value of $\mu$ ($T$) and $\xi_{0}$. In
Figs. \ref{fig13} and \ref{fig14}, the $T$ and $\mu$ dependence of
$\Gamma_{+}$ (panel a), $\Gamma_{-}$ (panel b) as well as
$\Gamma_{f}^{-}$ (panel c) are plotted for $\omega_{f}=300$ MeV and
$\mu=150$ MeV (Fig. \ref{fig13}), as well as for $\omega_{f}=300$
MeV and $T=150$ MeV (Fig. \ref{fig14}). The red, gray and blue solid
and dashed lines correspond to $m_{b}^{0}=300,450,600$ MeV and
$m_{f}^{0}=5$ MeV. The dashed lines correspond to $\Gamma_{\pm}$ and
$\Gamma_{f}^{-}$ as functions of $(T,\mu)$ independent
$\xi_{0}=60,90,120$, and the solid lines correspond to the same
quantities, including the thermal masses of bosons and fermions,
with $\xi_{0}^{T}=m_{b}^{0}/m_{f}^{0}=60,90,120$. According to the
results in Figs. \ref{fig13} and \ref{fig14}, it turns out that the
absolute value of the difference between $\Gamma_{+}$ and
$\Gamma_{-}$, $|\Gamma_{f}^{-}|$, increases with increasing $T$ and
constant $\mu$ (Fig. \ref{fig13}), as well as with increasing $\mu$
and constant $T$ (Fig. \ref{fig14}). It decreases with increasing
$\xi_{0}$ and $\xi_{0}^{T}$. Moreover, for small value of $\xi_{0}$
or $\xi_{0}^{T}$ and fixed $(T,\mu)$, $\Gamma_{-}$ is always larger
than $\Gamma_{+}$. 
\par
To compare $\Gamma_{+}$ and $\Gamma_{-}$ more directly, their $T$
and $\mu$ dependence are plotted in Fig. \ref{fig15} for constant
$\omega_{f}=300$ MeV and $\mu=150$ MeV (panel a) and $T=150$ MeV.
Here, $\Gamma_{\pm}$ include only thermal bosonic and fermionic
masses. The dashed (solid) lines correspond to $\Gamma_{+}$
($\Gamma_{-}$). The red, gray and blue dashed and solid correspond
to $\xi_{0}^{T}=60,90,120$, respectively.
As it turns out, whereas for smaller $\xi_{0}^{T}=m_{b}^{0}/m_{f}^{0}$, $\Gamma_{+}$, the spectral width of normal fermion excitations, decreases with $T$ or $\mu$, for larger $\xi_{0}^{T}$, it increases with increasing $T$ or $\mu$. In contrast, $\Gamma_{-}$, the spectral width of the plasmino excitations, increases with $T$ or $\mu$, independent of $\xi_{0}^{T}$. Assuming, in analogy to the bosonic case, that the spectral widths $\Gamma_{+}$ and $\Gamma_{-}$ are inversely proportional to the mean free paths of the normal and plasmino excitations of the fermions, $\lambda_{+}$ and $\lambda_{-}$, the above results suggest that  at higher temperature or chemical potential, plasminos have smaller $\lambda_{-}$, while for normal fermions, the thermal behavior of $\lambda_{+}$ depends strongly on the relation between the masses of the fermions and bosons included in our Yukawa-Fermi gas. Heavier (normal) fermions have smaller $\lambda_{+}$, as expected.  Let us eventually mention that, according to the plots in Figs. \ref{fig13} and \ref{fig14}, $|\Gamma_{f}^{-}|=|\Gamma_{+}-\Gamma_{-}|$ increases with
increasing $T$ ($\mu$) and fixed $\mu$ ($T$), as suggested from the fact that holes (plasminos) are more significant at higher temperature \cite{weldon1989}. In what follows, we
will study the impact of this difference on the fermionic part of
the shear viscosity.
\subsubsection{Fermionic part of the shear viscosity}
\par\noindent
In Sec. \ref{sec3b}, the fermionic part of the shear viscosity,
$\eta_{f}$, is computed in terms of $\Gamma_{+}$ and $\Gamma_{-}$
for vanishing chemical potential [see (\ref{A49b})]. In App.
\ref{appC}, we presented $\eta_{f}$ for non-vanishing chemical
potential [see (\ref{appD1a})]. Neglecting the contribution of
$\mathfrak{Re}[\Sigma_{R}^{f}]$ in $E_{\pm}$ from (\ref{A38b}) and
in ${\cal{E}}_{\pm}$ from (\ref{appD2a}), and replacing $E_{\pm}$
and ${\cal{E}}_{\pm}$, appearing in (\ref{A49b}) and (\ref{appD1a}),
with $\omega_{f}$ and $\omega_{\pm}=\omega_{f}\pm \mu$,
respectively, we have plotted the $T$ dependence of $\eta_{f}$ for
fixed $\mu=120$ MeV and $\xi_{0}=40,80$ in Fig. \ref{fig16}(a) and
for $\mu=120$ MeV and $\xi_{0}^{T}=40,80$ in Fig. \ref{fig16}(b). In
contrast to the $T$ dependence of $\eta_{b}$ from Fig. \ref{fig10},
we observe that $\eta_{f}$ decreases with increasing $T$, $\eta_{f}$
is in general larger than $\eta_{b}$, and at a fixed temperature and
for a fixed chemical potential, $\eta_{f}$ increases with increasing
$\xi_{0}$ [Fig. \ref{fig16}(a)] as well as $\xi_{0}^{T}$ [Fig.
\ref{fig16}(b)]. The fact that for a fixed $T$ and $\mu$, $\eta_{f}$
decreases with increasing $\xi_{0}$ is compatible with the results
arising from Fig. \ref{fig12}, where it is shown that $\Gamma_{\pm}$
increases with increasing $\xi_{0}$, and confirms the fact that for
small values of $\xi_{0}$ (or $\xi_{0}^{T}$), $\eta_{f}\sim\Gamma_{\pm}^{-1}$. But, in general, it seems that the thermal property of $\eta_{f}$ is dominated by the thermal behavior of $\Gamma_{-}$. The fact that $\eta_{f}$ is inversely proportional to the fermionic spectral width coincides with the results presented in \cite{plasmino-NJL}, and indicates that $\eta_{f}$ increases with increasing the mean free path.\footnote{In \cite{plasmino-NJL}, no difference is made between the mean free paths of normal and plasmino excitations.}
\par
In Fig. \ref{fig17}, the $\mu$ dependence of $\eta_{f}$ is plotted
for $T=120$ MeV and $\xi_{0}=40$ (blue solid line) and
$\xi_{0}^{T}=40$ (red dashed line). In contrast to the $\mu$
dependence of $\eta_{b}$ from Fig. \ref{fig11}, $\eta_{f}$ decreases
with increasing $\mu$ at a fixed temperature. Moreover, at a fixed
$T$ and $\mu$, $\eta_{f}$ decreases when the thermal corrections to
the bosonic and fermionic masses are taken into account. This is
again in contrast with the observed results for $\eta_{b}$ in Fig.
\ref{fig11}. 
\par
As we have shown in Figs. \ref{fig13}, \ref{fig14} and \ref{fig15},
$\Gamma_{+}$ and $\Gamma_{-}$ have different thermal properties. To
study how this difference can affect $\eta_{f}$, we define a
quantity $\Delta$, as the difference between $\eta_{f}$ as a
functional of $\Gamma_{+}=\Gamma_{-}$, and $\eta_{f}$ as a
functional of $\Gamma_{+}\neq \Gamma_{-}$,
\begin{eqnarray}\label{H4}
\Delta=\eta_{f}[\Gamma_{+}=\Gamma_{-}]-\eta_{f}[\Gamma_{+}\neq
\Gamma_{-}].
\end{eqnarray}
Let us remind, that in the literature the difference between
$\Gamma_{+}$ and $\Gamma_{-}$ is often neglected, and in so far,
$\Delta\simeq 0$ is assumed. In Figs. \ref{fig18}(a) and (b), the
$T$ and $\mu$ dependence of $\Delta$ is plotted for constant
$\mu=120$ MeV (panel a) and $T=120$ MeV (panel b), and for
$\xi_{0}=40$ (blue solid lines) and $\xi_{0}^{T}=40$ (red dashed
lines). It turns out that in the whole range of $T$ and $\mu$,
$\Delta$ is positive. This means that the value of $\eta_{f}$
increases, when the difference between $\Gamma_{+}$ and $\Gamma_{-}$
is neglected. Moreover, for fixed $\mu$ (T) and constant $\xi_{0}$
or $\xi_{0}^{T}$, $\Delta$ decreases (increases) with $T$ ($\mu$).
In other word, as it is shwon in Fig. \ref{fig18}(a), whereas at
lower temperature and for an intermediate value of $\mu$, the
difference between $\eta_{f}[\Gamma_{+}=\Gamma_{-}]$ and
$\eta_{f}[\Gamma_{+}\neq \Gamma_{-}]$ is relatively large, and
becomes larger by including the thermal corrections to the bosonic
and fermionic masses, it can be neglected at higher temperature. In
contrast, the difference between $\eta_{f}[\Gamma_{+}=\Gamma_{-}]$
and $\eta_{f}[\Gamma_{+}\neq \Gamma_{-}]$ is negligible at fixed
temperature and for small value of chemical potential. It increases
with increasing $\mu$ and is enhanced by adding the thermal
corrections to the bosonic and fermionic masses.
\section{Summary and outlook}\label{sec6}
\par\noindent
The shear viscosity $\eta$ is a transport coefficient, that
characterizes the diffusion of momentum transverse to the direction
of propagation. It plays an important r\^{o}le in the physics of
QGP. In the past few years, there have been several attempts to
explore its thermal properties, in particular in the vicinity of QCD
chiral transition point. The aim is to determine the position of the
transition temperature of QCD, using the thermal properties of
$\eta$, in addition to and independently of the equation of state
\cite{kapusta2008}. In this paper, we studied the thermal properties
of the shear viscosity of an interacting boson-fermion system with
Yukawa coupling. We followed the method presented in \cite{lang2012}
to derive the bosonic part of the shear viscosity of this theory in
terms of the bosonic spectral width, $\Gamma_{b}$. The latter is
then determined in a one-loop perturbative expansion in the orders
of the Yukawa coupling. Using $\eta_{b}[\Gamma_{b}]$, it was then
possible to study the thermal properties of $\eta_{b}$, in addition
to its dependence to the masses of bosons and fermions.
\par
We took the method used in \cite{lang2012}, as our guideline, and
determined the fermionic part of the shear viscosity of the Yukawa
theory in terms of the fermionic widths $\Gamma_{+}$ and
$\Gamma_{-}$. The expression $\eta_{f}[\Gamma_{\pm}]$ from
(\ref{A49b}) and (\ref{appD1a}) for vanishing and non-vanishing
chemical potential, build the central analytical results of the
present paper. Here, $\Gamma_{+}$ and $\Gamma_{-}$ are the spectral
widths, corresponding to the normal and collective (plasmino)
excitations of fermions. They are studied very intensively in the
literature and lead e.g. to structures in the low mass dilepton
production rate, which might provide a unique signature for the QGP
formation at relativistic heavy ion collisions \cite{plasmino-app}.
However, to the best of our knowledge, the difference between their
spectral widths is often neglected (see e.g. in
\cite{nardi2009,kiessig2010,plasmino-yukawa-1}), and, as in
\cite{jeon2007,lang2013}, the fermionic spectral density function,
$\rho_{f}$, is given in terms of one and the same fermionic spectral
width. We, however, used the structure of $\rho_{f}$ presented in
\cite{plasmino-NJL}, including both $\Gamma_{+}$ and $\Gamma_{-}$,
and following the method presented in \cite{lang2012}, determined
$\eta_{f}[\Gamma_{\pm}]$ in an appropriate Laurent expansion.
Moreover, we completed the results presented in \cite{plasmino-NJL},
and evaluated first $\Gamma_{\pm}$ in one-loop perturbative
expansion in the orders of the Yukawa coupling, and studied their
thermal properties. Plugging then $\Gamma_{\pm}$ in the proposed
relation for the fermionic shear viscosity, $\eta_{f}[\Gamma_{\pm}]$
from (\ref{A49b}) and (\ref{appD1a}), we determined the thermal
properties of $\eta_{f}$, and studied its mass dependence. Apart from various results on the thermal properties of $\Gamma_{b},\Gamma_{\pm}$ as well as $\eta_{b}$ and $\eta_{f}$, discussed in the previous section, we showed
that, depending on temperature and/or chemical potential,
$\eta_{f}[\Gamma_{+}\neq \Gamma_{-}]$ is smaller than
$\eta_{f}[\Gamma_{+}=\Gamma_{-}]$.
\par
It shall be noted that our one-loop computation, including bare fermion and boson masses, is incomplete and
can be improved, for instance, by considering the full HTL
correction to the fermion propagator. The latter plays a crucial
r\^{o}le in determining $\Gamma_{b}$ and $\Gamma_{\pm}$, and
consequently $\eta_{b}$ and $\eta_{f}$. This drawback is partly
compensated in the present paper by adding thermal corrections to
the bosonic and fermionic masses. This ad-hoc treatment of thermal
masses seems to be natural, since, as it is also discussed in
\cite{kiessig2009, kiessig2010}, it equals the HTL treatment with an
approximate fermion propagator. Moreover, since it is known that the
HTL/HDL treatment are only valid for soft momenta $p\lesssim gT,
g\mu$, even the HTL treatment can be improved by studying the
ultra-soft fermionic excitations, with $p\lesssim g^{2}T, g^{2}\mu$.
They are recently discussed in \cite{plasmino-yukawa-2,balizot2014},
in the framework of the Yukawa theory. 
An important question related to the perturbative treatment of transport coefficients, in general, and shear viscosity, in particular, is the appearance of  the so called pinch singularities, that would break the perturbation theory based on loop expansion. A useful description of these singularities is presented in \cite{hidaka2010}: In the quasiparticle approximation, where the propagators are  given by the energy and spectral widths of the quasiparticles, the pinch singularity is essentially related to the IR behavior of the product of retarded and advanced propagators, that appears in the perturbative loop calculations. Once the spectral width is zero, the above mentioned product becomes IR divergent. 
The consequence is that higher loop diagrams, if they are sufficiently IR sensitive, become as important as the one-loop contribution, and a resummation of an infinite number of ladder diagrams will be necessary.
In \cite{jeon1994}, a detailed power-counting is presented for $\lambda\varphi^{3}$ and $\lambda\varphi^{4}$ theories, and it is shown that all ladder diagrams contribute in the same leading-order. In \cite{plasmino-yukawa-2}, similar power-counting is performed for the ladder diagrams contributing to the fermion self-energy of a Yukawa theory, and it is shown that in contrast to the above mentioned scalar theories with cubic and quartic interactions, and also in sharp contrast to QED and QCD, the ladder diagrams are indeed suppressed, and consequently the one-loop self-energy diagram with dressed propagators (including the thermal masses) gives the leading-order contribution to the fermion self-energy. The main reason for this suppression is the fact that the Yukawa coupling constant receives no correction in the leading order HTL approximation. Or, as is stated in \cite{plasmino-yukawa-2}, ``the ladder diagrams giving a vertex correction do not contribute in the leading order in the scalar coupling''. As concerns higher loop contributions to the spectral width and shear viscosity of the Yukawa theory,  it seems therefore that no ladder resummation may be necessary, and the one-loop computation, including the thermal masses, may build the leading order contribution to these quantities. A recent perturbative computation of the shear viscosity of the Yukawa theory up to two-loop order confirms this conclusion \cite{ghosh2014}. It is, in particular, shown that the two-loop diagrams, having same power of coupling as the one-loop diagram, is substantially suppressed comparing to one-loop contribution. According to the arguments presented in \cite{ghosh2014}, it is indeed expected that by increasing the number of loops, the suppression successively grows, so that the one-loop results of the shear viscosity of the Yukawa-Fermi gas can be considered as the leading order. A more detailed analysis of ladder resummation corresponding to the shear viscosity of the Yukawa theory will be postponed to future publication.
\par
In Sec. \ref{sec4}, the leading order contributions to the bosonic and fermionic spectral widths of the Yukawa theory are determined by  computing the imaginary part of two one-loop bosonic and fermionic self-energy diagrams [see Figs. \ref{fig3} and \ref{fig4}]. Let us notice at this stage, that these one-loop contributions correspond to 1$\to$2 scattering processes (Landau damping), which seem to build the leading order contribution to the spectral widths of the Yukawa theory. This is again in contrast to the situation apprearing in QED, where, as it is argued by Gagnon and Jeon in  \cite{jeon2007}, apart from the special case of 1$\to$2 collinear scatterings including massless electrons, the perturbative series of the spectral widths starts from the leading 2$\to$2 scattering processes, arising from two-loop self-energies. This is because of the fact that in QED, in contrast to the Yukawa theory, the imaginary parts of the one-loop boson (photon)  and fermion (electron) self-energies vanish, as can be easily checked, and as it is also stated in \cite{jeon2007}. Hence, an on mass-shell massless excitations cannot decay into two on mass-shell excitations, as expected. We can therefore conclude that in the Yukawa theory,  the  2$\to$2 scattering processes, arising from two-loop contributions to the bosonic and fermionic self-energies build 
the subleading contribution to the spectral widths of this theory relative to 1$\to$2 scattering processes, arising from one-loop self-energy diagrams demonstrated in Figs. \ref{fig3} and \ref{fig4} of the present paper. 
\par
Let us finally notice that one of the possibilities to extend the present computation is to apply it for QCD-like model, e.g.
quark-meson or NJL models, including spontaneous or dynamical chiral
symmetry breaking, and study the behavior of $\eta$ in the vicinity
of chiral transition point. The latter project is currently under
investigation. The results will be reported elsewhere.
\begin{appendix}
\section{Spectral density function of fermions}\label{appA}
\setcounter{section}{1}\setcounter{equation}{0}
\par\noindent
In this appendix, we will apply the method presented in
\cite{blaizot1996} for massive fermions, and will show that the
spectral density function of fermions is given by (\ref{A36b}). To
start, let us consider the K\"allen-Lehmann representation of free
fermion propagator in terms of free spectral density function
$\rho_{f}^{0}$
\begin{eqnarray}\label{App1a}
S_{0}(\mathbf{p},\omega)=\int_{-\infty}^{+\infty}\frac{dp_{0}}{2\pi}\frac{\rho_{f}^{0}(\mathbf{p},p_{0})}{p_{0}-\omega}.
\end{eqnarray}
Plugging
\begin{eqnarray}\label{App2a}
\hspace{-0.4cm}\rho_{f}^{0}(\mathbf{p},p_{0})=2\pi(p\cdot
\gamma+m_{f})\mbox{sgn}(p_{0})\delta(p_{0}^{2}-\omega_{f}^{2}),
\end{eqnarray}
with $\omega_{f}^{2}=\mathbf{p}^{2}+m_{f}^{2}$, in (\ref{App1a}),
and integrating over $p_{0}$, we arrive at the following
decomposition of $S_{0}$ in terms of two independent matrices
$\hat{g}_{\pm}$, defined in (\ref{A37b}),
\begin{eqnarray}\label{App3a}
S_{0}(\mathbf{p},\omega)=-\frac{1}{\omega-\omega_{f}}\hat{g}_{+}-\frac{1}{\omega+\omega_{f}}\hat{g}_{-}.
\end{eqnarray}
To determine the inverse propagator of free fermions, we introduce
new matrices
\begin{eqnarray}\label{App4a}
\hat{g}'_{\pm}\equiv
\frac{1}{2\omega_{f}}\big[\gamma_{0}\omega_{f}\mp\left(\gamma.\mathbf{p}+m_{f}\right)\big],
\end{eqnarray}
that satisfy
\begin{eqnarray}\label{App5a}
(\hat{g}'_{\pm})^{\dagger}=\hat{g}_{\mp},&\qquad&\hat{g}_{\pm}\hat{g}'_{\pm}=0,\nonumber\\
\hat{g}'_{\mp}\hat{g}_{\pm}=\hat{g}'_{\mp}\gamma_{0},&\qquad&
\hat{g}_{\pm}\hat{g}'_{\mp}=\gamma_{0}\hat{g}'_{\mp}.
\end{eqnarray}
The inverse propagator of free fermions is then given by
\begin{eqnarray}\label{App6a}
S_{0}^{-1}(\mathbf{p},\omega)=-(\omega+\omega_{f})\hat{g}'_{+}-(\omega-\omega_{f})\hat{g}'_{-}.
\end{eqnarray}
To determine the dressed spectral density $\rho_{f}(\omega,p_{0})$
for dressed fermion propagator $S(\mathbf{p},\omega)$, let us now
consider the inverse fermion propagator,
\begin{eqnarray}\label{App7a}
S^{-1}(\mathbf{p},\omega)=S_{0}^{-1}(\mathbf{p},\omega)+\Sigma^{f}(\mathbf{p},\omega),
\end{eqnarray}
where $\Sigma(\mathbf{p},\omega)$ is the fermion self-energy,
including all one-particle irreducible radiative corrections,
corresponding to the two-point Green's function of fermions.
Decomposing now $\Sigma^{f}$ as
\begin{eqnarray}\label{App8a}
\Sigma^{f}(\mathbf{p},\omega)=\hat{g}'_{-}\Sigma_{+}(\omega,\mathbf{p})-\hat{g}'_{+}\Sigma_{-}(\omega,\mathbf{p}),
\end{eqnarray}
and combining the resulting expression with (\ref{App6a}), we arrive,
according to (\ref{App7a}), at
\begin{eqnarray}\label{App9a}
\lefteqn{\hspace{-0.7cm}S^{-1}(\mathbf{p},\omega)}\nonumber\\
&&\hspace{-0.7cm}=-\hat{g}'_{+}\left(\omega+\omega_{f}+\Sigma_{-}\right)-\hat{g}'_{-}\left(\omega-\omega_{f}-\Sigma_{+}\right).
\end{eqnarray}
Using the identities (\ref{App5a}) for $\hat{g}_{\pm}$ and
$\hat{g}'_{\pm}$, it is easy to show that $\Sigma_{\pm}$ from
(\ref{App8a}) is given by
\begin{eqnarray}\label{App10a}
\Sigma_{\pm}=\pm\frac{1}{2}\mbox{tr}\left(\hat{g}_{\pm}\Sigma^{f}\right).
\end{eqnarray}
Inverting now (\ref{App9a}), by making use of the properties
(\ref{App5a}), the dressed fermion propagator reads
\begin{eqnarray}\label{App11a}
\lefteqn{\hspace{-0.8cm}S(\mathbf{p},\omega)}\nonumber\\
&&\hspace{-1cm}
=-\frac{1}{\omega-\left(\omega_{f}+\Sigma_{+}\right)}\hat{g}_{+}-\frac{1}{\omega+\left(\omega_{f}+\Sigma_{-}\right)}\hat{g}_{-}.
\end{eqnarray}
Using at this stage the definition
$\rho_{f}=-2~\mathfrak{Im}[S_{R}]$, and introducing
\begin{eqnarray}\label{App12a}
E_{\pm}\equiv\omega_{f}+\mathfrak{Re}[\Sigma_{\pm}^{R}],
\end{eqnarray}
as well as
\begin{eqnarray}\label{App13a}
\Gamma_{\pm}\equiv\mathfrak{Im}[\Sigma_{\pm}^{R}],
\end{eqnarray}
we arrive at $\rho_{f}(\mathbf{p},\omega)$ from (\ref{A36b}). Let us
finally notice that $E_{\pm}$ and $\Gamma_{\pm}$ defined in
(\ref{A38b}), arise by plugging (\ref{App10a}) in (\ref{App12a}) and
(\ref{App13a}) and neglecting the imaginary part of $\hat{g}_{\pm}$,
defined in (\ref{A37b}).
\section{Computation of (\ref{G14a}) and (\ref{G19a})}\label{appB}
\setcounter{equation}{0}
\par\noindent
In this appendix, we will perform analytically the three-dimensional
$k$-integration in (\ref{G13a}) and (\ref{G18a}) to arrive at
(\ref{G14a}) and (\ref{G19a}), respectively. We also present the
final result for $\Gamma_{f}^{-}$.
\par
Let us start by considering the integral
\begin{eqnarray}\label{appC1a}
{\cal{I}}=\int
\frac{d^{3}k}{(2\pi)^{2}2\omega_{1}2\omega_{2}}\delta(\omega_{b}-\omega_{1}-\omega_{2})f(\omega_{b},\omega_{1},\omega_{2}),\nonumber\\
\end{eqnarray}
where $f(\omega_{b},\omega_{1},\omega_{2})$ is a generic function of
$\omega_{i}, i=b,1,2$. According to the definitions in Sec.
\ref{sec4a}, $\omega_{b}^{2}=\mathbf{p}^{2}+m_{b}^{2}$,
$\omega_{1}^{2}=\mathbf{k}^{2}+m_{f}^{2}$ and
$\omega_{2}^{2}=(\mathbf{k}-\mathbf{p})^{2}+m_{f}^{2}$. Denoting the
angle between $\mathbf{k}$ and $\mathbf{p}$ with $\theta_{p}$, and
inserting
\begin{eqnarray}\label{appC2a}
1=\frac{1}{2}\int d(\cos\theta_{p}),
\end{eqnarray}
in the integration over $k$, appearing in (\ref{appC1a}), we arrive
at
\begin{eqnarray}\label{appC3a}
\lefteqn{\hspace{-0.8cm}{\cal{I}}=\frac{1}{2}\int
\frac{d^{3}k}{(2\pi)^{2}2\omega_{1}}\frac{d(\cos\theta_{p})}{2\omega_{2}}
}\nonumber\\
&&~~~\times
\delta(\omega_{b}-\omega_{1}-\omega_{2})f(\omega_{b},\omega_{1},\omega_{2})
\nonumber\\
&&\hspace{-1cm}=\frac{1}{2}\int
\frac{d^{3}k}{(2\pi)^{2}2\omega_{1}}\int d\omega_{2}\left(\frac{d
\omega_{2}^{2}}{d(\cos\theta_{p})}\right)^{-1}\nonumber\\
&&~~~\times\delta(\omega_{b}-\omega_{1}-\omega_{2})f(\omega_{b},\omega_{1},\omega_{2})\nonumber\\
&&\hspace{-1cm}=-\frac{1}{8\pi|\mathbf{p}|}\int
d\omega_{1}f(\omega_{b},\omega_{1},\omega_{2}=\omega_{b}-\omega_{1}).
\end{eqnarray}
To derive the above relation, the identity
\begin{eqnarray}\label{appC4a}
\omega_{2}^{2}=\omega_{1}^{2}+\mathbf{p}^{2}-2|\mathbf{p}||\mathbf{k}|\cos\theta_{p},
\end{eqnarray}
arising from the definition of $\omega_{2}$ in terms of $\mathbf{p}$
and $\mathbf{k}$, is used. The latter identity can also be used to
determine the range of integration over $\omega_{1}$ in
(\ref{appC3a}). Having in mind that
\begin{eqnarray}\label{appC5a}
-1\leq
\cos\theta_{p}=\frac{\omega_{1}^{2}+\mathbf{p}^{2}-\omega_{2}^{2}}{2
|\mathbf{k}||\mathbf{p}|}\leq +1,
\end{eqnarray}
we arrive at
\begin{eqnarray}\label{appC6a}
\omega_{1}^{2}-\omega_{1}\omega_{b}+\frac{m_{b}^{4}+4m_{f}^{2}\mathbf{p}^{2}}{4m_{b}^{2}}\leq
0,
\end{eqnarray}
whose solution yields $\alpha^{-}_{b}\leq
\omega_{1}\leq\alpha^{+}_{b}$, with
\begin{eqnarray}\label{appC7a}
\alpha^{\pm}_{b}\equiv
\frac{1}{2}\left(\omega_{b}\pm\frac{|\mathbf{p}|}{\xi}\sqrt{\xi^{2}-4}\right),
\end{eqnarray}
and $\xi=\frac{m_{b}}{m_{f}}$. Plugging
\begin{eqnarray}\label{appC8a}
\lefteqn{f(\omega_{b},\omega_{1},\omega_{2})}\nonumber\\
&&
=\frac{g^{2}(4m_{f}^{2}-m_{b}^{2})}{4\omega_{b}}\frac{\sinh(\frac{\beta\omega_{b}}{2})}{\cosh(\frac{\beta\omega_{1}}{2})
\cosh(\frac{\beta\omega_{2}}{2})},
\end{eqnarray}
from (\ref{G13a}) in the expression on the r.h.s. of (\ref{appC3a}),
we arrive after some straightforward manipulations at (IV.14).
\par
To derive (\ref{G19a}), let us now consider (\ref{G18a}), where in
contrast to the previous case two $\delta$-functions
$\delta(\omega_{f}\mp\omega_{1}\pm\omega_{2})$ contribute to
$\Gamma_{+}$. Having in mind that in the fermionic case
$\omega_{f}^{2}=\mathbf{p}^{2}+m_{f}^{2}$,
$\omega_{1}^{2}=\mathbf{k}^{2}+m_{f}^{2}$ and
$\omega_{2}^{2}=(\mathbf{k}-\mathbf{p})^{2}+m_{b}^{2}$, we obtain
$\omega_{2}^{2}=\omega_{1}^{2}+\mathbf{p}^{2}-
2|\mathbf{p}||\mathbf{k}|\cos\theta_{p}+m_{b}^{2}-m_{f}^{2}$.
Following now the same steps leading from (\ref{appC1a}) to
(\ref{appC3a}), we arrive at
\begin{eqnarray}\label{appC9a}
\lefteqn{\hspace{-0.5cm}\int
\frac{d^{3}k}{(2\pi)^{2}2\omega_{1}2\omega_{2}}\delta(\omega_{f}\mp\omega_{1}\pm\omega_{2})f(\omega_{f},\omega_{1},\omega_{2})
}\nonumber\\
&&\hspace{-0.2cm}=-\frac{1}{8\pi|\mathbf{p}|}\int
d\omega_{1}f(\omega_{f},\omega_{1},\omega_{2}=\omega_{1}\mp
\omega_{f}).
\end{eqnarray}
As concerns the range of integration over $\omega_{1}$, we can use
\begin{eqnarray*}
-1\leq
\cos\theta_{p}=\frac{\omega_{1}^{2}-\omega_{2}^{2}+\mathbf{p}^{2}+m_{b}^{2}-m_{f}^{2}}{2
|\mathbf{k}||\mathbf{p}|}\leq +1,\nonumber\\
\end{eqnarray*}
to get
\begin{eqnarray}\label{appC10a}
\hspace{-0.5cm}\omega_{1}^{2}\pm(\xi^{2}-2)\omega_{f}\omega_{1}+\mathbf{p}^{2}+\frac{m_{f}^{2}}{4}(\xi^{2}-2)^{2}\leq
0.
\end{eqnarray}
Here, the $\pm$ signs before the second term correspond to
$\omega_{2}=\omega_{1}\mp \omega_{f}$, respectively. Solving the
above equation, we arrive for $\omega_{2}=\omega_{1}-\omega_{f}$ at
$m_{f}\leq\omega_{1}\leq \alpha_{f}^{+}$, with
\begin{eqnarray}\label{appC11a}
\alpha_{f}^{+}\equiv
\frac{-\omega_{f}(\xi^{2}-2)+|\mathbf{p}|\xi\sqrt{\xi^{2}-4}}{2},
\end{eqnarray}
and for $\omega_{2}=\omega_{1}+\omega_{f}$ at $\beta_{f}^{-}\leq
\omega_{1}\leq \beta_{f}^{+}$, with
\begin{eqnarray}\label{appC12a}
\beta_{f}^{\pm}\equiv\frac{\omega_{f}(\xi^{2}-2)\pm|\mathbf{p}|\xi\sqrt{\xi^{2}-4}}{2}.
\end{eqnarray}
Plugging (\ref{appC11a}) and (\ref{appC12a}) in (\ref{appC9a}), and
using the resulting expression, the three-dimensional
$k$-integration in (\ref{G18a}) can be performed analytically. We
arrive after some algebra at (\ref{G19a}).
\par
To evaluate $\Gamma_{f}^{-}$ from (\ref{G18a}), we follow the same
procedure as above. Using
\begin{eqnarray}\label{appC13a}
\lefteqn{\int
\frac{d^{3}k}{(2\pi)^{2}2\omega_{2}}[\delta(\omega_{f}-\omega_{1}+\omega_{2})+\delta(\omega_{f}+\omega_{1}-\omega_{2})]}\nonumber\\
&&\times
f(\omega_{f},\omega_{1},\omega_{2})\nonumber\\
&&=-\frac{1}{4\pi|\mathbf{p}|}
\bigg[\int_{m_{f}}^{\alpha_{f}^{+}}d\omega_{1}\omega_{1}f(\omega_{f},\omega_{1},
\omega_{2}=\omega_{1}-\omega_{f})\nonumber\\
&&+\int_{\beta_{f}^{-}}^{\beta_{f}^{+}}d\omega_{1}\omega_{1}
f(\omega_{f},\omega_{1},\omega_{2}=\omega_{1}+\omega_{f})\bigg],
\end{eqnarray}
with
\begin{eqnarray*}
f(\omega_{f},\omega_{1},\omega_{2})=\frac{g^{2}}{4}\frac{\cosh(\frac{\beta\omega_{f}}{2})}{\cosh(\frac{\beta\omega_{1}}{2})\sinh(\frac{\beta\omega_{2}}{2})},
\end{eqnarray*}
and
\begin{eqnarray*}
\lefteqn{\int du u (\coth u)^{\pm
1}}\nonumber\\
&&=\frac{1}{2}\bigg[u\left(u+2\ln (1\mp
e^{-2u})\right)-\mbox{Li}_{2}(\pm e^{-2u})\bigg],
\end{eqnarray*}
we arrive at
\begin{eqnarray}\label{appC14a}
\lefteqn{\Gamma_{f}^{-}=-\frac{g^{2}T}{8\pi\kappa_{f}\sqrt{1-\gamma_{f}^{2}}}
}\nonumber\\
&&\times\bigg\{
\kappa_{f}\ln\bigg[\frac{1-\cosh(2~\Xi_{-})}{\cosh(\Upsilon_{-}+\Xi_{+})-\cosh(\Upsilon_{-}-\Xi_{+})}\bigg]
\nonumber\\
&&+[u(u+2\ln(1-e^{-2u}))-\mbox{Li}_{2}(e^{-2u})]\bigg|_{\Upsilon_{-}}^{-\Xi_{-}}\nonumber\\
&&+[u(u+2\ln(1-e^{-2u}))-\mbox{Li}_{2}(e^{-2u})]\bigg|_{\Xi_{-}}^{\Xi_{+}}\nonumber\\
&&-[u(u+2\ln(1+e^{-2u}))-\mbox{Li}_{2}(-e^{-2u})]\bigg|_{\Upsilon_{-}+\frac{\kappa_{f}}{2}}^{-\Xi_{-}+\frac{\kappa_{f}}{2}}\nonumber\\
&&-[u(u+2\ln(1+e^{-2u}))-\mbox{Li}_{2}(-e^{-2u})]\bigg|_{\Xi_{-}-\frac{\kappa_{f}}{2}}^{\Xi_{+}-\frac{\kappa_{f}}{2}}\bigg\},\nonumber\\
\end{eqnarray}
where $\kappa_{f}$, $\Xi_{\pm}$ and $\Upsilon_{\pm}$ are defined
below (\ref{G19a}) and in (\ref{G20a}).
\section{Shear viscosity and spectral width of fermions for non-vanishing chemical potential}\label{appC}
\setcounter{equation}{0}
\par\noindent
In this appendix, we will first determine the fermionic spectral
widths $\Gamma_{\pm}$ and shear viscosity $\eta_{f}$ for
non-vanishing chemical potential. To do this, we will follow the
same method, described in Sec. \ref{sec3b} and App. \ref{appA}. We
will then use the method presented in Sec. \ref{sec4} and App.
\ref{appB}, and derive the one-loop contribution to the bosonic and
fermionic spectral widths for non-vanishing temperature and chemical
potential.
\subsection{Fermionic contribution to $\eta_{f}$ for $\mu\neq 0$}\label{appCseca}
\par\noindent
In what follows, we will show that in the one-loop skeleton expansion, the fermionic part of the shear viscosity $\eta_{f}$, is given by
\begin{eqnarray}\label{appD1a}
\eta_{f}&\sim&\frac{2\beta}{15\pi^{2}}\int_{0}^{\infty} dp
\frac{\mathbf{p}^{4}}{\omega_{f}^{2}}\sum_{s=\pm}\bigg\{\frac{e^{\beta
({\cal{E}}_{s}-s\mu)}}{(e^{\beta ({\cal{E}}_{s}-s\mu)}+1)^{2}}\nonumber\\
&& \times
\bigg[\frac{\mathbf{p}^{2}}{\Gamma_{s}}-\frac{4m_{f}^{2}(\Gamma_{f}^{+}-\Gamma_{s})}{[{\cal{E}}_{f}+is\Gamma_{f}^{+}][{\cal{E}}_{f}+i\Gamma_{f}^{-}]}\bigg]\bigg\},
\end{eqnarray}
where ${\cal{E}}_{f}={\cal{E}}_{+}+{\cal{E}}_{-}$ and
$\Gamma_{f}^{\pm}=\Gamma_{+}\pm\Gamma_{-}$, similar to the
definitions in (\ref{A48b}). Here, in contrast to $E_{\pm}$ defined
in (\ref{A38b}), ${\cal{E}}_{\pm}$ appearing in ${\cal{E}}_{f}$ are
given by
\begin{eqnarray}\label{appD2a}
{\cal{E}}_{\pm}(\mathbf{p},\omega_{\pm})\equiv \omega_{\pm}\pm\frac{1}{2}
\mbox{tr}\left(\hat{g}_{\pm}(\mathbf{p},\omega_{f})\mathfrak{Re}[\Sigma_{R}^{f}(\mathbf{p},\omega_{f})]\right),\nonumber\\
\end{eqnarray}
where $\omega_{\pm}\equiv \omega_{f}\pm\mu$. To derive
(\ref{appD1a}), we start, as in App. \ref{appA}, with the
K\"allen-Lehmann representation of the free fermion propagator in
terms of the free spectral density function, $\rho_{f}^{0}$,
\begin{eqnarray}\label{appD3a}
S_0(\mathbf{p},\omega)=
\int_{-\infty}^{\infty}\frac{dp_0}{2\pi}\frac{\rho_{f}^{0}(\mathbf{p},p_{0})}{p_0+\mu
-\omega},
\end{eqnarray}
where $\rho_{f}^{0}(\mathbf{p},p_{0})$ is defined in (\ref{App2a}).
Integrating over $p_{0}$, we arrive at a decomsposition, similar to
what is demonstrated in (\ref{App3a}),
\begin{eqnarray}\label{appD4a}
S_0(\mathbf{p},\omega)=-\frac{1}{\omega -\omega_{+}}\hat{g}_{+}
-\frac{1}{\omega +\omega_{-}}\hat{g}_{-}.
\end{eqnarray}
Here, $\omega_{\pm}=\omega_{f}\pm\mu$ and $\hat{g}_{\pm}$ are
defined in (\ref{A37b}). Following now the same steps as described
in App. \ref{appA}, we arrive first at the dressed fermion
propagator for non-vanishing $\mu$,
\begin{eqnarray}\label{appD5a}
S(\mathbf{p},\omega)=-\frac{1}{\omega -\left(\omega_+ +\Sigma_+\right)}\hat{g}_{+}-\frac{1}{\omega +\left(\omega_{-} +\Sigma_{-}\right)}\hat{g}_{-},\nonumber\\
\end{eqnarray}
where $\Sigma_{\pm}$ are given in (\ref{App10a}). Using at this stage $\rho_{f}=-2~\mathfrak{Im}[S_{R}]$, we arrive at
\begin{eqnarray}\label{appD6a}
\lefteqn{\hspace{-0.8cm}\rho_{f}(\mathbf{p},\omega)=\frac{2\Gamma_{+}(\mathbf{p},\omega_{f})}{[\omega-{\cal{E}}_{+}(\mathbf{p},\omega_{f}]^{2}+\Gamma_{+}^{2}
(\mathbf{p},\omega_{f})}\hat{g}_{+}(\mathbf{p},\omega_{f})
}\nonumber\\
&&\hspace{-1cm}-\frac{2\Gamma_{-}(\mathbf{p},\omega_{f})}{[\omega+{\cal{E}}_{-}(\mathbf{p},\omega_{f})]^{2}+\Gamma_{-}^{2}(\mathbf{p},\omega_{f})}\hat{g}_{-}(\mathbf{p},\omega_{f}),
\end{eqnarray}
with  ${\cal{E}}_{\pm}$ defined in (\ref{appC2a}) and $\Gamma_{\pm}$
in (\ref{A38b}). Plugging then the standard representation
\begin{eqnarray}\label{appD7a}
S_{T}(\mathbf{p},\omega_{n})=\frac{1}{2\pi}\int_{-\infty}^{+\infty}d\omega\frac{\rho_{f}(\mathbf{p},\omega)}{i\omega_{n}-\omega+\mu},
\end{eqnarray}
in (\ref{A29b}) and performing the summation over Matsubara
frequencies $\omega_{n}$, we arrive at
\begin{eqnarray}\label{appD8a}
\lefteqn{S_{T}(\mathbf{p},\tau)=\frac{1}{2\pi}\int_{-\infty}^{+\infty}
d\omega e^{(\mu -\omega) \tau}\rho_{f}(\mathbf{p},\omega)
}\nonumber\\
&&\times
\bigg(\theta(-\tau)n_f^+(\omega)-\theta(\tau)(1-n_f^+(\omega))
\bigg),
\end{eqnarray}
that replaces (\ref{A30b}). Here, fermionic distribution functions,
including $\mu$, are defined by
\begin{eqnarray}\label{appD9a}
n_{f}^{\pm}(\omega)\equiv \frac{1}{e^{\beta(\omega\mp\mu)}+1}.
\end{eqnarray}
Plugging now $S_{T}(\mathbf{p},\tau)$ from (\ref{appD8a}) in
(\ref{A9b}), and following the same steps leading from (\ref{A33b})
to (\ref{A49b}), we arrive at $\eta_{f}[\Gamma_{\pm}]$ from
(\ref{appD1a}).
\subsection{Bosonic and fermionic spectral widths for $\mu\neq 0$}\label{appCsecb}
\par\noindent
To determine the one-loop contributions to $\Gamma_{b}$ and
$\Gamma_{\pm}$ for non-vanishing chemical potential, we will follow
the same method as described in Sec. \ref{sec4},  and will compute
the imaginary part of the one-loop bosonic and fermionic self-energy
diagrams, using Schwinger-Keldysh real-time formalism
\cite{schwinger1961}. Since the chemical potential is only
introduced for fermions, the free propagator of scalar bosons
remains unchanged [see (\ref{G1a}) and (\ref{G2a})]. As concerns the
free fermion propagator, it is given for non-vanishing $\mu$ by
\begin{eqnarray}\label{appD10a}
{\cal{S}}=\left(
\begin{array}{cc}
S_{++}&S_{+-}\\
S_{-+}&S_{--}
\end{array}
\right),
\end{eqnarray}
with $S_{ab}, a,b=\pm$ slightly different from (\ref{G4a}),
\begin{widetext}
\begin{eqnarray}\label{appD11a}
S_{++}(p)&=&=\left(\gamma\cdot p+m_{f}\right)\left(-\frac{i}{p^2-m_{f}^2+i\epsilon}+2\pi \delta(p^2-m_{f}^2)[\theta(p_0)n_f(x_p)+ \theta(-p_0)n_f(-x_p)]\right),\nonumber\\
S_{+-}(p)&=&-2\pi \left(\gamma\cdot p+m_{f}\right)\big[\theta(-p_0)\left(1-n_f(-x_p)\right)-\theta(p_0)n_f(x_p)\big],\nonumber\\
S_{-+}(p)&=&-2\pi
\left(\gamma\cdot p+m_{f}\right)\big[\theta(p_0)\left(1-n_f(x_p)\right)-\theta(-p_0)n_f(-x_p)\big],\nonumber\\
S_{--}(p)&=&=\left(\gamma\cdot
p+m_{f}\right)\left(\frac{i}{p^2-m_{f}^2-i\epsilon}+2\pi
\delta(p^2-m_{f}^2)[\theta(p_0)n_f(x_p)+
\theta(-p_0)n_f(-x_p)]\right),
\end{eqnarray}
where $x_{p}$ is defined by $x_{p}\equiv p_{0}+\mu$ and
$n_{f}(\omega)$ is given in (\ref{A31b}). According to (\ref{G9a}),
the bosonic spectral width, $\Gamma_{b}$, is given by the imaginary
part of $\Sigma_{R}^{b}$. At one-loop level,
$\mathfrak{Im}[\Sigma_{R}^{b}(p)]$ is given in (\ref{G10a}). Using
$S_{ab}, a,b=\pm$ from (\ref{appD11a}), we arrive  at
\begin{eqnarray}\label{appD12a}
\lefteqn{\Gamma_b(\mathbf{p},\omega_{b})=\frac{g^{2}}{8\omega_{b}}\int
\frac{d^{3}k}{(2\pi)^{2}}\frac{(4m_{f}^{2}-m_{b}^{2})}{\omega_{1}\omega_{2}}
}\nonumber\\
&&\times\bigg\{\delta(\omega_{b}-\omega_{1}-\omega_{2})[1-n_{f}^{-}(\omega_{1})-n_{f}^{+}(\omega_{2})]+\delta(\omega_{b}-\omega_{1}+\omega_{2})[n_{f}^{-}(\omega_{1})-n_{f}^{-}(\omega_{2})]\nonumber\\
&&~~-\delta(\omega_{b}+\omega_{1}-\omega_{2})[n_{f}^{+}(\omega_{1})-n_{f}^{+}(\omega_{2})]-\delta(\omega_{b}+\omega_{1}+\omega_{2})[1-n_{f}^{+}(\omega_{1})-n_{f}^{-}(\omega_{2})]\bigg\}.
\end{eqnarray}
Here, $n_{f}^{\pm}$ are defined in (\ref{appD9a}). Following the
same steps leading from (\ref{G11a}) to (\ref{G13a}), we arrive
after some work first at
\begin{eqnarray}\label{appD13a}
\Gamma_b(\mathbf{p},\omega_{b})=\frac{g^{2}(4m_{f}^{2}-m_{b}^{2})}{16\omega_{b}}\int
\frac{d^{3}k}{(2\pi)^{2}}\frac{\sinh(\frac{\beta
\omega_{b}}{2})}{\cosh(\frac{\beta(\omega_1+\mu)}{2})\cosh(\frac{\beta(\omega_2-\mu)}{2})}
\frac{\delta(\omega_{b}-\omega_1 -\omega_2)}{\omega_{1}\omega_{2}},
\end{eqnarray}
and finally, after performing the integration over $k$, using the
method demonstrated in App. \ref{appB},  at
\begin{eqnarray}\label{appD14a}
\Gamma_b(\mathbf{p},\omega_{b})=\frac{g^{2}T}{16\pi}\frac{\gamma_{b}^{2}(\xi^{2}-4)}{\xi^{2}\sqrt{1-\gamma_{b}^{2}}}
\ln\bigg[\frac{\cosh(\tau_{f})+\cosh\frac{\kappa_{b}}{2}(1+\frac{1}{\xi}\sqrt{(\xi^{2}-4)(1-\gamma_{b}^{2})})}
{\cosh(\tau_{f})+\cosh\frac{\kappa_{b}}{2}(1-\frac{1}{\xi}\sqrt{(\xi^{2}-4)(1-\gamma_{b}^{2})})}\bigg],
\end{eqnarray}
where apart from $\xi,\kappa_{b}, \gamma_{b}$ which are defined
below (\ref{G14a}), $\tau_{f}\equiv \mu/T$.
\par
As concerns the one-loop contribution to the fermionic spectral
widths $\Gamma_{\pm}$ from (\ref{A38b}), let us consider
$\mathfrak{Im}[\Sigma_{R}^{f}]$ from (\ref{G15a}). Using $G_{ab}$
and $S_{ab}$, $a,b=\pm$ from (\ref{G2a}) and (\ref{appC11a}), we
arrive first at
\begin{eqnarray}\label{appD15a}
\lefteqn{\hspace{-1cm}\Gamma_{\pm}(\mathbf{p},\omega_{f})=\pm\frac{g^2}{8\omega_{f}}\int
\frac{d^3 k}{(2\pi)^2}\frac{1}{\omega_1 \omega_2} }\nonumber\\
&&\hspace{-0.9cm}\times\bigg[[\omega_{f}\omega_1
\mp\mathbf{p}\cdot\mathbf{k}\pm
m_f^2]\big\{\delta (\omega_{f}-\omega_1 -\omega_2)[1-n_f^{-}(\omega_1)+n_b(\omega_2)]+\delta(\omega_{f}-\omega_1 +\omega_2)[n_f^{-}(\omega_1)+n_b(\omega_2)]\big\}\nonumber\\
&&\hspace{-0.9cm} +[\omega_{f}\omega_1
\pm\mathbf{p}\cdot\mathbf{k}\mp m_f^2] \big\{\delta
(\omega_{f}+\omega_1 +\omega_2)[1-n_f^{+}(\omega_1)+n_b(\omega_2)]
+\delta(\omega_{f}+\omega_1-\omega_2)[n_f^{+}(\omega_1)+n_b(\omega_2)]\big\}
\bigg],
\end{eqnarray}
\end{widetext}
with $n_{f}^{\pm}(\omega)$ and $n_{b}(\omega)$ defined in
(\ref{appC9a}) and (\ref{A15b}), respectively. Following the same
arguments, as described in Sec. \ref{sec4b},
 the relevant expression of $\Gamma_{+}$ for non-vanishing $\mu$ is given by
\begin{eqnarray}\label{appD16a}
\lefteqn{\hspace{-0.5cm}\Gamma_{+}(\mathbf{p},\omega_{f}) }\nonumber\\
&&\hspace{-0.5cm}=\frac{g^{2}}{32\omega_{f}}\int \frac{d^3 k}{(2
\pi)^2}\frac{(4m_{f}^{2}-m_{b}^{2})}{\omega_{1}\omega_{2}}\frac{\cosh(\frac{\beta
(\omega_{f}+\mu)}{2})}{\sinh(\frac{\beta \omega_2}{2})}
\nonumber\\
&&\hspace{-0.5cm}\times\bigg\{\frac{\delta(\omega_{f}-\omega_{1}+\omega_{2})}{\cosh(\frac{\beta
(\omega_1+\mu)}{2})}
-\frac{\delta(\omega_{f}+\omega_{1}-\omega_{2})}{\cosh(\frac{\beta
(\omega_1-\mu)}{2})}\bigg\}.
\end{eqnarray}
Performing the three-dimensional integration over $k$, using the
method described in App. \ref{appB}, we finally arrive at
$\Gamma_{+}$ in term of dimensionless variables
$\xi,\gamma_{f},\kappa_{f}$ and $\tau_{f}$, defined in Sec.
\ref{sec4b},
\begin{eqnarray}\label{appD17a}
\lefteqn{\Gamma_{+}(\xi, \gamma_{f},\kappa_{f},
\tau_{f};T)=\frac{g^{2}T}{32\pi}\frac{\gamma_{f}^{2}(\xi^{2}-4)}{\sqrt{1-\gamma_{f}^{2}}}
}\nonumber\\
&&\times
\bigg\{\ln\bigg[\frac{1-\cosh(2~\Xi_{-})}{\cosh(\Upsilon_{-}+\Xi_{+})-
\cosh(\Upsilon_{-}-\Xi_{+})}\bigg]\nonumber\\
&&-\ln\bigg[
\frac{1+\cosh(2~\Xi_{-}-(\kappa_{f}+\tau_{f}))}{\cosh(\Upsilon_{-}+\Xi_{+})+
\cosh(\Upsilon_{+}-\Xi_{+}+\tau_{f})}
\bigg]\bigg\}.\nonumber\\
\end{eqnarray}
Here, $\Xi_{\pm}$ and $\Upsilon_{\pm}$ are defined in (\ref{G20a}).
The difference between $\Gamma_{+}$ and $\Gamma_{-}$ is, according
to (\ref{A48b}), defined by $\Gamma_{f}^{-}=\Gamma_{+}-\Gamma_{-}$.
For non-vanishing $\mu$, $\Gamma_{f}^{-}$ is first given by
\begin{eqnarray}\label{appD18a}
\lefteqn{ \hspace{-1cm}
\Gamma_{f}^{-}(\mathbf{p},\omega_{f})=\frac{g^2}{8}\int \frac{d^3
k}{(2\pi)^{2}\omega_{2}} \frac{\cosh(\frac{\beta
(\omega_{f}+\mu)}{2})}{\sinh(\frac{\beta\omega_{2}}{2})}
}\nonumber\\
&&\hspace{-0.8cm}\times
\bigg\{\frac{\delta(\omega_{f}-\omega_{1}+\omega_{2})}{\cosh(\frac{\beta
(\omega_1+\mu)}{2})}
+\frac{\delta(\omega_{f}+\omega_{1}-\omega_{2})}{\cosh(\frac{\beta
(\omega_1-\mu)}{2})}\bigg\},
\end{eqnarray}
and, after integrating over the three-momentum $k$, using the method
described in App. \ref{appB}, it reads
\begin{widetext}
\begin{eqnarray}\label{appD19a}
\lefteqn{\Gamma_{f}^{-}=-\frac{g^{2}T}{8\pi\kappa_{f}\sqrt{1-\gamma_{f}^{2}}}\bigg\{
\kappa_{f}\ln\bigg[\frac{1-\cosh(2~\Xi_{-})}{\cosh(\Upsilon_{-}+\Xi_{+})-\cosh(\Upsilon_{-}-\Xi_{+})}\bigg]
}\nonumber\\
&&+\tau_{f}\ln\bigg[\frac{1+\cosh(2~\Xi_{-}-(\kappa_{f}+\tau_{f}))}{\cosh(\Upsilon_{-}+\Xi_{+})+
\cosh(\Upsilon_{+}-\Xi_{+}+\tau_{f})}\bigg]\nonumber\\
&&+[u(u+2\ln(1-e^{-2u}))-\mbox{Li}_{2}(e^{-2u})]\bigg|_{\Upsilon_{-}}^{-\Xi_{-}}+[u(u+2\ln(1-e^{-2u}))-\mbox{Li}_{2}(e^{-2u})]\bigg|_{\Xi_{-}}^{\Xi_{+}}\nonumber\\
&&-[u(u+2\ln(1+e^{-2u}))-\mbox{Li}_{2}(-e^{-2u})]\bigg|_{\Upsilon_{-}+\frac{(\kappa_{f}+\tau_{f})}{2}}^{-\Xi_{-}+\frac{(\kappa_{f}+\tau_{f})}{2}}
-[u(u+2\ln(1+e^{-2u}))-\mbox{Li}_{2}(-e^{-2u})]\bigg|_{\Xi_{-}-\frac{(\kappa_{f}+\tau_{f})}{2}}^{\Xi_{+}-\frac{(\kappa_{f}+\tau_{f})}{2}}\bigg\}.\nonumber\\
\end{eqnarray}
\end{widetext}

\end{appendix}

\end{document}